\documentclass[aps,prb,twocolumn,groupedaddress,showpacs,floatfix,altaffilletter,longbibliography,includegraphics]{revtex4-2}
\usepackage{graphicx}
\usepackage{grffile}
\usepackage{amsmath}
\usepackage{amssymb}
\usepackage{bm}
\usepackage{color}
\usepackage[dvipsnames]{xcolor}
\usepackage{amsmath,amssymb,amsfonts}
\usepackage{epsfig}
\usepackage{times}
\usepackage[colorlinks,bookmarks=false,citecolor=blue,linkcolor=red,urlcolor=blue]{hyperref}
\usepackage{gensymb}
\usepackage[toc,page]{appendix}
\usepackage{float}
\usepackage{comment}

\counterwithout{table}{section}

\newcommand{\secref}[1]{Sec.~(\ref{#1})}

\newcommand{\fref}[1]{Fig.~\ref{#1}}
\newcommand{\eref}[1]{Eq.~(\ref{#1})}

\begin{document}

\title{True vs false Fermi surfaces in the Pseudogap regime and their transformation with doping and temperature in the Hubbard Model.} 

\author{ Y.M. Vilk}
\address{Retired}
\date{\today}

\begin{abstract}
The recent availability of exact diagrammatic quantum Monte Carlo results in Matsubara frequency for the nearest-neighbor Hubbard model motivates further studies to clarify the physics and the real-frequency behavior of the pseudogap in that model. Thus, using the improved two-particle self-consistent approach (TPSC+), we analyze real-frequency Energy (EDC) and Momentum Distribution Curves (MDC) simultaneously. We demonstrate that standard Fermi liquid terminology breaks down in the pseudogap regime, necessitating a distinction between true and false Fermi surfaces and zero-energy quasiparticle (ZEQ) lines. A false Fermi surface exhibits a momentum-space spectral maximum but a frequency-space depression at zero energy, while false ZEQ lines violate the standard quasiparticle condition $\partial \Sigma'(\mathbf{k},\omega)/\partial \omega \vert{}_{\omega=0} < 0$.

The pseudogap is driven by critical thermal spin fluctuations, which occur in two dimensions because the Mermin-Wagner theorem prohibits a finite-temperature phase transition. Commensurate fluctuations open an antinodal pseudogap first, leaving true Fermi arcs behind. As $T$ decreases, the Fermi surface (FS) evolves into distinct hole- and electron-like false Fermi surfaces at low temperatures. Incommensurate fluctuations generate hot spots near the diagonal, where the pseudogap persists down to the quantum critical point (QCP). By contrast, at the antinodal point $\mathbf{k}_{AN}$, the pseudogap ceases to exist before the QCP doping is reached. The spectral function at $\mathbf{k}=\mathbf{k}_{AN}$ remains non-trivial, featuring two precursor antiferromagnetic (AFM) bands, both of which are located in the unoccupied (invisible) part of the spectrum ($\omega > 0$).

We benchmark our results against the diagrammatic quantum Monte Carlo (DiagMC) studies of the Matsubara spectral function proxy $-\text{Im} [\mathcal{G}(\mathbf{k}, i\pi T)]/\pi$. Relative to DiagMC \cite{vsimkovic2024origin}, TPSC+ tends to underestimate the severity of pseudogap suppression in the Matsubara proxy in the strong-interaction regime, reproducing the DiagMC behavior at somewhat lower temperatures or reduced doping. We also demonstrate that the Matsubara proxy is equivalent to the spectral function evaluated with a large thermal broadening $\eta=\pi T$. This broadening obscures important real-frequency features in the pseudogap regime, where $\pi T$ is not the smallest energy scale. By evaluating the spectral function in the zero-frequency limit, $A(\mathbf{k},0)$, we show that hole-like Fermi surfaces emerge at any interaction strength, provided the temperature is sufficiently low, rather than exclusively for strong interactions.
\end{abstract}

\maketitle

The pseudogap phenomenon, namely a momentum-dependent loss of spectral weight at the Fermi surface, has now been observed in many quasi-two-dimensional strongly correlated materials \cite{Kordyuk2015}, including high-$T_c$ superconductors \cite{Proust_Taillefer_2019, Vishik2018}. 
%In this paper, we investigate the evolution of the pseudogap (PG) state as a function of doping and temperature in the nearest-neighbor single-band Hubbard model using the TPSC+ approach. Notably, 
%The TPSC approach was the first to predict that {\it in two dimensions} a pseudogap driven by critical thermal spin fluctuations can appear in a wide temperature and doping range in the Hubbard model~\cite{Vilk1996, Vilk1997}. 

Recent advances in cold-atom quantum simulations of the Hubbard model \cite{chalopin2026observation,kendrick2026pseudogap} and diagrammatic quantum Monte Carlo (DiagMC) calculations \cite{Schaefer2021,vsimkovic2024origin} have convincingly verified the presence of the pseudogap in the nearest-neighbor Hubbard model. These recent advances make it timely to revisit the pseudogap problem {\it for this model} using the TPSC+ approach, which improves upon standard TPSC by incorporating feedback from single-particle spectra into the spin and charge susceptibilities. The original TPSC approach was the first to predict that, {\it in two dimensions}, a pseudogap driven by critical thermal spin fluctuations can emerge over a wide range of temperatures and dopings in the Hubbard model~\cite{Vilk1996,Vilk1997}. The improved TPSC+ approach was recently used to explain \cite{vilk2026pseudogap} cold-atom simulation results showing a maximum in the isothermal compressibility $\kappa(\delta)$ as a function of doping $\delta$ \cite{kendrick2026pseudogap}. Furthermore, TPSC+ successfully captured the temperature-dependent maximum in uniform magnetic susceptibility (Knight shift) observed in cold-atom experiments and quantum Monte Carlo simulations \cite{chalopin2026observation}. Both phenomena were shown to be tied to the spectral transformation from the pseudogapped to the metallic regime. Additionally, the TPSC+ approach has been benchmarked against Monte Carlo simulations and cold-atom experiments, demonstrating quantitative agreement at weak-to-intermediate interaction strengths ($U \leq 3$) and qualitative agreement for stronger interactions. \cite{Schaefer2021,gauvin2023improved,vilk2024antiferromagnetic}. We observe a comparable level of accuracy in the present study. But the physics of the Mott transition within TPSC+ has not been investigated yet. 

Here, we limit our investigation to the model with only nearest-neighbor hopping. This allows us to benchmark the TPSC+ results directly against DiagMC calculations and to make predictions that can be tested in future cold-atom experiments or Monte Carlo simulations. 

For high-temperature superconductors, numerous theories regarding the origin and properties of the pseudogap have been proposed: 
Spin fluctuation theories~\cite{kampf1990pseudogaps,Vilk1996, Vilk1997, schmalian_microscopic_1999, Kyung_Hankkevych_2004, Hankevych_Kyung_2006, gauvin2022disorder, ye_location_2023, Ye_Chubukov_2023, kokkinis2025pseudogap, Yu_Krein_2025};
Resonating valence bonds~\cite{Anderson_1989, Baskaran_2000, Anderson_Lee_RVB_2004};
Effect of strong interactions, treated with cluster generalizations of dynamical mean-field theory for the Hubbard model~\cite{Maier_Jarrell_Pruschke_Keller_2000, jarrell2001phase, Civelli_Capone_Kancharla_Parcollet_Kotliar_2005, kyung2006pseudogap, Maier_2006, Macridin_Maier_2006, Imada_2007, LiebschTong:2009, Gull_Millis_2013, Gull_Parcollet_Millis_2013, sakai2023nonperturbative, Meixner_Menke_Klett_Heinzelmann_Andergassen_Hansmann_Schafer_2024, Sordi_Reaney_2025, tanaka2026pseudogap, fournier2026frenkel};
Orbital selective Mott transition~\cite{Ferrero_Cornaglia_De_Leo_Parcollet_Kotliar_Georges_2009, Gull_Ferrero_Parcollet_Georges_Millis_2010};
Signatures of the Mott transition away from half-filling~\cite{Sordi_Haule_Tremblay_2010, sordi2012pseudogap, Sordi_Semon_Haule_tremblay_2013, Fratino_Semon_Sordi_Tremblay_2016, Braganca_Sakai_Aguiar_Civelli_2018,  Reymbaut_2019, Fratino_Bag_Camjayi_Civelli_Rozenberg_2021, Walsh_Charlebois_Sémon_Sordi_Tremblay_2022}; 
The effect of frustration is important for the former physics~\cite{Schafer_Geles_Rost_Rohringer_Toschi_2015, Downey_Sordi_2024};
Gauge theories~\cite{Baskaran_1989, lee2006, Chowdhury_Sachdev, Wu_Scheurer_Chatterjee_Sachdev_Georges_Ferrero_2018, Bonetti_Christos_Nikolaenko_Patel_Sachdev_2025,Sachdev2020ancillaQbits};
Phenomenological approaches~\cite{norman1998phenomenology, Onoda_Imada_2000, YRZ_2006, Rice_Yang_Zhang_2011, Fisher_Nicol_Carbotte_2011, Verret_2017, Verret_Simard_2017, Singh_Huang_Xie_Okamoto_Chen_Watanabe_Fujimori_Imada_Huang_2022};
Hidden-Fermions~\cite{Yamaji_Imada_2011, Sakai_Imada_2016, Imada_Sakai_Civelli_2016, Imada_2019};
Broken symmetries~\cite{Gallais_Devereaux_2005,Hashimoto_He_Devereaux_2010, Vishik_Lee_He_Hashimoto_Hussain_Devereaux_Shen_2010, Pepin_de_Carvalho_Kloss_Montiel_2014, Hashimoto_Vishik_He_Devereaux_Shen_2014, atkinson2015charge, atkinson2015charge,  Hashimoto_Devereaux_2015, Kivelson_Symmetry_breaking_2019, White_Scalapino_2004}, including loop currents~\cite{varma1997non, Varma_2014, Varma2019} and nematicity~\cite{Kivelson_Fradkin_Emery_1998, Kivelson_nematic_2014, Scalapino_stripes_2014, Pseudogap_Huang_Devereaux_2023};
Phase fluctuations~\cite{franz1998phase,chen1998pairing, Scalapino_Hanke_2002, Efetov_Meier_Pepin_2013, dai2020modeling, wang2023phase, Maier_Scalapino_2019};
Direct quantum Monte Carlo simulations of the Hubbard model~\cite{Bulut_Scalapino_1994, White_Vekic_1993, Wang_Devereaux_2026, scalapino2007numerical}.
Monte Carlo simulations are in good agreement with the TPSC~\cite{Moukouri2000} pseudogap at half-filling for small interactions. 
Within our TPSC+ approach, a more realistic model for high-$T_c$ superconducting materials, incorporating second-neighbor hopping $t'$, will be addressed in a separate work. 

Our focus is on qualitative changes in the topology and shape of the Fermi surface (FS) and the zero-energy quasiparticle (ZEQ) line as functions of doping and temperature. The DiagMC method \cite{vsimkovic2024origin} provides highly reliable results in the Matsubara representation. To analyze physical properties, the authors of \cite{vsimkovic2024origin} utilized the Matsubara proxy at the lowest fermionic Matsubara frequency $ik_0=i\pi T$
\begin{equation}
    -\frac{1}{\pi}\text{Im} [\mathcal{G}(\mathbf{k}, ik_0)]=-\frac{1}{\pi}\text{Im}\left[\int d\omega \frac{A(\mathbf{k},\omega)}{ik_0-\omega}\right], \nonumber
\end{equation} 
that we call $A(\mathbf{k}, ik_0)$ for short, as a substitute for the zero-frequency spectral function $A(\mathbf{k},\omega=0)$. Using this proxy, they obtained several virtually exact results regarding the doping dependence of momentum distribution curves (MDC) (heat maps), the Fermi surface derived from the maxima of $A(\mathbf{k}, ik_0)$, and the ZEQ line obtained from $\epsilon(\mathbf{k}) + \Sigma'(\mathbf{k},ik_0) - \mu = 0$. However, they did not analyze the corresponding energy distribution curves (EDC), which, as we argue here, provide a novel and significant perspective on the evolution of the PG state with respect to doping and temperature. EDC analysis requires the analytical continuation of Matsubara-frequency data to real frequencies—a notoriously challenging task for Monte Carlo methods due to the extreme sensitivity of continuation procedures to statistical noise. In this work, we employ the TPSC+ approach \cite{Schaefer2021,gauvin2023improved,vilk2024antiferromagnetic,vilk2026pseudogap} to simultaneously analyze the evolution of MDC and EDC curves in the pseudogap state. A key advantage of this relatively simple self-consistent approach is its minimal computational error in the Matsubara representation, which allows for reliable analytical continuation to the real-frequency domain using the Padé approximation. 

Our complementary MDC and EDC analysis demonstrates the necessity of concepts that go beyond standard Fermi liquid terminology, such as ``Fermi surfaces'' and ``ZEQ lines''. While in a Fermi liquid state a maximum in the momentum dependence of the spectral function $A(\mathbf{k},0)$ implies a corresponding maximum in the frequency dependence $A(\mathbf{k}_{max},\omega)$ at $\omega=0$, this relationship generally breaks down in the PG state. Similarly, the ZEQ line—defined by

\begin{equation}
\label{eq:zero_energy_line}
\epsilon(\mathbf{k})+ \Sigma'(\mathbf{k},0) -\mu=0.
\end{equation}
—can exhibit a positive partial derivative of the real part of the self-energy within the PG regime, i.e., it violates the quasiparticle condition:

\begin{equation}
\label{eq:Sigma_deriv_neg}
 \frac{\partial \Sigma'(\mathbf{k},\omega)}{\partial \omega} \bigg|_{\omega=0} < 0 .
\end{equation}
Even when that condition is satisfied, one may argue that $\Sigma^{\prime\prime}\sim (\omega^2+(\pi T))^2$ should also be satisfied for a quasiparticle in the Landau sense. We do not require this, loosely using the term quasiparticle when conditions~\eref{eq:zero_energy_line} and \eref{eq:Sigma_deriv_neg} are satisfied. 

The violation of condition \eref{eq:Sigma_deriv_neg} implies an unphysical quasiparticle weight that is either larger than unity or negative when viewed through the lens of Fermi liquid theory. To address these complexities, we introduce the concepts of false Fermi surfaces and false ZEQ lines. We define ``false Fermi surfaces'', called gossamer Fermi surfaces in Refs.~\cite{xu2023bogoliubov,kokkinis2025pseudogap}, as curves in momentum space where the spectral function $A(\mathbf{k},0)$ exhibits a maximum, yet the frequency-dependent function $A(\mathbf{k}_{max},\omega)$ shows a minimum or depression at $\omega=0$. Furthermore, we define the ``false ZEQ'' line as the set of points where \eref{eq:zero_energy_line} is satisfied, but where the partial derivative of the real part of the self-energy is positive.

In addition to the analysis of real-frequency MDC and EDC curves as a function of doping and temperature, we compare our results with DiagMC results \cite{vsimkovic2024origin} (\secref{sec:Comp_with_DiagMC}). For the spectral function, the DiagMC results are only available for the Matsubara proxy $A(\mathbf{k}, ik_0)$. We compare them to the TPSC+ results for the same quantity. We also analyze the relationship between the Matsubara proxy $A(\mathbf{k}, ik_0)$ and the actual spectral function at zero real frequency $A(\mathbf{k}, 0)$. To better understand this relationship, we recall that the standard method for analytical continuation is to replace $ik_n \to \omega + i\eta$. For zero real frequency, this implies $ik_0 \to i\eta$. In other words, the proxy quantities can be viewed as an analytical continuation with a finite broadening $\eta = \pi T$. In particular, we found that analytical continuation via Padé approximation with $\eta=\pi T$ gives a virtually identical spectral function to the Matsubara proxy spectral function, $A(\mathbf{k}, 0)|_{\eta=\pi T}=A(\mathbf{k}, ik_0)$.

To analyze real-frequency physical properties, such as the spectral function $A(\mathbf{k}, \omega)$ and the density of states $N(\omega)$, we chose $\eta$ to be as small a fraction of $\pi T$ as possible ($\eta \ll \pi T$). Furthermore, in the PG regime, there exist other small energy scales, including $\gamma \xi^{-2} \ll \pi T$, $v \xi^{-1} \ll \pi T$, and $\Delta/\ln{\xi}$ (see \secref{sec:Analytical_results}), where $\gamma$ is the Landau damping, $v$ is the electron velocity, $\xi$ is the correlation length, and $\Delta$ is the gap parameter. In fact, the conditions $\gamma \xi^{-2} \ll T$ and $v_F \xi^{-1} \ll T$ define the criteria for the precursors of AFM bands at the antinodal Van Hove point \cite{vilk2023criteria,vilk2025_crossovers} and at regular points ($v_F > \gamma \xi^{-1}$) \cite{Vilk1996, Vilk1997}, respectively. {\it As a result, in the PG regime, $k_0=\pi T$ is not the smallest energy scale, and some of the results for the Matsubara-frequency proxy of the spectral function $A(\mathbf{k}, ik_0)$ are qualitatively different from their real-frequency counterpart $A(\mathbf{k}, 0)$.}

Most of the results in this paper were obtained numerically by solving the set of TPSC+ self-consistent equations. However, where feasible, we also provide analytical results based on the asymptotic expression for the self-energy in the pseudogap (PG) regime \cite{Vilk1997,vilk1997shadow}. These results, along with their implications for the evolution of true/false Fermi surfaces and ZEQ lines, are presented in \secref{sec:Analytical_results}.

The main advantage of the TPSC+ approach in this paper is that it enables the simultaneous analysis of energy and momentum distribution curves at various dopings and temperatures. Our main findings are as follows:

\begin{enumerate}

\item
The precursors of AFM/SDW bands and the pseudogap originate from thermal critical spin fluctuations below the renormalized classical temperature $T_{RC}(\delta)$, which decreases with doping. Because the Mermin-Wagner theorem forbids a finite-$T$ transition in 2D, no true AFM/SDW gap forms; instead, a pseudogap appears below $T_{RC}$ \cite{Vilk1996,Vilk1997,vilk1997shadow}. This is a low-dimensionality effect present for both weak and strong interactions, favored by larger $U$. For intermediate interactions, $U>6$, the antinodal pseudogap requires only $\xi\sim 1$ at half-filling, while for intermediate interaction strength, $U=3.5$, $\xi \sim 4$ \cite{vilk2025_crossovers,vilk2026pseudogap}.

\item
The criterion for the pseudogap at hot spots is satisfied when the spin correlation length exceeds the thermal de Broglie wavelength of the electron:

	\begin{equation} \label{eq:Criteria_R}
		\xi> \xi_{th\_db}=\frac{\hbar v_{\mathbf{k}_F +\mathbf{Q}}}{\pi k_B T}
	 \end{equation}

The conditions for the PG are most favorable at the momentum points with the smallest electron velocity, which correspond to the shortest thermal wavelength of the electron. Away from hot spots, the criterion \eref{eq:Criteria_R} is necessary but not sufficient. In addition, the gap parameter $\Delta$ must be large enough to push the lower AFM band to negative frequencies. Otherwise, AFM bands are present, but both are located in the invisible part of the spectrum, $\omega >0$.

\item
Since $T_{RC} \ll W$, Luttinger's theorem approximately holds for $T_{RC} < T \ll W$: the interacting FS closely matches the non-interacting one, $\epsilon(\mathbf{k})=\mu_1$, and $\Sigma'(\mathbf{k},0)$ is weakly momentum-dependent (the quantum fluctuation regime).

\item
Below $T_{RC}$, classical thermal fluctuations dominate, leading to the formation of AFM/SDW precursor bands and the opening of the pseudogap. In this regime, the spectral function $A(\mathbf{k},0)$ can exhibit a maximum in momentum space while showing a minimum or strong suppression at $\omega=0$ in the frequency domain, $A(\mathbf{k}_{\mathrm{max}},\omega)$. In addition, the maximum in $A(\mathbf{k},0)$ decreases with decreasing temperature, rather than sharpening as it would for a true Fermi surface. Finally, the quasiparticle condition \eref{eq:Sigma_deriv_neg} is violated in this case. We designate such ridge lines in $A(\mathbf{k},0)$ as false Fermi surfaces. Based on these criteria, an analogous distinction can be made for the ZEQ line (\eref{eq:zero_energy_line}). When there is a true segment of  ZEQ line, it practically coincides with the segment of the FS for $T \ll W$; however, when the FS and the ZEQ line become false, they differ.

\item 	
The effects of spin fluctuations on the single-particle spectra differ qualitatively depending on whether the fluctuations are commensurate or incommensurate.

\item
In the commensurate case, the antinodal pseudogap always precedes the development of the pseudogap along the Brillouin zone diagonal. As temperature decreases, the portion of the true Fermi surface near the BZ boundary disappears, leaving behind Fermi arcs centered on the diagonals. At even lower temperatures, two false Fermi surfaces emerge: one electron-like and one hole-like. These two ridge lines in $A(\mathbf{k},0)$ arise because the asymptotic expression of the spectral function depends on both $\epsilon(\mathbf{k})$ and the product $\epsilon(\mathbf{k})\epsilon(\mathbf{k}+\mathbf{Q})=-\epsilon^2(\mathbf{k})$ (see \secref{sec:Analytical_results} for details). Consequently, the condition for a local maximum is non-linear in $\epsilon(\mathbf{k})$, yielding two solutions: one hole-like (tracking the zero-energy line) and one electron-like.

\item
The ZEQ line systematically evolves from an electron-like contour—which practically coincides with the non-interacting FS—into a hole-like shape, eventually coinciding with the shadow FS. The latter is obtained from the FS by a shift along the vector(s) $\mathbf{q}_{max}$ at which the spin susceptibility is maximal. This behavior is driven by the discontinuity in the real part of the self-energy at $\omega=0$ along the shadow FS in the limit $\xi \to \infty$.

\item
In the incommensurate case, the antinodal pseudogap ceases to exist before the quantum critical doping $\delta_{\mathrm{QCP}}$ is reached. Although AFM/SDW precursor bands persist at the antinodal Van Hove point, both are located in the unoccupied sector of the spectrum ($\omega > 0$) and are thus inaccessible to ARPES.

\item
Incommensurate fluctuations give rise to four shadow Fermi surfaces, along which the imaginary part of the self-energy develops a singularity and the real part exhibits a discontinuity as $\xi \to \infty$. These four shadow surfaces are governed by four incommensurate peaks in the spin susceptibility. Two of them intersect the underlying Fermi surface (FS), generating hot spots near the zone diagonals. At these hot spots, the pseudogap persists at sufficiently low temperatures all the way to the quantum critical point $\delta_{\mathrm{QCP}}$. 

\item
The picture described above applies directly to $A(\mathbf{k},0)$ at real zero frequency. As discussed above, the Matsubara proxy $A(\mathbf{k},ik_0)$ used in Ref.~\cite{vsimkovic2024origin} ($k_0=\pi T$) corresponds to an analytic continuation with an effective finite broadening $\eta=\pi T$, such that $A(\mathbf{k}, ik_0)=A(\mathbf{k}, \omega=0)\vert{}_{\eta=\pi T}$. Because $\eta=\pi T$ is not the smallest energy scale in the pseudogap regime, $A(\mathbf{k},ik_0)$ differs significantly from the true real-frequency quantity $A(\mathbf{k},\omega=0)\vert{}_{\eta\to 0}$. Our TPSC+ results indicate that this broadening weakens the manifest signatures of the pseudogap, leading to the following key differences:

\begin{enumerate}
\item
The two ridge lines identified in $A(\mathbf{k},0)$ largely overlap when viewed through the Matsubara proxy $A(\mathbf{k},ik_0)$, requiring a very fine momentum grid to resolve. In the DiagMC study \cite{vsimkovic2024origin}, only a single FS was identified, using a $64\times64$ mesh and a radial search for the absolute maximum along each ray.

\item 
The authors of Ref.~\cite{vsimkovic2024origin} identified every ridge line in $A(\mathbf{k},ik_0)$ as a Fermi surface; our EDC analysis reveals that parts (or all) of such contours become false Fermi surfaces at low temperatures.

\item
Using DiagMC with $A(\mathbf{k},ik_0)$, Ref.~\cite{vsimkovic2024origin} observed a hole-like FS only at large $U$, whereas we find a hole-like ridge line in $A(\mathbf{k},0)$—as well as a hole-like ZEQ line \eref{eq:zero_energy_line}—at arbitrary $U$, provided the temperature is sufficiently low. This discrepancy is an artifact of the finite broadening $i\eta=i\pi T$ inherent to the Matsubara proxies $A(\mathbf{k},ik_0)$ and $\Sigma(\mathbf{k},ik_0)$, and it vanishes in the $\eta\to 0$ limit.

\item
TPSC+ tends to underestimate the severity of pseudogap suppression in the Matsubara proxy $A(\mathbf{k},ik_0)$ in the strong-interaction regime relative to DiagMC \cite{vsimkovic2024origin}. It reproduces the DiagMC behavior at somewhat lower temperatures or reduced doping. Alternatively, TPSC+ yields spectral-function heat maps that are similar to DiagMC at the same temperature and doping when evaluated with a somewhat reduced broadening parameter (e.g., $\eta=0.5 k_0=0.5\pi T$).

\end{enumerate}

\end{enumerate}

The paper is organized as follows. In \secref{sec:model}, we briefly describe the model and the TPSC+ approach. In \secref{sec:PG_definition}, we define the pseudogap, including the case where hot spots are absent. In \secref{sec:Criterion_definition}, we establish the necessary and sufficient physical conditions for the existence of the pseudogap. In \secref{sec:Ph_diag_0}, we present the ground-state phase diagram and compare it with auxiliary-field quantum Monte Carlo results. In \secref{sec:Phase_diagram}, we present and discuss the finite-temperature $\delta - T$ pseudogap phase diagram for intermediate and strong interactions. In \secref{sec:Characteristic_cases}, we analyze in detail four characteristic regimes identified in the finite-temperature phase diagram. In \secref{sec:Comp_with_DiagMC}, we compare the TPSC+ results with DiagMC data and discuss the differences between the real-frequency spectral function and its Matsubara proxy. In \secref{sec:Analytical_results}, we present analytical derivations that explain key aspects of our numerical findings. Finally, in \secref{sec:Summary}, we summarize our conclusions.

\section{Model and the two-particle self-consistent plus (TPSC+) approach:
TPSC+ Classical approach}
\label{sec:model}
We first define the model, then the TPSC+ approach, and finally the asymptotic form of the spin susceptibility. 
\subsection{The Hubbard Model}

We study the one-band Hubbard model on a two-dimensional square lattice, 
given by
\begin{equation}
    H = \sum_{\mathbf{k},\sigma} \epsilon_{\mathbf{k}}\, 
    c^{\dagger}_{\mathbf{k}\sigma} c_{\mathbf{k}\sigma} 
    + U \sum_{i} n_{i\uparrow} n_{i\downarrow} 
    - \mu \sum_{i,\sigma} n_{i\sigma},
    \label{eq:hubbard}
\end{equation}
where $c^{\dagger}_{\mathbf{k}\sigma}$ ($c_{\mathbf{k}\sigma}$) creates 
(annihilates) an electron with spin $\sigma$ and wavevector $\mathbf{k}$, 
$n_{i\sigma}$ is the number operator for electrons with spin $\sigma$ 
at site $i$, $U$ is the on-site repulsive interaction, 
$\epsilon_{\mathbf{k}}$ is the bare electronic dispersion, 
and $\mu$ is the chemical potential.

Throughout this paper, we consider a square lattice with only 
nearest-neighbor hopping $t$. We set $t=1$, the lattice spacing $a = 1$, 
as well as $\hbar = 1$ and $k_B = 1$. In these units, 
the dispersion relation takes the form
\begin{equation}
    \epsilon_{\mathbf{k}} = -2(\cos k_x + \cos k_y).
    \label{eq:dispersion}
\end{equation}

\subsection{The TPSC+ Classical approach}
\label{sec:TPSCpl}

In this subsection, we present the TPSC+ classical framework and how it is applied to the two-dimensional Hubbard model. TPSC+ builds on the Two-Particle Self-Consistent (TPSC) theory, a non-perturbative method that satisfies the Pauli principle, the Mermin-Wagner theorem, and the fluctuation-dissipation theorem (FDT) for both spin and charge susceptibilities \cite{vilk1994theory, Vilk1997}.

TPSC was the first theoretical framework to predict, and to establish a criterion \eref{eq:Criteria_R} for, the opening of an antiferromagnetic pseudogap in the two-dimensional Hubbard model within the renormalized classical regime \cite{Vilk1996,Vilk1997}.

Although TPSC correctly describes the onset of the pseudogap in the renormalized classical regime, it breaks down deep in that regime. TPSC+ \cite{Schaefer2021,gauvin2023improved,vilk2024antiferromagnetic} addresses this by feeding single-particle information back into the two-particle quantities, extending the theory's reach to much lower temperatures and larger correlation lengths.

TPSC+ has already been presented in detail elsewhere \cite{Schaefer2021,gauvin2023improved,vilk2024antiferromagnetic, vilk2026pseudogap} and thoroughly benchmarked. For $U < 3 - 3.5$, it agrees quantitatively with Monte Carlo simulations and cold-atom experiments. At larger $U$, the agreement becomes qualitative, and TPSC+ tends to underestimate pseudogap effects: comparable behavior is recovered in TPSC+ either at lower temperature for fixed doping, or at lower doping for fixed temperature.

We now summarize the TPSC+ approach. The spin and charge susceptibilities have RPA-like forms, but with the bare interaction $U$ replaced by renormalized spin ($U_{sp}$) and charge ($U_{\text{ch}}$) vertices, and with the bare bubble replaced by a semi-dressed bubble, defined below.

%%%%%%%%%%%%%%%%
\begin{align}
    \chi_{sp}(\mathbf{q},iq_n) = \frac{\chi^{(2)}(\mathbf{q},iq_n)}{1-\frac{U_{sp}}{2}\chi^{(2)}(\mathbf{q},iq_n)},
    \label{eq:chisp2}\\
    \chi_{ch}(\mathbf{q},iq_n) = \frac{\chi^{(2)}(\mathbf{q},iq_n)}{1+\frac{U_{ch}}{2}\chi^{(2)}(\mathbf{q},iq_n)}.
    \label{eq:chich2}
\end{align}
%%%%%%%%%%%%%%%%

Here $\mathbf{q}$ denotes momentum and $q_n=2\pi n $ is the bosonic Matsubara frequency. The spin and charge irreducible vertices are obtained just as in TPSC, via self-consistency with the local sum rules (obtained from the fluctuation-dissipation theorem), which read~\cite{vilk1994theory}:

%%%%%%%%%%%%%%%
\begin{align}
\frac{T}{N}\sum_{\mathbf{q},q_n} \chi_{sp}(\mathbf{q},iq_n) &= n-2 \langle n_{\uparrow}n_{\downarrow} \rangle, \label{eq:sumrule_sp}\\
\frac{T}{N}\sum_{\mathbf{q},q_n} \chi_{ch}(\mathbf{q},iq_n) &= n+2 \langle n_{\uparrow}n_{\downarrow} \rangle - n^2.  \label{eq:sumrule_ch}
\end{align}
%%%%%%%%%

Note that the right-hand sides follow from applying the Pauli principle $n_{\sigma}^2=n_{\sigma}$.

Determining the vertices $U_{sp}$ and $U_{ch}$ self-consistently from these sum rules requires a third relation. TPSC supplies this via the following \emph{ansatz}~\cite{vilk1994theory,hedayati1989ground} for $U_{sp}$:

%%%%%%%%%%%
\begin{equation}
U_{sp} = U \frac{\langle n_{\uparrow}n_{\downarrow} \rangle}{\langle n_{\uparrow} \rangle \langle n_{\downarrow} \rangle}  \label{eq:ansatz_tpsc}
\end{equation}
%%%%%%%%%%%%%%

Here the renormalized vertex $U_{sp}$ is reduced relative to the bare interaction by the pair correlation function $g_{\uparrow \downarrow} (0) = \langle n_{\uparrow} n_{\downarrow} \rangle / \langle n_{\uparrow} \rangle \langle n_{\downarrow} \rangle$.

In the TPSC+ classical scheme, the semi-dressed bubble $\chi^{(2)}(\mathbf{q}, i q_n)$ is given by:
\begin{align}
    \chi^{(2)}(\mathbf{q},iq_n&) =  \nonumber \\
    -\frac{T}{N}&\sum_{\mathbf{k},ik_n}\left((\mathcal{G}_{\sigma}^{(2)}(\mathbf{k},ik_n)\mathcal{G}_{\sigma}^{(1)}(\mathbf{k}+\mathbf{q},ik_n+iq_n)\right.\nonumber \\
     &+\left. \mathcal{G}_{\sigma}^{(2)}(\mathbf{k},ik_n)\mathcal{G}_{\sigma}^{(1)}(\mathbf{k}-\mathbf{q},ik_n-iq_n) \right ).
    \label{eq:chi2}
\end{align}
%%%%%%%%%%%%%%%%

In this expression, $\mathcal{G}_{\sigma}^{(1)}(\mathbf{k} - \mathbf{q}, i k_n - i q_n)$ is the non-interacting Green function, while $\mathcal{G}_{\sigma}^{(2)}(\mathbf{k}, i k_n)$ is the interacting Green function, obtained iteratively from the self-energy generated by classical spin fluctuations that drive the pseudogap:

\begin{align}
    \Sigma_{cl}(\mathbf{k},ik_n) = \frac{T}{N} \tilde{g}_{fb} U U_{sp} \sum_{\mathbf{q}} \chi_{sp}(\mathbf{q},0) \mathcal{G}_\sigma^{(1)}(\mathbf{k}+\mathbf{q},ik_n).		
    \label{eq:selfEnergy_Cl}
\end{align} 

Here $\tilde{g}_{fb}$ is a numerical factor discussed further below, and $k_n=(2n+1) \pi$ is the fermionic Matsubara frequency. This construction parallels the pairing approximation (GG0 theory) originally introduced by Kadanoff and Martin for the pair susceptibility~\cite{Kadanoff_Martin_1961, chen2005bcs, Boyack_2018}.

The full self-energy is then obtained in a second step, using the spin and charge susceptibilities from the first step, which now capture collective modes~\cite{Vilk1996, Vilk1997, Moukouri2000}.

\begin{align}
    \Sigma_\sigma^{(2)}(\mathbf{k},&ik_n) = Un_{-\sigma}\nonumber+\frac{T}{N} U\sum_{\mathbf{q},iq_n}\left [\tilde{g} U_{sp}\chi_{sp}(\mathbf{q},iq_n)\right.\nonumber \\
     &+\left.(1/2 - \tilde{g}) U_{ch}\chi_{ch}(\mathbf{q},iq_n)\right ] \mathcal{G}_\sigma^{(1)}(\mathbf{k}+\mathbf{q},ik_n+iq_n).		
    \label{eq:selfEnergy2}
\end{align} 

Unlike the electron-phonon problem, there is no Migdal theorem for electron-electron interactions, so the spin and charge vertices must be renormalized.

The original TPSC papers~\cite{Vilk1996, Vilk1997} adopted $\tilde{g} = 1/4$, while a later study~\cite{Moukouri2000} argued for $\tilde{g} = 3/8$ instead. This difference arises because a constant vertex $U_{\text{sp}}$ violates rotational invariance, producing separate expressions in the longitudinal and transverse spin channels; $\tilde{g} = 3/8$ results from averaging the two channels to approximately restore crossing symmetry (rotational invariance). Comparison with Monte Carlo data showed that for $U <3$ the coupling constants for feedback $\tilde{g}_{fb} = 3/8$, and for the full self-energy expression $\tilde{g} = 1/4$ are good choices, whereas for $U \geq 3$ the pair $\tilde{g}_{fb} = 1/4$, $\tilde{g} = 3/8$ performs best. %Here $\tilde{g}_{fb}$ enters the feedback self-energy \eref{eq:selfEnergy_Cl}, and $\tilde{g}$ enters the full self-energy expression \eref{eq:selfEnergy2}.

\subsection{Asymptotic Susceptibility at Low Momentum and Frequency}
\label{sec:Chis_sp_asymptotic}

In the renormalized classical (RC) regime, the main contribution comes from 
small-momentum and low-energy spin fluctuations. These can be described by the 
Ornstein-Zernike (OZ) form of the spin susceptibility, which in the commensurate case has the form:

 \begin{equation} \label{eq:OZ_s}
\chi_{oz}(\mathbf{q},\omega)=\frac{A}{\xi^{-2} +(\mathbf{q}-\mathbf{Q})^2-i\omega/\gamma}
\end{equation}
  Here $A$ is the prefactor, or amplitude, of the susceptibility, $\xi$ is the correlation length, $\gamma$ is the diffusion coefficient (Landau damping), and $\mathbf{Q}=\mathbf{q}_{max}$. In TPSC+, the parameters of the OZ susceptibility can be calculated as:
	
\begin{equation} \label{eq:PrefactorA}
    A =\frac{1}{2 U_{sp} \xi_0^2}
\end{equation}
where the particle-hole characteristic length is: 

%%%%%%%%%%%%%%%%
\begin{equation} \label{eq:xi_0}
    \xi_0^2 = \frac{-1}{2 \chi^{(2)}(\mathbf{Q}, 0)}\left .\frac{\partial^2\chi^{(2)}(\mathbf{q},0)}{\partial q_x^2}\right|_{\mathbf{q}=\mathbf{Q}}.
\end{equation} 
%%%%%%%%%%%%%%%
while the Landau damping is
%%%%%%%%%%%%%%%
\begin{equation}
    \gamma = \frac{\xi_0^2}{\tau_0}
	 \label{eq:gamma_L}
\end{equation} 
with the microscopic relaxation time
%
%%%%%%%%%%%%%%%%
\begin{equation} 
    \tau_0 = \frac{1}{\chi^{(2)}(\mathbf{Q}, 0)} \frac{\partial\chi^{(2)}(\mathbf{Q},\omega)}{\partial \omega}|_{\omega=0}.
\end{equation} 

%%%%%%%%%%%%%%
 \begin{equation} 
    \xi = \xi_0 \sqrt{\frac{U_{sp}}{U_c-U_{sp}}},
		\label{eq:xi_sp}
\end{equation}
%%%%%%%%%%%%%
%
where the critical $U_c$ is:

%%%%%%%%%%%%%%
 \begin{equation} \label{eq:U_c}
    U_c =\frac{2}{\chi^{(2)}(\mathbf{Q}, 0)} 
\end{equation}
%%%%%%%%%%%%%

In the case of incommensurate spin fluctuations, the asymptotic form of the susceptibility is the sum of four terms like \eref{eq:OZ_s}, with the AFM vector $\mathbf{Q}$ replaced by one of the four incommensurate vectors $\mathbf{Q}_i$. In principle, the value of the particle-hole characteristic length $\xi_0$ differs along the $x$ and $y$ directions, but in practice they are very close numerically (See supplemental material of Ref.~\cite{leblanc2026resilient}). The asymptotic form of the spin susceptibility is used in TPSC+ only to accelerate convergence when this is justified \cite{vilk2024antiferromagnetic}. 

The program for the self-consistent solution of the above TPSC+ equations runs on a laptop, thanks in large part to fast Fourier transforms for momentum convolutions and sparse-IR for Matsubara-frequency convolutions~\cite{Shinaoka_2017, Shinaoka_sparse_2022, Li_2020}. The initial Python program was developed by Chloé Gauvin-Ndiaye and Camille Lahaie \cite{gauvin2023improved}. Further improvements were made in \cite{vilk2024antiferromagnetic}, where the convergence problem at low temperatures was resolved, and the capability to calculate very large correlation lengths ($\xi < 10^6$) was added.

\section{The definition of the pseudogap}
\label{sec:PG_definition}

When the Fermi surface has so-called hot spots, the definition of the pseudogap is straightforward: the spectral function $A(\mathbf{k}_{\text{hot}},\omega)$ develops a minimum at $\omega=0$, signaling a pseudogap. Away from the hot spots, however, this definition must be generalized. In those regions, the spectral function is suppressed but does not necessarily exhibit a local minimum at $\omega=0$. Note that in the $t'=0$ model there are no hot spots away from half-filling when the spin fluctuations are commensurate, $q_{\text{max}} = Q = (\pi,\pi)$.

\begin{figure}
    \centering
    \includegraphics[width=\columnwidth]{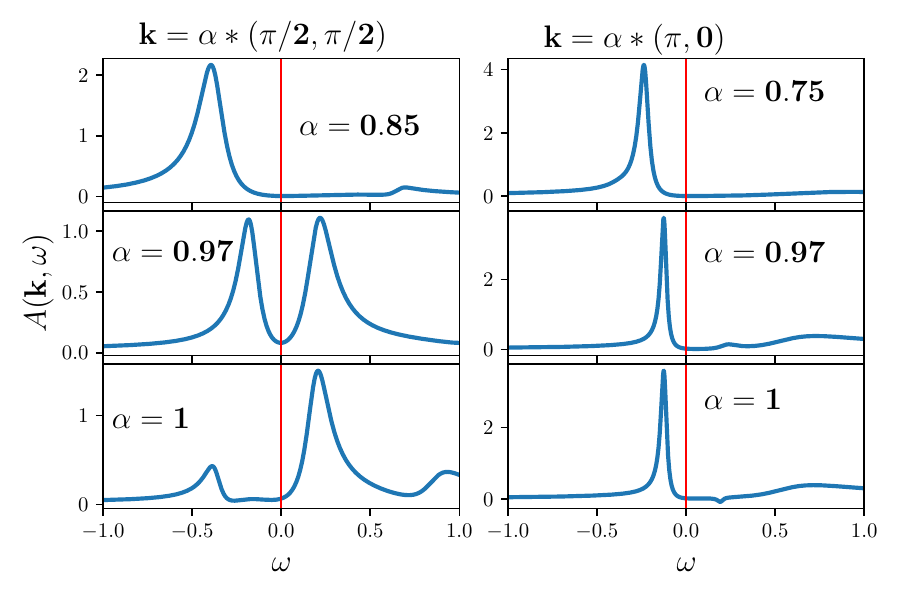} 
  \caption{EDC curves along the diagonal (left panel) and edge $(0,0)$–$(\pi,0)$ (right panel) of the BZ. The parameters are $U=6$, $\delta=0.12$, $T = 0.025$. The correlation length $\xi=130$. At this temperature, the EDC curves in both directions clearly exhibit a pseudogap. As momentum approaches the Fermi surface in the diagonal direction, the quasiparticle peak first moves toward zero energy, then stops at a finite energy and reverses direction. This is a characteristic signature of the precursor to the lower SDW band in this direction. For the EDC curves along the $(0,0)$–$(\pi,0)$ direction, the peak position remains at negative energy even when it reaches the end of the BZ at $(\pi,0)$.}
    \label{fig:Akw_by_k_U_6_T_0_0250}
\end{figure}

\begin{figure}
    \centering
    \includegraphics[width=\columnwidth]{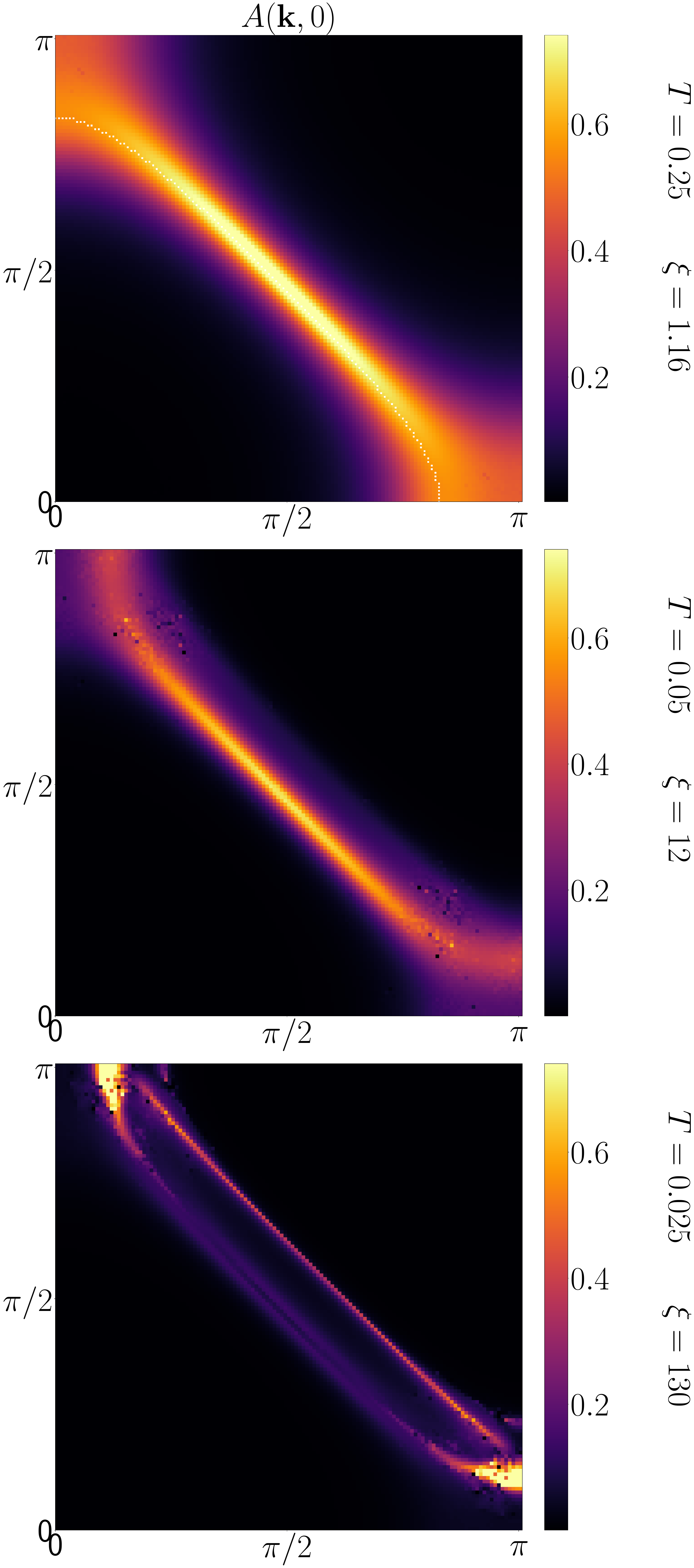} 
  \caption{Evolution of the MDC heatmaps with decreasing temperature. The parameters $U$ and $\delta$ are the same as in the EDC panel \fref{fig:Akw_by_k_U_6_T_0_0250}. The lowest temperature $T=0.025$ is the same as in \fref{fig:Akw_by_k_U_6_T_0_0250}. At the temperature $T_{RC} < T=0.25 \ll W$, the maximum intensity of the spectral function $A(\mathbf{k},0)$ is along the non-interacting FS (white line). As temperature drops to $T=0.05 < T_{RC}$, the intensity of $A(\mathbf{k},0)$ drops in the direction $(0,0)$–$(\pi,0)$, indicating a PG in this direction. At the even lower temperature $T=0.025 <T_D$, the intensity of the spectral function drops around the diagonal direction $(0,0)$–$(\pi,\pi)$ as well, indicating a PG in this direction as well. While the true FS along the diagonal and $(0,0)$–$(\pi,0)$ direction is gone, a new small segment of the FS appears close to $(\pi,0)$–$(\pi,\pi)$.}
    \label{fig:Akw_panel_heatmap_vers_U_6_T_n_0_88}
\end{figure}

In this work, we adopt an ARPES-centric perspective on the pseudogap. We focus on the evolution of the Energy Distribution Curve (EDC) $A({\mathbf{k}},\omega)$ with momentum (see \fref{fig:Akw_by_k_U_6_T_0_0250}) and on the evolution of the Momentum Distribution Curve (MDC) at real zero frequency $A(\mathbf{k},0)$ with temperature (see \fref{fig:Akw_panel_heatmap_vers_U_6_T_n_0_88}). We define the pseudogap along a specific direction (cut) in the Brillouin zone (BZ). Consider the evolution of the energy distribution curves (EDCs) as the momentum moves from $(0,0)$ to $(0,\pi)$ in the hole-doped Hubbard model with nearest-neighbor hopping. If the quasiparticle peak fails to cross $\omega=0$ even when the momentum reaches the Brillouin zone (BZ) boundary at $\mathbf{k}_{AN}=(\pi,0)$, we define this as a pseudogap in the $(0,0)-(0,\pi)$ direction. This occurs when the precursor of the lower AFM/SDW band at the antinodal Van Hove point lies at negative frequencies. We call this the antinodal PG. Similarly, a pseudogap in the diagonal direction—from $(0,0)$ to $(\pi,\pi)$—is defined by the absence of an $\omega=0$ peak in the EDCs for all momenta along that path. More generally, we define a pseudogap in a given direction if there is no quasiparticle peak at $\omega=0$ at low temperatures, despite such a peak being present at higher temperatures. Equivalently, the pseudogap can be defined using MDC spectra $A(\mathbf{k},0)$ as the disappearance of the true Fermi surface in a specific direction as the temperature is lowered.

\fref{fig:Akw_by_k_U_6_T_0_0250} illustrates our definition of the pseudogap using EDC spectra for $U=6$, $\delta=0.12$, $T = 0.025$. At this temperature, the EDC curves in both directions clearly exhibit a pseudogap. As momentum approaches the Fermi surface in the diagonal direction, the quasiparticle peak first moves toward zero energy, then stops at a finite energy and reverses direction. This is a characteristic signature of the precursor to the lower SDW band in this direction. For the EDC curves along the $(0,0)$–$(\pi,0)$ direction, the peak position remains at negative energy even when it reaches the end of the BZ at $(\pi,0)$.

\fref{fig:Akw_panel_heatmap_vers_U_6_T_n_0_88} illustrates the PG definition by showing the evolution of the MDC heatmaps $A(\mathbf{k},0)$ with decreasing temperature. The parameters $U$ and $\delta$ are the same as in the EDC panel \fref{fig:Akw_by_k_U_6_T_0_0250}. The lowest temperature, $T=0.025$, is the same as in \fref{fig:Akw_by_k_U_6_T_0_0250}. At the temperature $T_{RC} < T=0.25 \ll W$, the maximum intensity of the spectral function $A(\mathbf{k},0)$ is along the non-interacting FS (white line). This is the quantum regime in which Luttinger's theorem is approximately satisfied. As temperature drops to $T=0.05 < T_{RC}$, the intensity of $A(\mathbf{k},0)$ drops in the direction $(0,0)$–$(\pi,0)$, indicating a PG in this direction. Around the diagonal direction, a Fermi arc remains. At the even lower temperature $T=0.025 <T_D$, the intensity of the spectral function also drops around the diagonal direction $(0,0)$–$(\pi,\pi)$, indicating a PG in this direction too. Note that there are still regions of relatively higher intensity in $A(\mathbf{k},0)$. Along the diagonal, such a region is located approximately where the true FS was at higher temperature. However, this is a false FS segment because: a) the intensity of the maximum drops with decreasing temperature, rather than increasing, as it would for a true FS; and b) the corresponding EDC curve in the left panels of \fref{fig:Akw_by_k_U_6_T_0_0250} shows a minimum at $\omega=0$. While the true FS along the diagonal and along the $(0,0)$–$(\pi,0)$ direction is gone, a new small segment of the FS appears close to the $(\pi,0)$–$(\pi,\pi)$ direction. As we will see in more detail in \secref{sec:Case_P2}, this is a non-trivial effect of incommensurate spin fluctuations.

There are other definitions of the pseudogap in the literature, including those based on the maximum in the temperature dependence of the uniform magnetic susceptibility $\chi_{sp}(0,0)$ (Knight shift) \cite{chalopin2026observation} and on the imaginary part of the Matsubara self-energy \cite{Schaefer2021, vsimkovic2024origin}. We discuss these definitions in detail in \secref{sec:Comp_PG_def}. Here, we summarize the main conclusions.

The temperature $T_{max}$ at which $\chi_{sp}(0,0)(T)$ is maximal lies between the diagonal and antinodal pseudogap temperatures, $T_D< T_{max} < T_{AN}$. In other words, the maximum occurs when a significant portion of the Fermi surface is gapped, and it therefore differs from the first appearance of the pseudogap at the antinode.

The Matsubara self-energy criterion is convenient because it does not require analytical continuation, which is challenging when there is noticeable noise in the Matsubara results. However, it can in some cases yield a false negative, and in other cases a false positive, compared to the real-frequency definition. In addition, it focuses on the quasi-singularity in the self-energy, which is a necessary but not sufficient condition for the PG except at the hot spots. For example, for the antinodal PG in the $t'=0$ model, the real-frequency analysis also requires that the gap parameter $\Delta$ be large enough, relative to the chemical potential, to push the precursor of the lower AFM band below zero frequency. Otherwise, precursors of the AFM band are present but are located in the invisible part of the spectrum, $\omega >0$.
	
\section{Mechanism and condition for creation of precursors of antiferromagnetic bands and the pseudogap in TPSC+. }
\label{sec:Criterion_definition}

The precursors of antiferromagnetic bands originate in TPSC+ from thermal critical fluctuations below the renormalized classical (RC) regime temperature $T_{RC}$, which decreases with doping $\delta$. This regime exists because the Mermin-Wagner theorem forbids a finite-temperature phase transition in two dimensions. Consequently, instead of an ordered AFM/SDW state with true AFM bands, precursors of AFM/SDW bands appear below $T_{RC}$ \cite{Vilk1996, Vilk1997,vilk1997shadow}. The effect is rooted in low dimensionality and occurs for both weak and strong interactions, with more favorable conditions at larger interaction strength.

The critical thermal fluctuations in the RC regime lead to a quasi-singularity in the imaginary part of the self-energy $\Sigma''(\mathbf{k},0)$ on the shadow Fermi surface \cite{vilk1997shadow}. The latter is obtained from the regular Fermi surface by a shift along the vectors $\mathbf{q}_{max}$ at which the spin susceptibility has a maximum.

Two main cases differ qualitatively: the commensurate case, with a single spin-susceptibility maximum at $\mathbf{q}_{max}=\mathbf{Q}=(\pi, \pi)$, and the incommensurate case, with four maxima in the spin susceptibility $\chi_{sp}(\mathbf{q},0)$.

\subsection{Commensurate case}
\label{Cond_commens_case}

At low doping and not-too-low temperatures, there is only one maximum in $\chi_{sp}(\mathbf{q}, 0)$, at the commensurate antiferromagnetic vector $\mathbf{q}_{max}=\mathbf{Q}=(\pi,\pi)$. The imaginary part of the self-energy is dominated by classical spin fluctuations and has the form \cite{Vilk1997}, \cite{vilk1997shadow}, \secref{sec:Analytical_results}:

\begin{equation} \label{eq:ImSigma_R}
\Sigma''(\mathbf{k},\omega)=-\frac{\Delta^2}{2 \ln \xi \sqrt{(\omega -\epsilon_{\mathbf{k}+\mathbf{Q}} +\mu_1)^2 +v^2_{\mathbf{k}+\mathbf{Q}} \xi^{-2}}}.
\end{equation}

Here $v$ is the electron velocity, $\xi$ is the correlation length, and $\Delta$ is the gap parameter. This asymptotic form is valid when the correlation length is significantly larger than the thermal wavelength of the electron \eref{eq:Criteria_R}. The expression \eref{eq:ImSigma_R} has its maximum absolute value along the shadow Fermi surface:

\begin{equation}
\label{eq:Shadow_FS}
\epsilon(\mathbf{k} +\mathbf{Q})=-\epsilon(\mathbf{k}) =\mu_1
\end{equation}
where $\mu_1$ is the non-interacting chemical potential. It follows that, in the $t'=0$ model, the shadow FS is a mirror image of the non-interacting FS relative to the line $(\pi,0)-(0,\pi)$.

\begin{figure}
    \centering
    \includegraphics[width=\columnwidth]{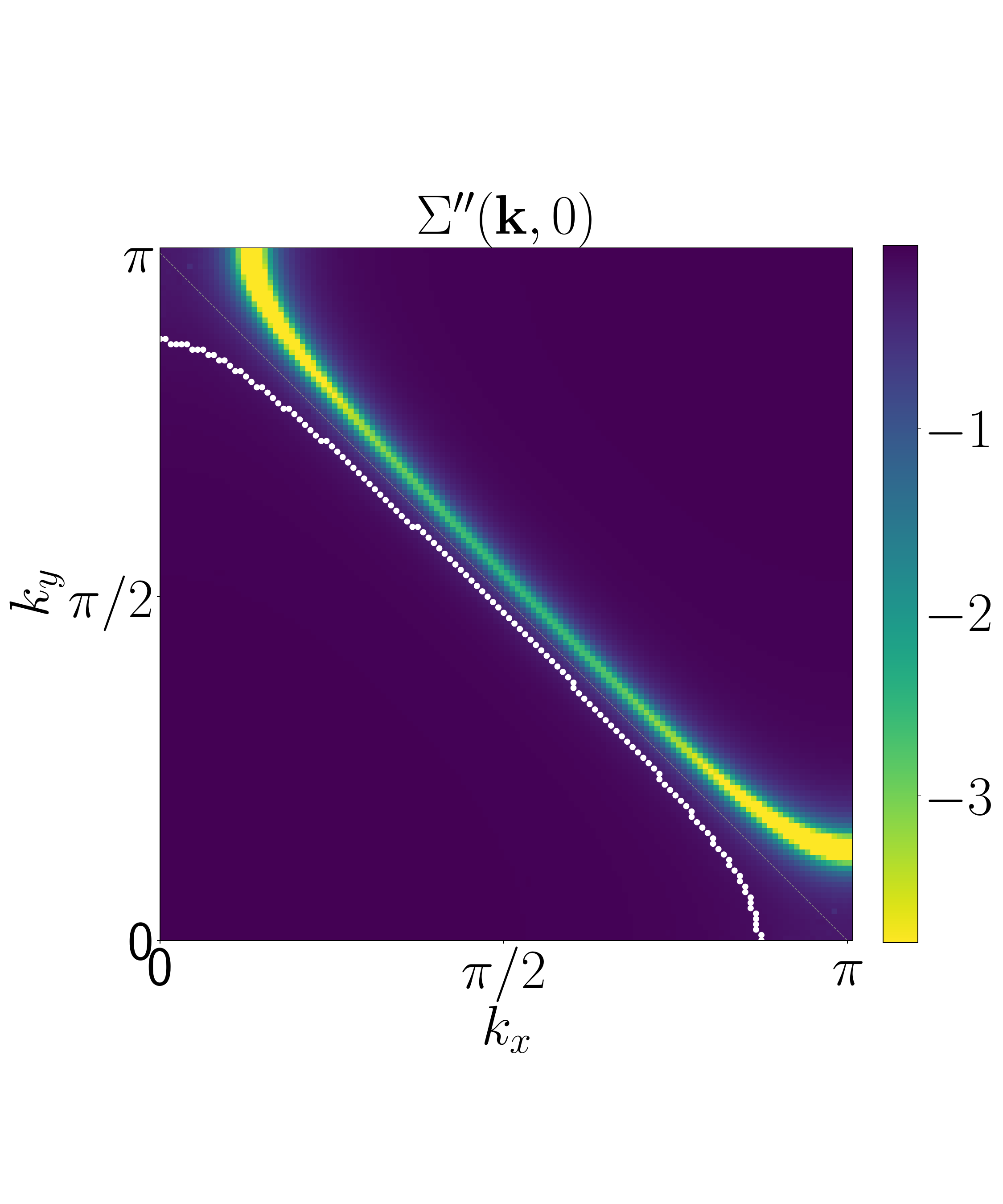} 
    \caption{Heat map of the imaginary part of the self-energy at zero frequency $\Sigma''(\mathbf{k},0)$ obtained using TPSC+. The parameters are $U=6$, $T=0.06$, $\delta=0.087$. $\xi=42.5$. The white line shows the non-interacting Fermi surface. The shadow Fermi surface \eref{eq:Shadow_FS} is a mirror image of the non-interacting FS relative to the line $(\pi,0)-(0,\pi)$ for this model. The imaginary part of the self-energy $\Sigma''(\mathbf{k},0)$ shows a quasi-singularity along the shadow FS with maximum intensity close to the antinodal Van Hove point $(\pi,0)$.}
    \label{fig:Im_Sigma_heatmap_vers_T_0_06_n_0_913}
\end{figure}

\fref{fig:Im_Sigma_heatmap_vers_T_0_06_n_0_913} shows a heat map of the imaginary part of the self-energy at zero frequency, $\Sigma''(\mathbf{k},0)$. It displays a quasi-singularity along the shadow FS, with maximum intensity close to the edge of the Brillouin zone, in the vicinity of the antinodal Van Hove point $(\pi,0)$. The larger intensity near the antinodal Van Hove point arises from the very small electron velocity $v$ in this region.

The criterion \eref{eq:Criteria_R} and the asymptotic expression \eref{eq:ImSigma_R} are, of course, invalid at the antinodal Van Hove point, where $v=0$, and in its vicinity. In that case, a different criterion and asymptotic expression apply \cite{vilk2023criteria}. However, in this work we are primarily interested in the doped model, where the antinodal point is located far from the shadow Fermi surface.

The quasi-singularity in the imaginary part of the self-energy $\Sigma''(\mathbf{k},\omega)$ corresponds to a quasi-discontinuity in the real part $\Sigma'(\mathbf{k},\omega)$ at the shadow FS. In the limit $\xi \to \infty$, the asymptotic form of $\Sigma'(\mathbf{k},\omega)$ is:

\begin{equation}
\label{Real_Sigma_assymp}
\Sigma'(\mathbf{k},\omega)=\frac{\Delta^2}{\omega - \epsilon(\mathbf{k} +\mathbf{Q}) +\mu_1}
\end{equation}

It is easy to see that this real part of the self-energy gives rise to two precursors of AFM bands at $\omega=\pm \Delta$ for all points on the shadow FS. At half-filling, for $t'=0$, the real FS and the shadow FS coincide (perfect nesting), and the appearance of the precursors of AFM bands is concomitant with the appearance of the pseudogap at the Fermi surface ($A(\mathbf{k}_F,0) \to 0$ as $T \to 0$). A similar situation occurs when there are isolated hot spots where the real FS and the shadow FS intersect.

In general, however, the existence of precursors of AFM bands does not guarantee the existence of a pseudogap on the FS. For example, for the antinodal PG at finite doping, the gap parameter $\Delta$ must be large enough to push the precursor of the lower AFM band below zero energy. Analytical results are possible in TPSC+ assuming commensurate spin fluctuations (see \secref{sec:Analytical_results}). In this case, the pseudogap exists when

\begin{equation} \label{eq:Criteria_PG_commens}
\Delta > \left|\frac{\tilde{\mu}_2 + \mu_1}{2}\right| .
\end{equation}

Here $\tilde{\mu}_2=\mu_2 - \Sigma'_{l}(0)$, where the local real part of the self-energy is $\Sigma'_{l}(\omega)=\frac{1}{N} \sum _{\mathbf{k}} \Sigma'(\mathbf{k},\omega)$ and is mainly determined by quantum fluctuations. $\mu_2$ is the chemical potential that gives the occupation appearing in the sum rules when it is computed with the Green's function that includes $\Sigma^{(2)}_\sigma$. 

Above $T_{RC}$, the interacting Fermi surface practically coincides with the non-interacting FS (see \secref{sec:Characteristic_cases}), and $\tilde{\mu}_2 \approx \mu_1$. Below $T_{RC}$, $|\tilde{\mu}_2| > |\mu_1|$, but it remains comparable to $\mu_1$.

As doping increases, the values of $|\tilde{\mu}_2|$ and $|\mu_1|$ increase, while the gap parameter $\Delta$ decreases. The PG exists in the doping range defined by \eref{eq:Criteria_PG_commens}. Importantly, it ceases to exist before the doping reaches the quantum critical point, where $\Delta_{QCP}=0$.

The above analysis is based on the assumption of commensurate critical spin fluctuations. In reality, the spin fluctuations at $T_{RC}$ become incommensurate at some doping $\delta_{inc}$ before \eref{eq:Criteria_PG_commens} is violated. As we will see in the next section, incommensurate fluctuations have a profound effect on the single-particle spectra: they lead to the appearance of hot spots close to the diagonal. At the hot spots, the PG exists at all dopings below the QCP, $\delta < \delta_{QCP}$. This is because, at a hot spot, the singularity in the self-energy occurs directly on the Fermi surface. By contrast, the antinodal point is located far from the hot spots, and the PG there behaves qualitatively similarly to the commensurate case: it ceases to exist at a doping $\delta_{PG}$ significantly lower than $\delta_{QCP}$.

To summarize, when the PG is driven by commensurate spin fluctuations, the antinodal PG always precedes the diagonal PG. However, this does not hold in the incommensurate case. There, $T_{AN}$ goes to zero well before the QCP doping is reached, $\delta_{AN} < \delta_{QCP}$, while the diagonal pseudogap continues to exist for dopings $\delta > \delta_{AN}$. The phase diagram in the $T$–$\delta$ plane is discussed in detail in \secref{sec:Phase_diagram}.

\subsection{Incommensurate case}
\label{Cond_incommens_case}

The above picture becomes more complex and qualitatively different when the spin fluctuations become incommensurate with increasing doping. In this case, four peaks in the spin susceptibility $\chi_{sp}(\mathbf{q},0)$ lead to four shadow Fermi surfaces along which $\Sigma''(\mathbf{k},0)$ develops quasi-singularities. The peaks in $\chi_{sp}(\mathbf{q},0)$ are located at the momentum values $Q_1 = (\pi,(1-\alpha)\pi)$, $Q_2 = (\pi,(1+\alpha)\pi)$, $Q_3 = ((1-\alpha)\pi,\pi)$, and $Q_4 = ((1+\alpha)\pi,\pi)$, where $\alpha < 1$. The conditions for these shadow FS are given by \eref{eq:Shadow_FS} with the vector $\mathbf{Q}$ replaced by $\mathbf{Q}_{i}$.

\begin{figure}
    \centering
    \includegraphics[width=\columnwidth]{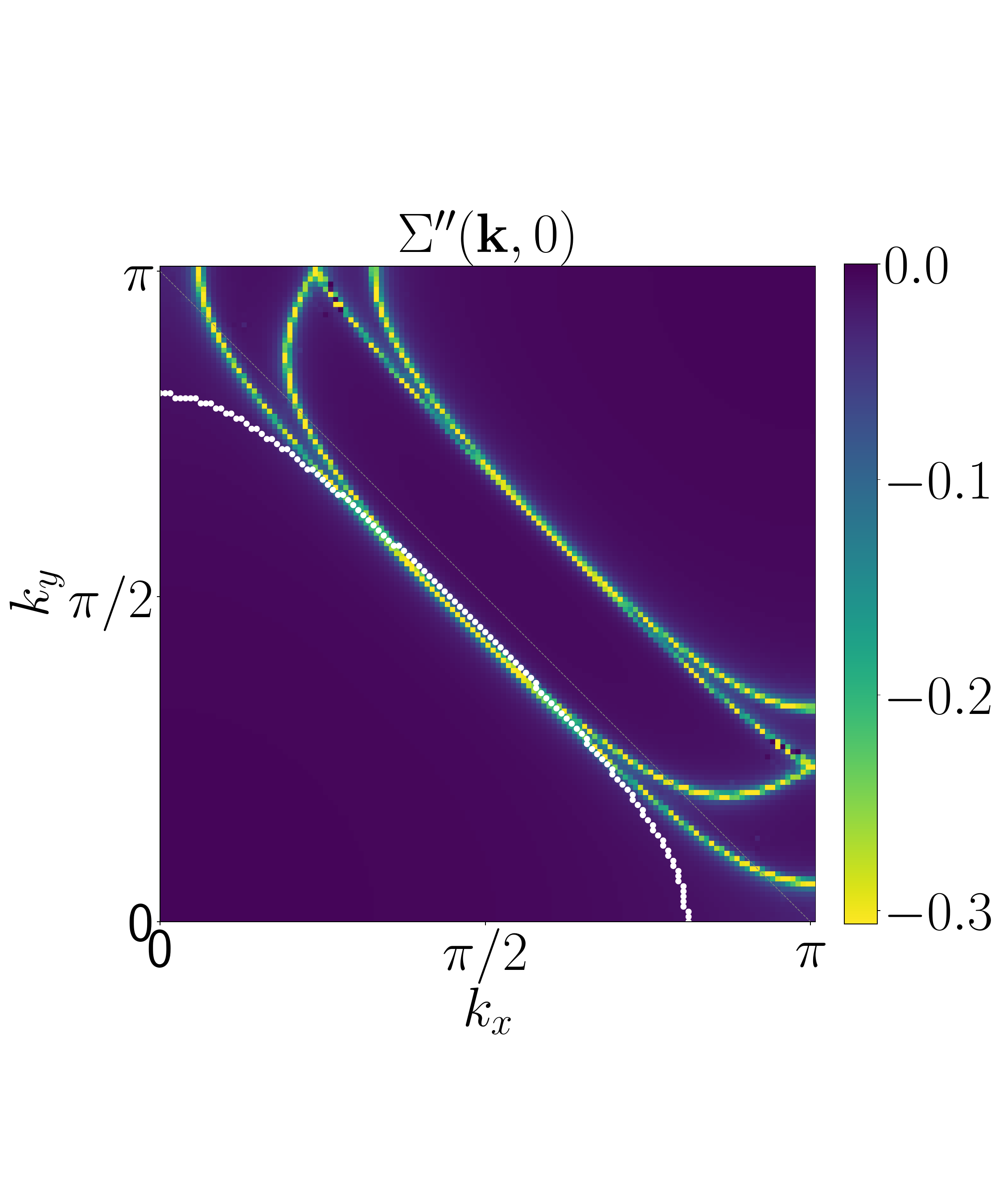} 
    \caption{Heat map of the imaginary part of the self-energy at zero frequency $\Sigma''(\mathbf{k},0)$ in the case of strongly incommensurate spin fluctuations. The parameters are $U=6$, $T=0.015$, $\delta=0.17$. $\xi=130$. The white line shows the non-interacting Fermi surface. There are four shadow Fermi surfaces that are driven by four peaks in the incommensurate spin susceptibility $\chi_{sp}(\mathbf{q},0)$. Two of these shadow FS cross the non-interacting FS, creating hot spots close to the diagonal. As a result, the pseudogap in the strongly incommensurate case forms close to the nodal direction, rather than at the antinode as in the commensurate case.}
\label{fig:Im_Sigma_heatmap_vers_T_0_015_n_0_830}
\end{figure}

\fref{fig:Im_Sigma_heatmap_vers_T_0_015_n_0_830} shows a heat map of the imaginary part of the self-energy at zero frequency, $\Sigma''(\mathbf{k},0)$, in the case of strongly incommensurate spin fluctuations. The parameters are $U=6$, $T=0.015$, $\delta=0.17$. One can clearly see four shadow Fermi surfaces driven by the four peaks in the incommensurate spin susceptibility $\chi_{sp}(\mathbf{q},0)$.

Note that in the commensurate case, the shadow FS lies entirely to the right of the AFM boundary $(\pi,0)$–$(0,\pi)$, and the non-interacting FS does not exhibit hot spots. By contrast, in the incommensurate case, two of the shadow FS shift to the left of the AFM boundary and intersect the non-interacting FS, creating hot spots close to the diagonal. Moreover, there is an extended region around the diagonal where the non-interacting FS and the shadow FS approach each other very closely. This has an important implication: in the strongly incommensurate case, the pseudogap persists in the nodal direction up to the QCP doping (see \secref{sec:Case_P3} for details).

\section{Phase diagram at almost zero temperature}
\label{sec:Ph_diag_0}

The central tenet of the spin-fluctuation theory regarding the pseudogap behavior in the two-dimensional Hubbard model is that critical thermal spin fluctuations drive these phenomena. Such fluctuations play a unique role in two dimensions because finite-temperature phase transitions are prohibited by the Mermin-Wagner theorem; consequently, the transition to the ordered spin-density wave (SDW) state occurs only at zero temperature. Thus, to determine the parameter range in which a pseudogap can exist, it is important to estimate the ground-state SDW phase diagram.

\begin{figure}
    \centering
    \includegraphics[width=\columnwidth]{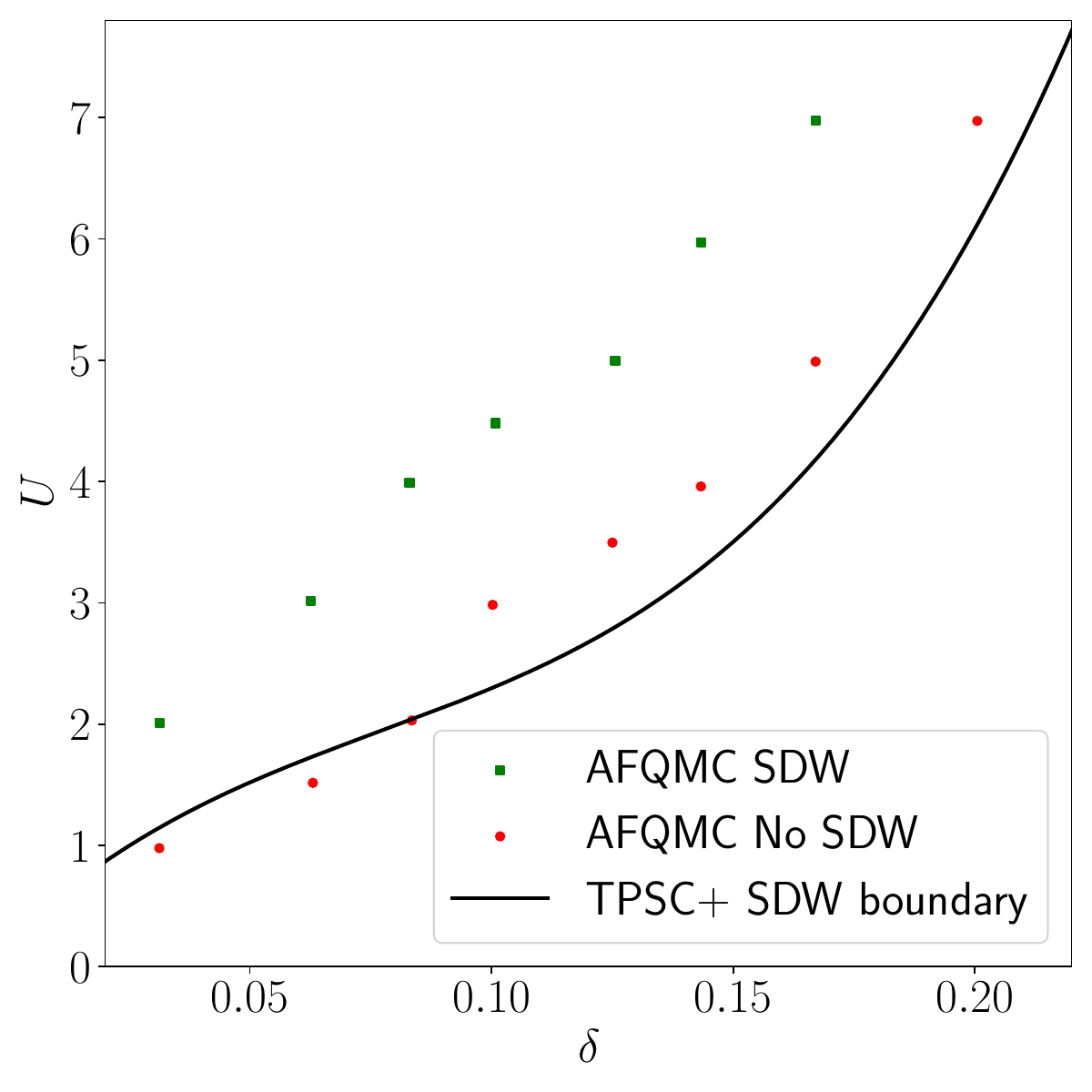} 
    \caption{Estimate of the ground-state SDW boundary in TPSC+ compared with ground-state calculations using constrained-path auxiliary field quantum Monte Carlo (AFQMC) \cite{xu2022stripes}. The green squares represent points from \cite{xu2022stripes} where SDW order definitely exists, while red squares indicate points where SDW order is absent. The TPSC+ estimate of the SDW boundary was performed at a finite but very low temperature, $T=0.005$.}   
    \label{fig:Phase_diag_T_0}
\end{figure}

In \fref{fig:Phase_diag_T_0}, we present an estimate of the ground-state SDW boundary using the TPSC+ approach and compare it with ground-state calculations using constrained-path auxiliary-field quantum Monte Carlo (AFQMC) \cite{xu2022stripes}. To estimate the SDW boundary in TPSC+, we calculated the correlation length $\xi$ at a very low temperature, $T=0.005$, and identified the quantum critical point as the location where the correlation length falls within the range $\xi = 30-40$ lattice spacings. We then fit a smoothing spline to the data obtained in this way. The advantage of this simplified approach is computational speed. To validate the approach, we checked the power-law scaling $\chi_{sp}(\mathbf{q}_{max},0)(T) \sim T^{-\alpha}$ at the QCP for two interaction strengths, $U=3.5$ and $U=6$. This method was used in \cite{leblanc2026resilient} for $U=6$. We discuss the power-law scaling of our results later in this section.

As illustrated in \fref{fig:Phase_diag_T_0}, in the weak-to-intermediate interaction strength ($U$) regime, the TPSC+ estimate of the SDW boundary is consistent with Monte Carlo results. However, TPSC+ tends to underestimate the value of $U_{SDW}$ at larger interaction strengths. Note that, in both AFQMC and TPSC+, the critical doping as a function of interaction strength, $\delta_{SDW}(U_{SDW})$, saturates to a finite value as $U$ increases.

Previous studies at relatively high temperatures ($T=0.1-0.2$) \cite{gauvin2023improved,vilk2024antiferromagnetic} have shown that TPSC+ is in good quantitative agreement with Monte Carlo data in the weak-to-intermediate regime ($U \le 3$) and in qualitative agreement in the intermediate-to-strong regime ($U < 7$). Our current analysis indicates that this level of agreement persists even at very low temperatures.

\begin{figure}
    \centering
    \includegraphics[width=\columnwidth]{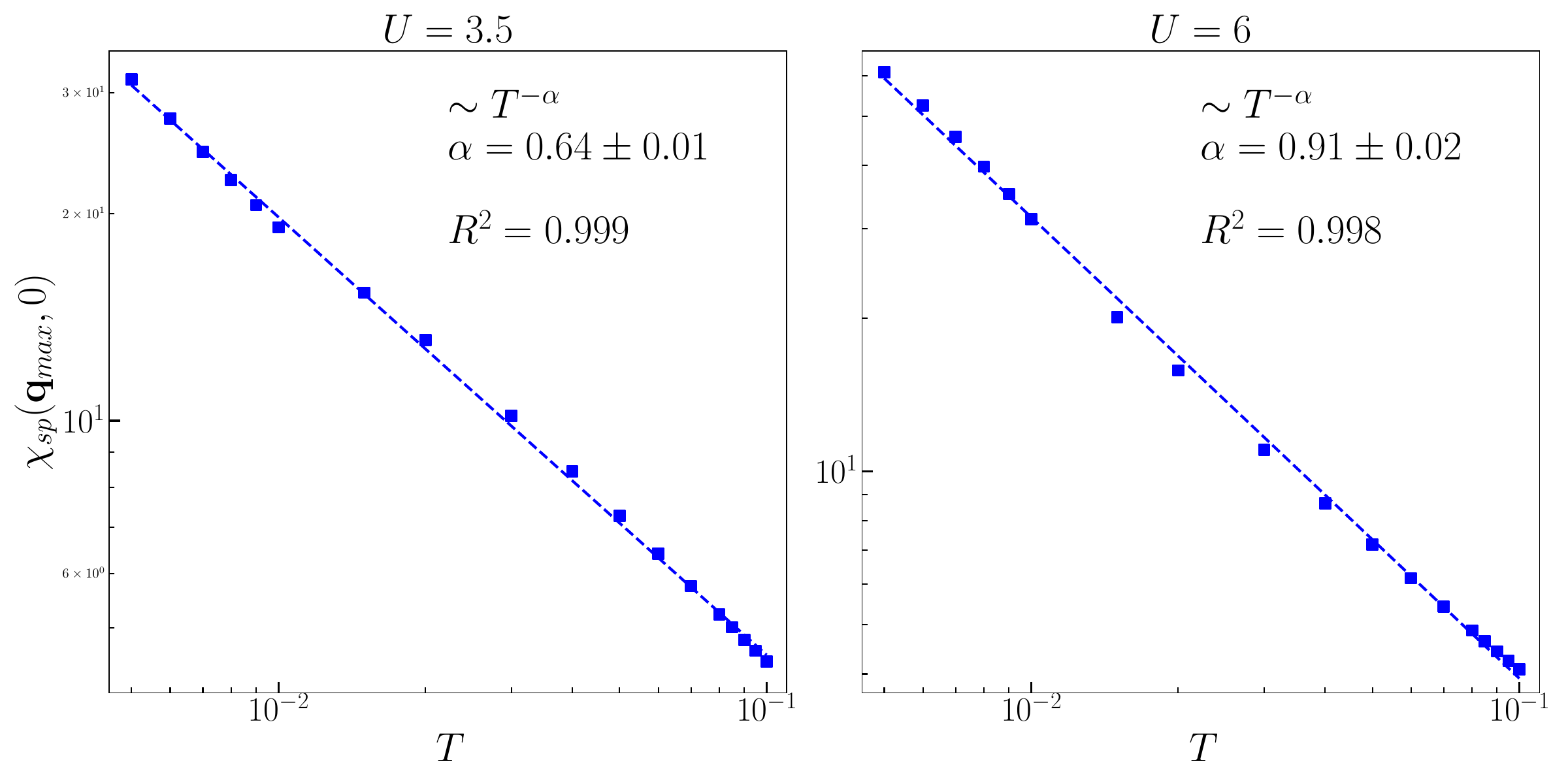} 
    \caption{The power-law scaling $\chi_{sp}(\mathbf{q}_{max},0)(T) \sim T^{-\alpha}$ at the QCP, defined with the simplified criteria described in the main text. For $U=3.5$, $\delta_{QCP} \approx 0.15$ and for $U=6$, $\delta_{QCP} \approx 0.2$. For both interaction strengths $U=3.5$ and $U=6$, the coefficient of determination $R^2$ is very high, indicating a very good fit. The power exponent $\alpha$ increases with $U$.}   
    \label{fig:chi_sp_vs_T}
\end{figure}

Let us now look at the power-law scaling $\chi_{sp}(\mathbf{q}_{max},0)(T) \sim T^{-\alpha}$ at $\delta_{QCP}$, defined using our simplified criteria described above. \fref{fig:chi_sp_vs_T} shows a log-log plot of $\chi_{sp}(\mathbf{q}_{max},0)(T)$. For both interaction strengths $U=3.5$ and $U=6$, the coefficient of determination $R^2$ is very high, indicating a very good linear fit. The power exponent $\alpha$ increases with $U$. For $U=6$, our estimate of the exponent $\alpha$ is within the margin of error of the value $\alpha=0.92$ found in \cite{leblanc2026resilient}. However, our value of $\delta_{QCP} \approx 0.2$ is larger than the value $0.183$ found in \cite{leblanc2026resilient}. The main source of this difference is the difference in the value of the TPSC+ parameter $\tilde{g}_{fb}$. We used $\tilde{g}_{fb}=1/4$, while \cite{leblanc2026resilient} used the value $\tilde{g}_{fb}=3/8$. The value of the parameter $\tilde{g}_{fb}$ does not affect the results qualitatively, but it does change them quantitatively. See \secref{sec:TPSCpl} for a discussion of this parameter.

The area above the SDW line in \fref{fig:Phase_diag_T_0} indicates the range of doping $\delta$ and interaction strength $U$ for which a renormalized classical regime exists below a certain crossover temperature $T_{RC}$. This crossover temperature $T_{RC}(\delta,U)$ decreases with increasing doping and decreasing interaction, and becomes zero on the SDW line. It is in the RC regime that precursors of the AFM/SDW bands develop, as the spin correlation length and susceptibility grow exponentially. In \secref{sec:Comp_spin_charge_DiagMC}, we present TPSC+ results for $\chi_{sp}(\mathbf{q}_{max}, 0)(T)$ and compare them with DiagMC benchmark results.

\section{Pseudogap phase diagram}
\label{sec:Phase_diagram}

\begin{figure}
\centering
\includegraphics[width=\columnwidth]{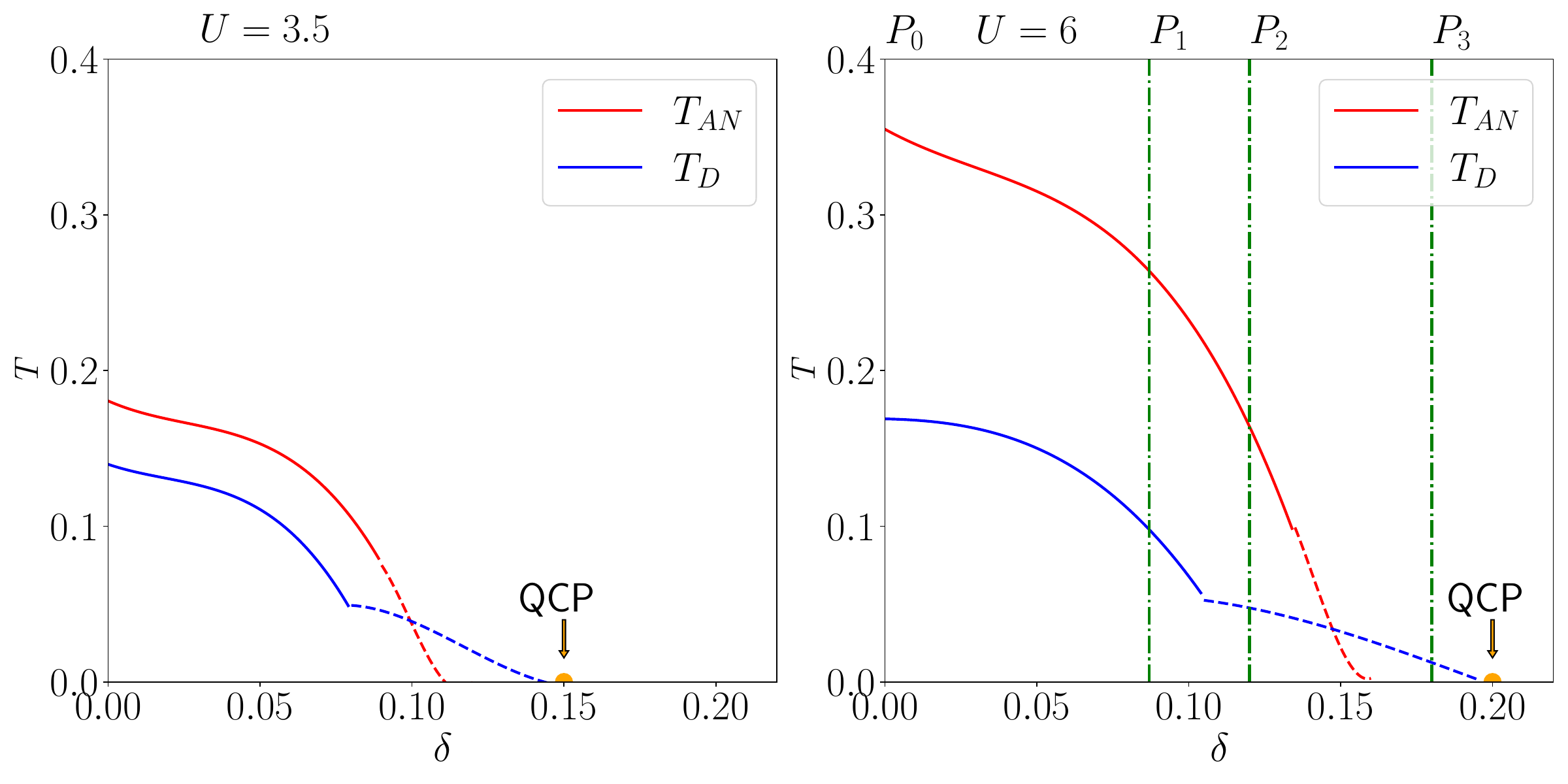}
\caption{Phase diagram for intermediate $U=3.5$ (left) and strong $U=6$ (right) interaction strengths. The red line shows the antinodal pseudogap temperature $T_{AN}(\delta)$, while the blue line indicates the pseudogap temperature in the diagonal direction $T_{D}(\delta)$. Solid lines correspond to commensurate spin fluctuations $\mathbf{q}_{max}=\mathbf{Q}=(\pi,\pi)$, while dashed lines indicate incommensurate fluctuations. In the commensurate case, $T_{AN}(\delta) > T_{D}(\delta)$ at all dopings. In the incommensurate case, $T_{AN}(\delta)$ vanishes well before the quantum critical point (QCP), whereas $T_{D}(\delta)$ extends up to the QCP. The physics along the vertical lines $P_0$ to $P_3$ is explained in the following section.}
\label{fig:PG_phase_diag_T_delta}
\end{figure}

\fref{fig:PG_phase_diag_T_delta} shows the pseudogap phase diagram for intermediate $U=3.5$ (left) and strong $U=6$ (right) interaction strengths. The red line denotes the crossover temperature for the antinodal pseudogap $T_{AN}(\delta)$ as a function of doping, while the blue line indicates the pseudogap temperature in the diagonal direction $T_{D}(\delta)$. Solid lines correspond to the transition into the pseudogap regime when spin fluctuations $\mathbf{q}_{max}=\mathbf{Q}=(\pi,\pi)$ are commensurate, whereas dashed lines indicate the incommensurate case.

As shown previously \cite{gauvin2023improved,vilk2024antiferromagnetic, vilk2026pseudogap}, TPSC+ provides a quantitative description of single-particle properties in the weak-to-intermediate interaction regime, $U < 3.5$, but only a qualitative description for stronger interactions. Our present study also finds that TPSC+ underestimates the pseudogap effects at larger $U$. The phase diagram for $U=6$ should therefore be regarded as a lower-bound estimate of the pseudogap temperature and doping.

Qualitatively, the phase diagrams for $U=3.5$ and $U=6$ within TPSC+ are similar, but they exhibit significant quantitative differences. For example, {\it at half-filling}, the correlation length required for the onset at $T_{AN}$ of the antinodal pseudogap is $\xi \sim 4$ for $U=3.5$, but only $\xi \sim 1$ for $U=6$.

In the commensurate regime, the antinodal pseudogap temperature is always higher than the diagonal one, i.e., $T_{AN}(\delta) > T_{D}(\delta)$. In contrast, in the incommensurate regime, $T_{AN}(\delta)$ vanishes well before reaching the quantum critical point (QCP), while the diagonal pseudogap temperature extends all the way to the QCP. This latter behavior arises from the emergence of hot spots near the diagonal in the incommensurate case, as discussed in \secref{sec:Criterion_definition}.

When the temperature drops below $T_{D}$, the diagonal pseudogap suppresses the portion of the Fermi surface with the largest electron velocity, leading to insulating-like behavior with increasing resistivity. This behavior may change in a more accurate theory that includes superconducting instabilities in the particle-particle channel. Although such effects are not considered here, the results may be relevant to high-$T_c$ experiments in strong magnetic fields that suppress superconductivity. In such experiments, an upturn in resistivity has been observed in several cuprates \cite{boebinger1996insulator,ono2000metal,laliberte2016origin}. A similar conclusion has also been reached using the dual fermion approximation (DFA) \cite{tanaka2026pseudogap}.

With DiagMC~\cite{vsimkovic2024origin}, a new phase
%an additional strange-metal phase 
was identified, but only in the strong-interaction limit. In this phase, the maximum line of the Matsubara spectral-function proxy $A(\mathbf{k},ik_0)$ has a hole-like shape. On the other hand, the ZEQ line defined from the Matsubara self-energy,

\begin{equation}
\label{eq:MF_zero_energy_line}
\epsilon(\mathbf{k})+\text{Re} \Sigma(\mathbf{k},ik_0) -\mu=0,
\end{equation}
was found to be hole-like in both the weak- and strong-interaction limits \cite{vsimkovic2024origin}. We argue below, and in more detail in \secref{sec:Comp_Ak_ik_0_with_DiagMC}, that the absence of a hole-like maximum line (Fermi surface) for weak interaction is an artifact of the spectral-function proxy $A(\mathbf{k},ik_0)$ used in \cite{vsimkovic2024origin}. The real-frequency spectral function $A(\mathbf{k},0)$ analysis using TPSC+ indicates the existence of the new  phase with hole-like FS~\cite{vsimkovic2024origin} in {\it both} the weak- and strong-interaction limits. By contrast, for the Matsubara proxies, we find using TPSC+ a hole-like ZEQ line but only electron-like maximum lines of $A(\mathbf{k},ik_0)$. To understand the reason for this difference, note that the Matsubara proxies $A(\mathbf{k},ik_0)$ and $\Sigma(\mathbf{k},ik_0)$ are equivalent to evaluating the real-frequency quantities at $\omega=0$ with a finite broadening $\eta=\pi T$, rather than in the limit $\eta \to 0$. Although the finite broadening $\eta=\pi T$ reduces the effect of critical spin fluctuations on both the maximum line of $A(\mathbf{k},ik_0)$ and the ZEQ line \eref{eq:zero_energy_line}, its impact is stronger on the spectral function (see \secref{sec:Comp_with_DiagMC} for details). This likely explains why Ref.~\cite{vsimkovic2024origin} finds a hole-like maximum line of $A(\mathbf{k},ik_0)$ only at large $U$, whereas the ZEQ line is already hole-like for weaker interaction, provided $T$ is sufficiently low.

Turning back to TPSC+, the method underestimates the pseudogap effects relative to the DiagMC results in the strong-interaction regime (see \secref{sec:Comp_Ak_ik_0_with_DiagMC}). This explains why TPSC+ does not produce a hole-like maximum line for the Matsubara proxy $A(\mathbf{k},ik_0)$, although such a feature does appear in the $\eta \to 0$ limit for all $U$.

The reference lines $P_0$–$P_3$ in the right panel of \fref{fig:PG_phase_diag_T_delta} indicate characteristic dopings that we analyze in more detail in the following sections, where we provide a simultaneous analysis of energy and momentum distribution curves and their evolution with temperature and doping. Below, we briefly describe each of these characteristic cases, $P_0$–$P_3$.

\begin{enumerate}

\item
Case $P_0$ corresponds to half-filling.

\item
Case $P_1$ ($\delta=0.087$) represents a situation in which both antinodal and diagonal pseudogaps occur while critical thermal spin fluctuations remain commensurate.

\item
Case $P_2$ ($\delta=0.12$) describes a situation in which the system enters the antinodal pseudogap regime while spin fluctuations are still commensurate, whereas the diagonal pseudogap develops only after they become incommensurate.

\item
Case $P_3$ ($\delta=0.18$), close to the QCP ($\delta_{QCP} \approx 0.2$), exhibits no antinodal pseudogap. The diagonal pseudogap emerges once critical thermal spin fluctuations become incommensurate. These incommensurate fluctuations give rise to hot spots near the diagonal of the Brillouin zone, and it is at these hot spots that the pseudogap persists, at sufficiently low temperatures, all the way up to the quantum critical point, i.e., for $\delta < \delta_{QCP}$.

\end{enumerate}

\section{Characteristic Momentum and energy distribution curves in different regions of the phase diagram}
\label{sec:Characteristic_cases} 

In this section, we present results of the numerical TPSC+ calculations for several cases representative of different regions of the phase diagram. We focus on the analysis of the momentum distribution curve at zero frequency $A(\mathbf{k},0)$ and the energy distribution curves $A(\mathbf{k}_i,\omega)$ for various wave vectors $\mathbf{k}_i$. All analytical continuations were performed using the Padé approximation, which is reliable here due to the negligible noise in the TPSC+ Matsubara data. The combined analysis of both MDC and EDC leads us to the concepts of the false Fermi surface and the false ZEQ line in the pseudogap regime, and allows us to demonstrate how the true and false Fermi surfaces evolve with doping and temperature.

Following standard terminology, we refer to the maximal line of the spectral function $A(\mathbf{k},0)$ as the Fermi surface and to the line defined by \eref{eq:zero_energy_line} as the ZEQ line. However, we distinguish between true and false realizations of these lines. We define true Fermi surfaces and true quasiparticle lines using the following criteria:

\begin{enumerate}
\item
The condition \eref{eq:Sigma_deriv_neg} must be satisfied, ensuring a positive quasiparticle weight less than unity,

$$ Z = \left[1 - \frac{\partial \Sigma'(\mathbf{k}, \omega)}{\partial \omega} \right]^{-1} < 1.$$

\item
The spectral weight at zero frequency must remain sufficiently large,
\begin{equation}
\label{eq:Near_maximum_FS_condition}
 A(\mathbf{k},0) \geq 0.75\, A(\mathbf{k},\omega_{\max}),
\end{equation}

where
$$ |\omega_{\max}| \leq |\Sigma''(\mathbf{k},0)| $$

This allows for a slight shift of the spectral peak away from zero frequency. Such shifts can occur at finite temperature even in a Fermi liquid, due to the imaginary part of the self-energy.

\end{enumerate}

\begin{figure}[htbp] % Force it to a single float page
\centering
\includegraphics[width=\columnwidth]{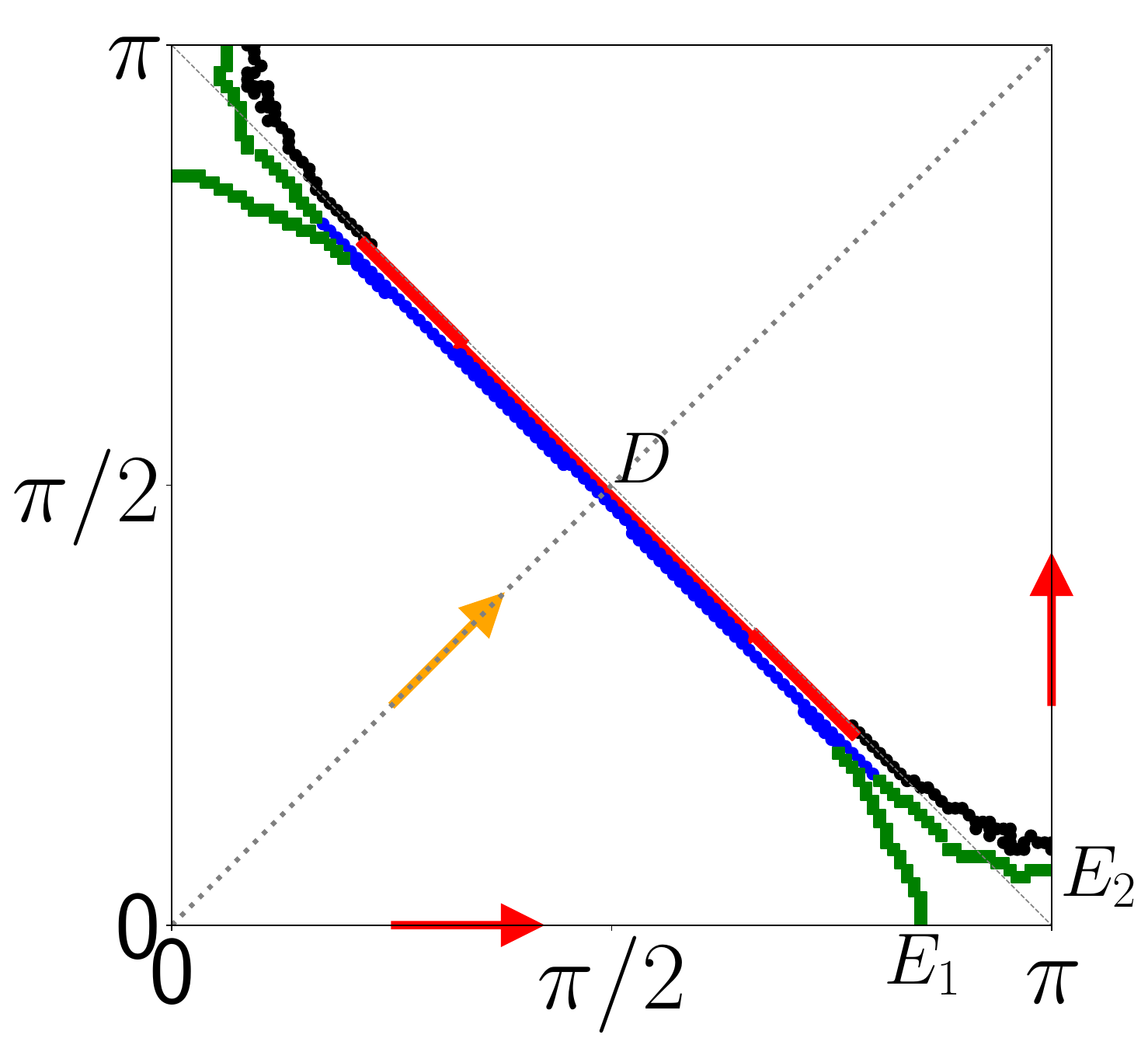}
\caption{One quadrant of the Brillouin zone and notation used in the following sections. Green lines denote false Fermi surface segments as defined in the main text, while blue lines indicate true Fermi surface segments. Red lines represent true ZEQ lines, and black lines represent false ZEQ lines. Points $D$, $E_1$, and $E_2$ are momentum points used in the EDC analysis that accompanies the MDC analysis. Point $D$ lies on the Fermi surface along the BZ diagonal $(0,0)$–$(\pi,\pi)$; points $E_1$ and $E_2$ lie on the Fermi surface along the BZ boundaries in the directions $(0,0)$–$(\pi,0)$ and $(\pi,0)$–$(\pi,\pi)$, respectively.}

\label{fig:FS_ZEQ_E_D_points_notation}
\end{figure}

In the following sections, we adopt the notation illustrated in \fref{fig:FS_ZEQ_E_D_points_notation}. Each of the following subsections focuses on one of the reference lines $P_0$–$P_3$ in the right panel of \fref{fig:PG_phase_diag_T_delta}.

\subsection{Case $P_0$ ($\delta=0$). Pseudogap and concepts of false Fermi surface and false ZEQ line}
\label{sec:Case_P_0}

In this section, we consider the half-filled case. The Energy Distribution Curves (EDC) have been extensively studied in this regime and are well understood. Here, we focus on the Momentum Distribution Curves at zero frequency (MDC) and their relationship to the EDC profiles.

This analysis allows us to demonstrate the new concepts of the false Fermi surface and the false ZEQ line in a context where there is consensus on the underlying physics.

\subsubsection{Simultaneous MDC and EDC analysis}
\label{Case_P_0a}
 
\begin{figure}[htbp] % Force it to a single float page
    \centering
    \includegraphics[width=\columnwidth]{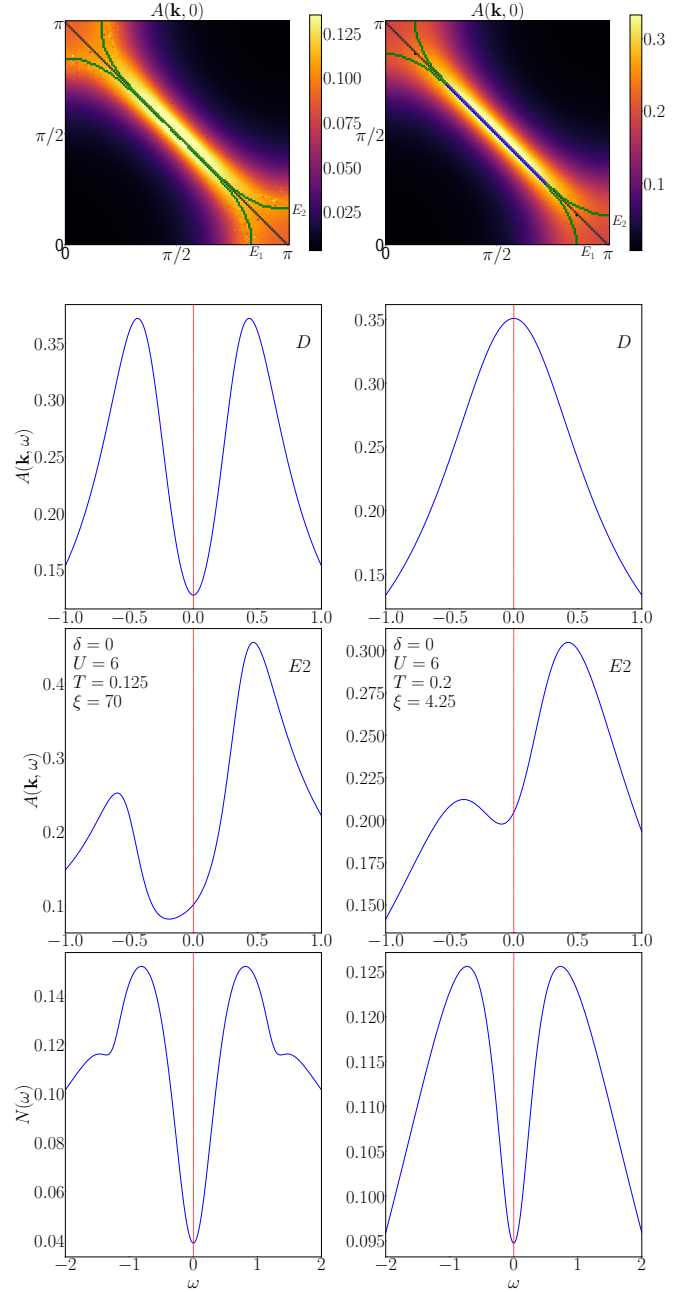} 
    \caption{Simultaneous MDC and EDC analysis at half-filling for $U=6$. Left: $T=0.125$; right: $T=0.2$. Top panels show $A(\mathbf{k},0)$. Green lines denote false Fermi surfaces (momentum maxima with pseudogapped EDCs), while blue lines indicate true Fermi surfaces (maxima at $\omega=0$). The second and third panels show EDCs at representative momentum points $D$ and $E_2$ defined in \fref{fig:FS_ZEQ_E_D_points_notation}. Lower panels show the density of states $N(\omega)$. At $T=0.125$, all EDCs are pseudogapped, indicating only false Fermi surfaces. At $T=0.2$, the EDC at $D$ peaks at $\omega=0$, revealing a Fermi arc near the diagonal, while the BZ edge remains pseudogapped. The density of states shows a pseudogap at both temperatures.}  
    \label{fig:Akw_panel_heatmap_n_1_T}
\end{figure}

In \fref{fig:Akw_panel_heatmap_n_1_T}, we present a combined MDC and EDC analysis at half-filling for $U=6$. The left panels correspond to $T=0.125$, and the right panels to $T=0.2$. The top panels show the momentum-resolved spectral function $A(\mathbf{k},0)$. Green lines denote false Fermi surfaces—loci where $A(\mathbf{k},0)$ is maximal in momentum space while the corresponding EDCs (lower panels) exhibit a pseudogap. Blue lines denote true Fermi surfaces, where $A(\mathbf{k},\omega)$ is maximal both in momentum at $\omega=0$ and in frequency at $\mathbf{k}=\mathbf{k}_F$.

The black line indicates the false ZEQ line, while the red line marks the true quasiparticle line. In this case, the ZEQ line coincides with the $(0,\pi)$–$(\pi,0)$ direction. For $T=0.125$, the red segment is obscured by the blue Fermi surface line.

Notably, two false Fermi surfaces are present, whereas there is only one zero-energy line. As shown analytically in \secref{sec:Analytical_results}, at low temperatures the Fermi surfaces are approximately given by $\epsilon(\mathbf{k})=\pm \Delta/\sqrt{3}$. These two lines are well separated near the BZ edge but nearly coincide along the diagonal. This difference arises from the strongly reduced electron velocity near the Van Hove point at the BZ edge ($v \to 0$).

The second and third rows in \fref{fig:Akw_panel_heatmap_n_1_T} show EDCs at points located on the maxima of $A(\mathbf{k},0)$: point $D$ lies along the Brillouin zone (BZ) diagonal $(0,0)$–$(\pi,\pi)$, while point $E_2$ lies along the BZ edge $(\pi,0)$–$(\pi,\pi)$. The fourth row shows the density of states $N(\omega)=1/N \sum_{\mathbf{k}} A(\mathbf{k},\omega)$.

At the lower temperature $T=0.125$, both EDCs exhibit a pseudogap, and the corresponding maxima of $A(\mathbf{k},0)$ are classified as false Fermi surfaces. At the higher temperature $T=0.2$, the EDC at point $D$ develops a peak at $\omega=0$, indicating the emergence of a true Fermi-surface arc near the diagonal, while the regions near the BZ edge remain characterized by false Fermi surfaces.

The density of states $N(\omega)$ (lower panels) exhibits a pseudogap at both temperatures. This suppression is primarily driven by states near the antinodal Van Hove point $(\pi,0)$.

\subsubsection{$A(\mathbf{k},0)$ along two high-symmetry directions. Suppression of $A(\mathbf{k}_{max},0)$ with temperature.}
\label{sec:Case_P_0b}

\begin{figure}[htbp] % Force it to a single float page
\centering
\includegraphics[width=\columnwidth]{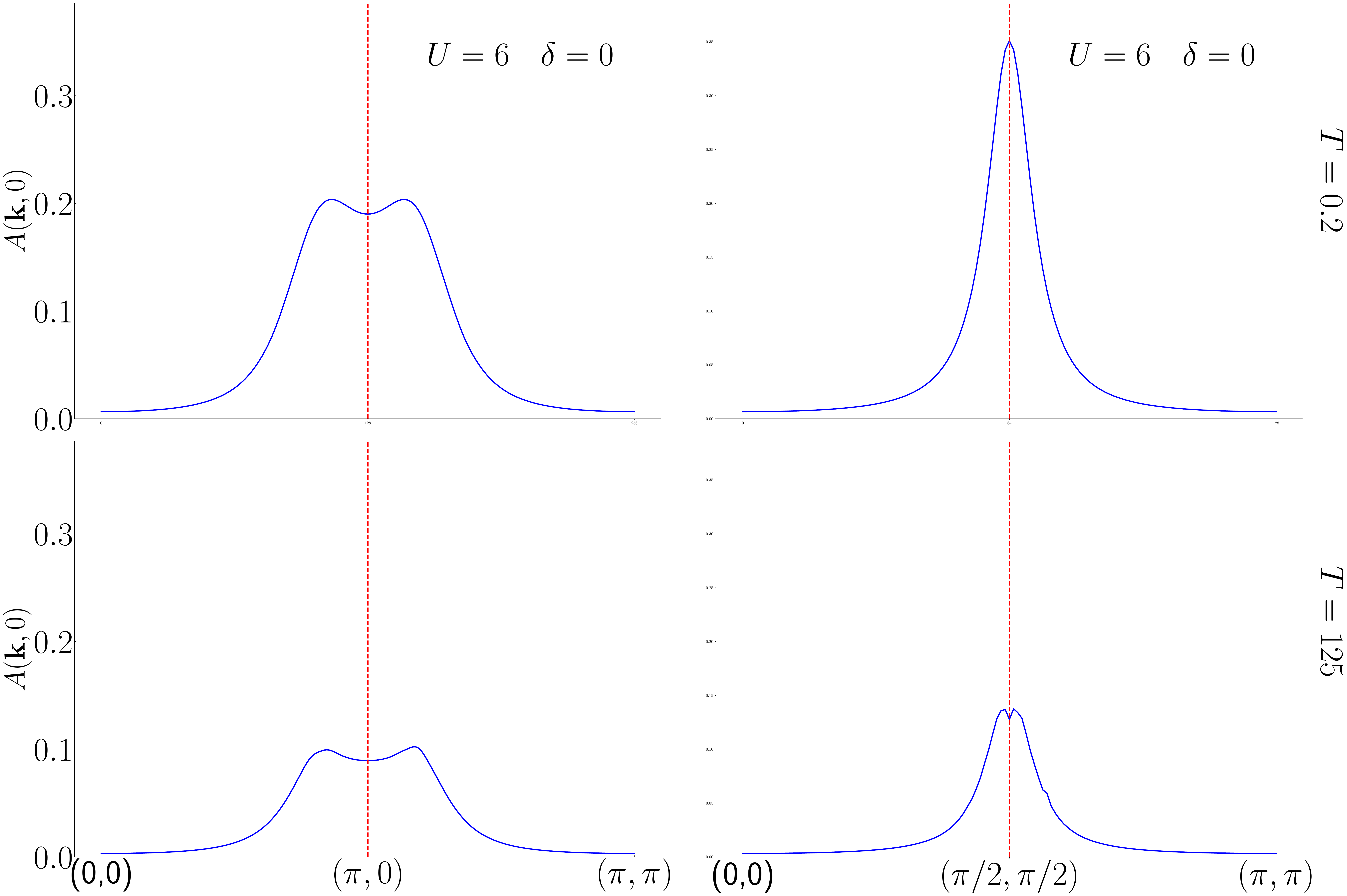}
\caption{Momentum dependence of the spectral function $A(\mathbf{k},0)$ along the BZ edge $(0,0)$–$(\pi,0)$–$(\pi,\pi)$ (left panels) and along the diagonal $(0,0)$–$(\pi/2,\pi/2)$–$(\pi,\pi)$ (right panels). Top and bottom rows correspond to $T=0.2$ and $T=0.125$, respectively. Along the BZ edge, the maxima are weak and broad. Along the diagonal, a well-defined peak at $T=0.2$ indicates a true Fermi surface, while at $T=0.125$ only two weak maxima associated with the false Fermi surface remain. The peak intensity decreases with decreasing temperature, opposite to the behavior expected for a true Fermi surface.}
\label{fig:Akw0_Edge_Diag_dir_high_low_T_n_1}
\end{figure}

Let us now examine in more detail the maxima of the zero-frequency spectral function $A(\mathbf{k},0)$ in momentum space. \fref{fig:Akw0_Edge_Diag_dir_high_low_T_n_1} shows $A(\mathbf{k},0)$ along two directions (see arrows in \fref{fig:FS_ZEQ_E_D_points_notation}): the Brillouin zone (BZ) edge $(0,0)$–$(\pi,0)$–$(\pi,\pi)$ (left panels) and the diagonal $(0,0)$–$(\pi/2,\pi/2)$–$(\pi,\pi)$ (right panels). The top and bottom rows correspond to $T=0.2$ and $T=0.125$, respectively.

Along the BZ edge, the maxima are weak and broad, corresponding to the false part of the Fermi surface and appearing as shallow features on an incoherent background. Along the diagonal, at $T=0.2$, a well-defined peak is present, located on the true Fermi surface. In contrast, at $T=0.125$, two weak maxima appear, both associated with the false Fermi surface. Notably, the peak intensity decreases as temperature is lowered, in contrast to the behavior expected for a true Fermi surface.

\subsubsection{Single-particle properties at characteristic momenta on the ZEQ line. True vs false ZEQ lines}
\label{sec:Case_P_0c}

\begin{figure}[htbp] % Force it to a single float page
\centering
\includegraphics[width=\columnwidth]{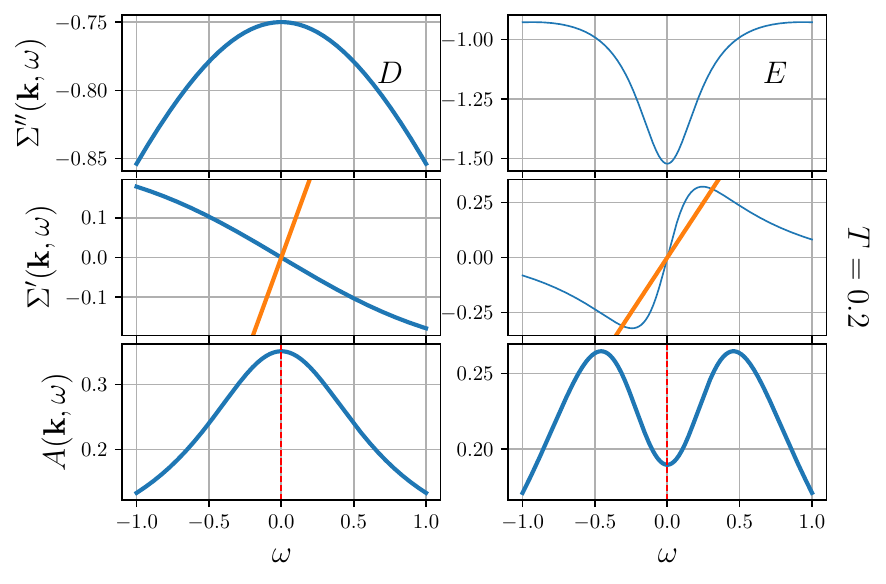}
\caption{Single-particle properties at two momentum points on the ZEQ line: the diagonal point $D$ and the edge point $E$ (see notations in \fref{fig:FS_ZEQ_E_D_points_notation}). Parameters are the same as in \fref{fig:Akw_panel_heatmap_n_1_T} at $T=0.2$: $U=6$, $\delta=0$. Top, middle, and bottom rows show $\Sigma''(\mathbf{k},\omega)$, $\Sigma'(\mathbf{k},\omega)$, and $A(\mathbf{k},\omega)$, respectively. The straight line $\omega - \epsilon(\mathbf{k}) + \mu$ is plotted in the middle panels; its intersection with $\Sigma'(\mathbf{k},\omega)$ gives the solution of \eref{eq:zero_energy_line}. At $E$, the positive slope at $\omega=0$ and the absence of a quasiparticle peak indicate a false quasiparticle line, while $D$ exhibits conventional quasiparticle behavior.}
\label{fig:Im_S_Re_S_Akw_n_1_U_6_T_0_2}
\end{figure}

We now examine single-particle properties at momenta on the ZEQ line. \fref{fig:Im_S_Re_S_Akw_n_1_U_6_T_0_2} shows results for two such points along $(0,\pi)$–$(\pi,0)$: the diagonal point $D$ and the edge point $E$. In this case, $D$ coincides with the nodal point $\mathbf{k}_N=(\pi/2,\pi/2)$, while $E$ corresponds to the antinodal point $\mathbf{k}_{AN}=(\pi,0)$. The parameters match those of \fref{fig:Akw_panel_heatmap_n_1_T} at $T=0.2$.

The top, middle, and bottom rows display $\Sigma''(\mathbf{k},\omega)$, $\Sigma'(\mathbf{k},\omega)$, and $A(\mathbf{k},\omega)$, respectively. In the middle panels, the line $\omega - \epsilon(\mathbf{k}) + \mu$ is also shown; its intersection with $\Sigma'(\mathbf{k},\omega)$ determines the solution of zero-energy equation \eref{eq:zero_energy_line}.

At point $E$, the slope of $\Sigma'(\mathbf{k},\omega)$ at $\omega=0$ is positive, and $A(\mathbf{k},\omega)$ exhibits a minimum rather than a quasiparticle peak, indicating that $E$ lies on the false quasiparticle line. In contrast, point $D$ shows conventional behavior, with a well-defined quasiparticle peak, placing it on the true quasiparticle line.

\subsection{Case $P_1$ ($\delta=0.087$): Pseudogaps occur along both antinodal and diagonal while critical thermal spin fluctuations remain commensurate}
\label{sec:Case_P1}

\subsubsection{Evolution of the MDC heatmap with decreasing temperature }
\label{Case_P_1a}

\begin{figure}[htbp] % Force it to a single float page
\centering
\includegraphics[width=\columnwidth]{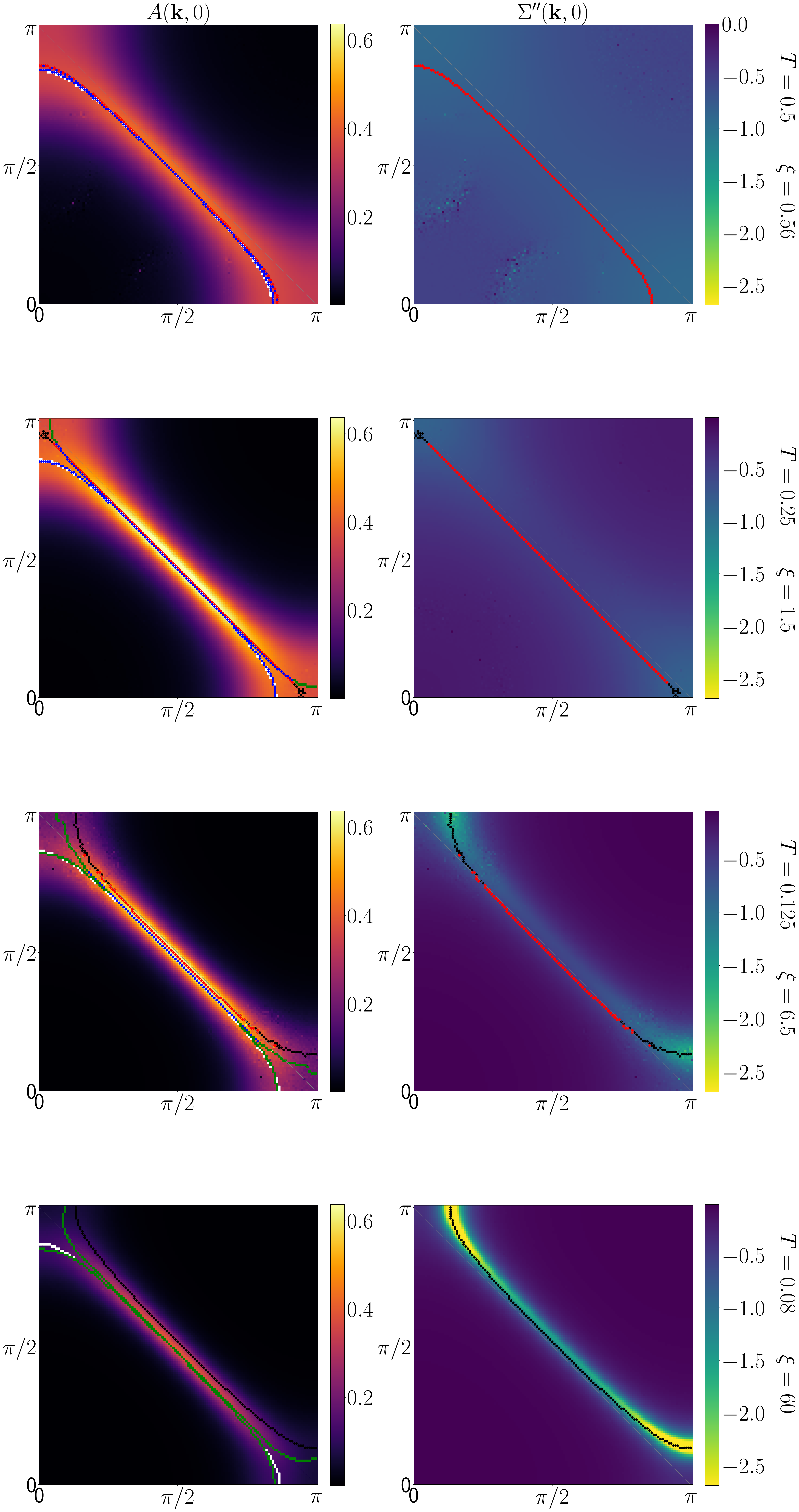}
\caption{Temperature evolution of $A(\mathbf{k},0)$ (left) and $\Sigma''(\mathbf{k},0)$ (right). $U=6$, $\delta=0.087$. Temperature decreases from top to bottom. Color coding is as in \fref{fig:Akw_panel_heatmap_n_1_T}; the white line marks the non-interacting Fermi surface. Red (black) lines indicate true (false) ZEQ lines. At $T=0.5$, the interacting and non-interacting Fermi surfaces, as well as the ZEQ line, nearly coincide. At $T=0.25$, two Fermi surfaces appear—one electron-like and one hole-like. The hole-like segment near the BZ edge is a false Fermi surface. The ZEQ line is nearly flat and lies close to the AFM boundary $(0,\pi)$-$(\pi,0)$. At $T=0.125$, true Fermi arcs emerge near the diagonal, while the segments near the BZ edge remain false. The ZEQ line acquires a hole-like shape. Finally, at $T=0.08$, both Fermi surfaces and the ZEQ line are entirely false, indicating a pseudogap across the entire Brillouin zone. Notably, the ZEQ line nearly coincides with the shadow Fermi surface, where $\Sigma''(\mathbf{k},0)$ exhibits a quasi-singular enhancement.}
\label{fig:Akw_panel_heatmap_vers_T_n_0_913}
\end{figure}

\fref{fig:Akw_panel_heatmap_vers_T_n_0_913} shows the temperature evolution of the MDC spectral function $A(\mathbf{k},0)$ (left) and the imaginary part of the self-energy $\Sigma''(\mathbf{k},0)$ (right). Temperature decreases from top to bottom. The color coding for true/false Fermi surfaces and ZEQ lines follows \fref{fig:Akw_panel_heatmap_n_1_T}; the white line denotes the non-interacting Fermi surface. True (red) and false (black) ZEQ lines are also shown on the $\Sigma''(\mathbf{k},0)$ maps.

At $T=0.5$, the interacting and non-interacting Fermi surfaces, as well as the ZEQ line, nearly coincide, indicating approximate validity of Luttinger's theorem.

At $T=0.25$, two Fermi surfaces emerge (electron-like and hole-like). The hole-like segment near the BZ edge is false, signaling the onset of an antinodal pseudogap, while the ZEQ line is nearly flat and lies close to the AFM boundary $(0,\pi)$–$(\pi,0)$.

At $T=0.125$, true Fermi arcs still exist near the diagonal, whereas segments near the BZ edge remain false. The ZEQ line acquires a hole-like shape. Near the diagonal, both the Fermi surface and the ZEQ line become locally flat and approach the AFM boundary.

Finally, at $T=0.08$, both Fermi surfaces and the ZEQ line are entirely false, indicating a pseudogap over the whole Brillouin zone. Notably, the false ZEQ line nearly coincides with the shadow Fermi surface, where $\Sigma''(\mathbf{k},0)$ exhibits a quasi-singular enhancement (right panels).

\subsubsection{EDC curves for some representative momentum points on the interacting Fermi surfaces and their transformation with temperature}
\label{sec:Case_P_1b}

Let us now look at the EDC curves for some representative momentum points on the interacting Fermi surfaces shown in \fref{fig:Akw_panel_heatmap_vers_T_n_0_913}.

\begin{figure}[htbp] % Force it to a single float page
\centering
\includegraphics[width=0.8\textwidth, height=0.65\textheight, keepaspectratio]{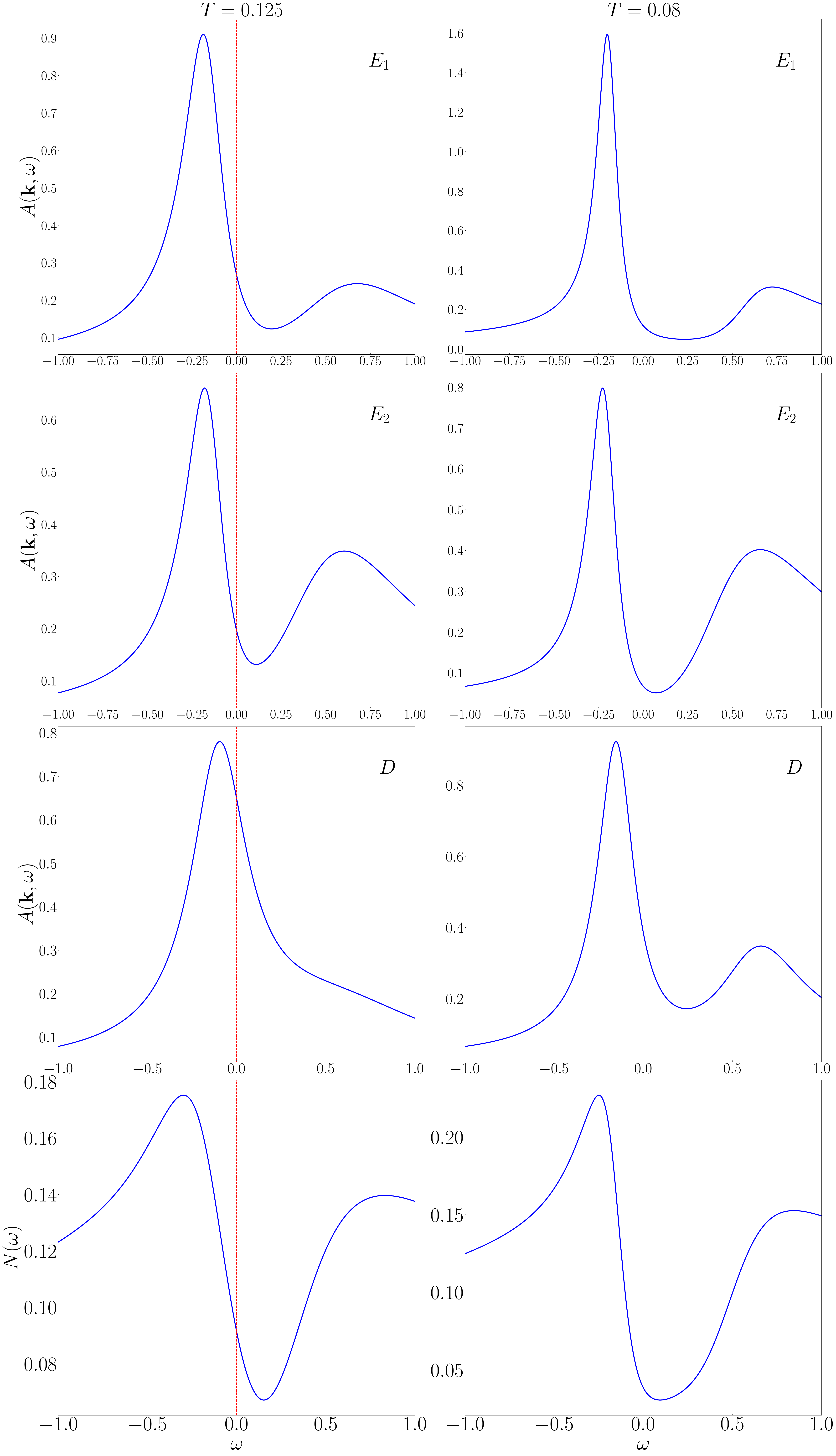}
\caption{EDCs $A(\mathbf{k},\omega)$ for three representative momentum points (top three rows) and the density of states $N(\omega)$ (bottom row). $U=6$, $\delta=0.087$. Left and right panels correspond to $T=0.125$ and $T=0.08$, respectively. Momentum points $E_1$, $E_2$, and $D$ are defined in \fref{fig:FS_ZEQ_E_D_points_notation}. At both temperatures, $A(\mathbf{k}_{E_1},\omega)$, $A(\mathbf{k}_{E_2},\omega)$, and $N(\omega)$ exhibit a pseudogap (suppression at $\omega=0$). The same holds for point $D$ at $T=0.08$. In contrast, at $T=0.125$ the spectrum at $D$ shows a peak slightly shifted from $\omega=0$ but still satisfies the true Fermi surface conditions \eref{eq:Sigma_deriv_neg} and \eref{eq:Near_maximum_FS_condition}}
\label{fig:EDC_panel_n_0_913_T_0_125_0_08}
\end{figure}

\fref{fig:EDC_panel_n_0_913_T_0_125_0_08} shows EDCs $A(\mathbf{k},\omega)$ for three representative momentum points (top three rows) and the density of states $N(\omega)$ (bottom row). Left and right panels correspond to $T=0.125$ and $T=0.08$, respectively. Momentum points $E_1$, $E_2$, and $D$ are defined in \fref{fig:FS_ZEQ_E_D_points_notation}. At both temperatures, $A(\mathbf{k}_{E_1},\omega)$, $A(\mathbf{k}_{E_2},\omega)$, and $N(\omega)$ exhibit a pseudogap (suppression at $\omega=0$). The same holds for point $D$ at $T=0.08$, consistent with all these points lying on false Fermi surfaces in \fref{fig:Akw_panel_heatmap_vers_T_n_0_913}. In contrast, at $T=0.125$ the spectrum at $D$ shows a peak slightly shifted from $\omega=0$ but still satisfies the true Fermi surface conditions \eref{eq:Sigma_deriv_neg} and \eref{eq:Near_maximum_FS_condition}, placing it on a true Fermi arc.

\subsubsection{Transformation of the ZEQ line from the electron-like Fermi surface at relatively high temperatures to the hole-like shadow Fermi surface at low temperatures}
\label{sec:Case_P_1c}
 
One of the key features of the above evolution is the transformation of the ZEQ line from the electron-like non-interacting Fermi surface at relatively high temperatures ($T_{RC}< T \ll W$, with bandwidth $W=8$) to the hole-like shadow Fermi surface at low temperatures ($T<T_D$). During this process, the quasiparticle line changes from true to false. The false segments first appear near the BZ edge (close to $(\pi,0)$) and then progressively extend toward the diagonal. We find this behavior to be universal across all dopings.

This transformation originates from the development of a quasi-singularity in the imaginary part of the self-energy on the shadow Fermi surface, which corresponds to a quasi-discontinuity in the real part. This mechanism is most clearly illustrated through the graphical solution of the zero-energy condition,

\begin{equation}
\label{eq:zero_energy_line_2}
\Sigma'(\mathbf{k},0) = \mu - \epsilon(\mathbf{k}).
\end{equation}

\begin{figure}[htbp] % Force it to a single float page
\centering
\includegraphics[width=\columnwidth]{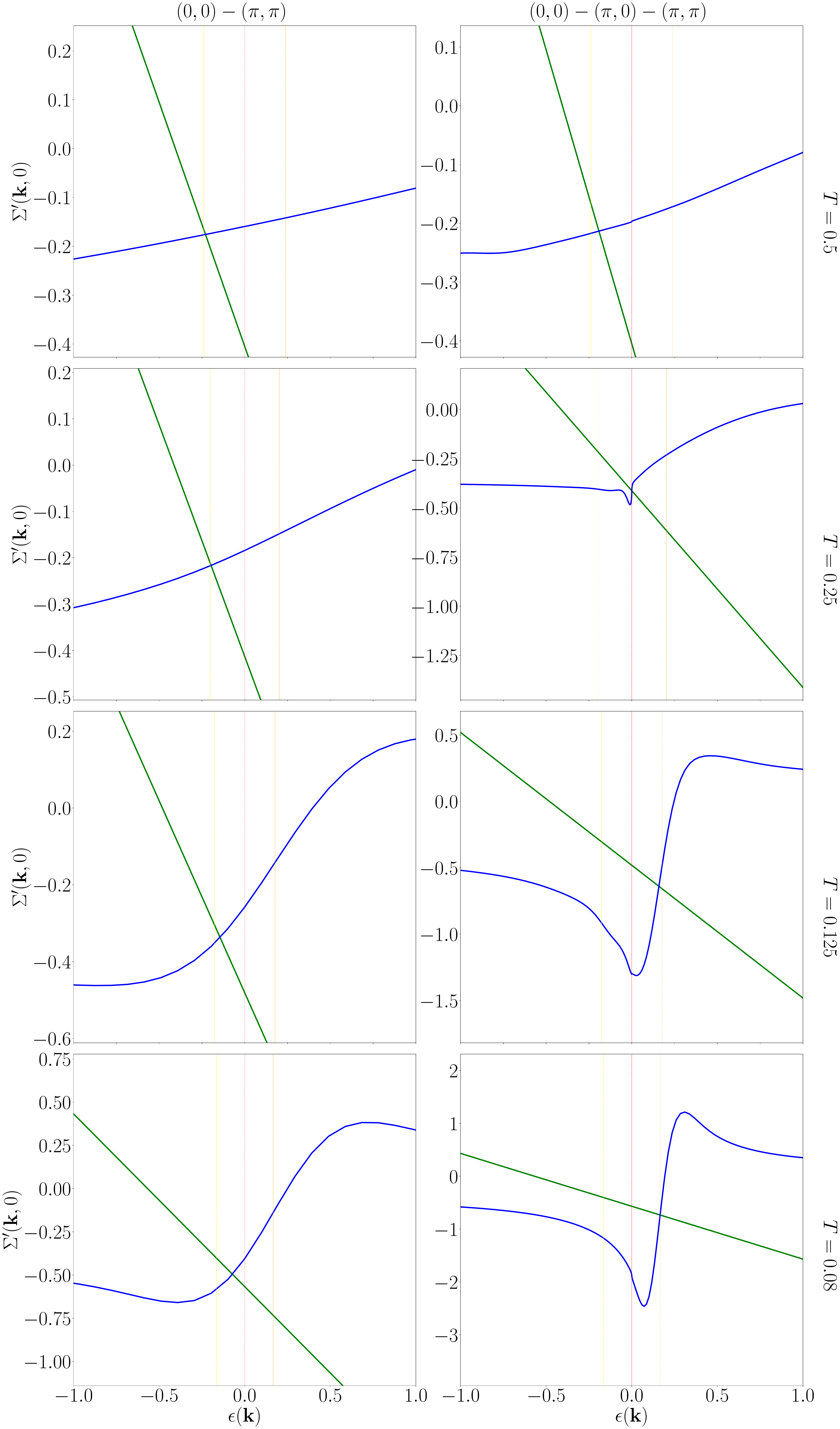}
\caption{Temperature evolution of the graphical solution of the ZEQ equation \eref{eq:zero_energy_line_2} along the BZ diagonal $(0,0)$–$(\pi,\pi)$ (left) and edge $(0,0)$–$(\pi,0)$–$(\pi,\pi)$ (right). $U=6$, $\delta=0.087$. The horizontal axis shows the bare dispersion $\epsilon(\mathbf{k})$, while the vertical axis shows $\Sigma'(\mathbf{k},0)$ (blue) and $\mu - \epsilon(\mathbf{k})$ (green). Their intersection gives the solution of ~\eref{eq:zero_energy_line_2}. The reference lines at $\pm \mu_1$ mark the non-interacting and shadow Fermi surfaces. At $\epsilon(\mathbf{k})=\mu_1$, the solution lies on the non-interacting Fermi surface (approximate Luttinger behavior), while at $\epsilon(\mathbf{k})=-\mu_1$, it lies on the shadow Fermi surface, where $\Sigma'(\mathbf{k},0)$ develops a quasi-discontinuity.}
\label{fig:Sigma_vs_ek_dif_T_n_0_913}
\end{figure}

In \fref{fig:Sigma_vs_ek_dif_T_n_0_913}, we show how this solution evolves with temperature along two symmetry directions (see arrows in \fref{fig:FS_ZEQ_E_D_points_notation}): the BZ diagonal $(0,0)$–$(\pi,\pi)$ (left panels) and the BZ edge $(0,0)$–$(\pi,0)$–$(\pi,\pi)$ (right panels). Temperature decreases from top to bottom. The horizontal axis represents the bare dispersion $\epsilon(\mathbf{k})$, while the vertical axis shows $\Sigma'(\mathbf{k},0)$ (blue) and $\mu - \epsilon(\mathbf{k})$ (green). Their intersection determines the solution of \eref{eq:zero_energy_line_2}. The reference lines at $\pm \mu_1$ mark the non-interacting and shadow Fermi surfaces.

At high temperature, the solution lies near $\epsilon(\mathbf{k})=\mu_1$, indicating that the interacting quasiparticle line closely follows the non-interacting Fermi surface, consistent with approximate Luttinger behavior. As temperature decreases, a quasi-discontinuity develops in $\Sigma'(\mathbf{k},0)$ at the shadow Fermi surface $\epsilon(\mathbf{k})=-\mu_1$ (using $\epsilon(\mathbf{k}+\mathbf{Q})=-\epsilon(\mathbf{k})$). In the limit $T\to 0$, this feature becomes nearly vertical, ensuring that the intersection shifts to $\epsilon(\mathbf{k})=-\mu_1$.

Thus, with decreasing temperature, the solution of the zero-energy equation moves from the non-interacting Fermi surface to the shadow Fermi surface. This shift occurs more rapidly along the BZ edge than along the diagonal.

\subsection{Case $P_2$ ($\delta=0.12$): Antinodal pseudogap with commensurate and diagonal pseudogap with incommensurate spin fluctuations}
\label{sec:Case_P2}

\subsubsection{Evolution of the zero-energy spectral function $A(\mathbf{k},0)$ and the imaginary part of the self-energy $\Sigma''(\mathbf{k},0)$ with decreasing temperature}
\label{Case_P_2a}

\begin{figure}[htbp] % Force it to a single float page
\centering
\includegraphics[width=\columnwidth]{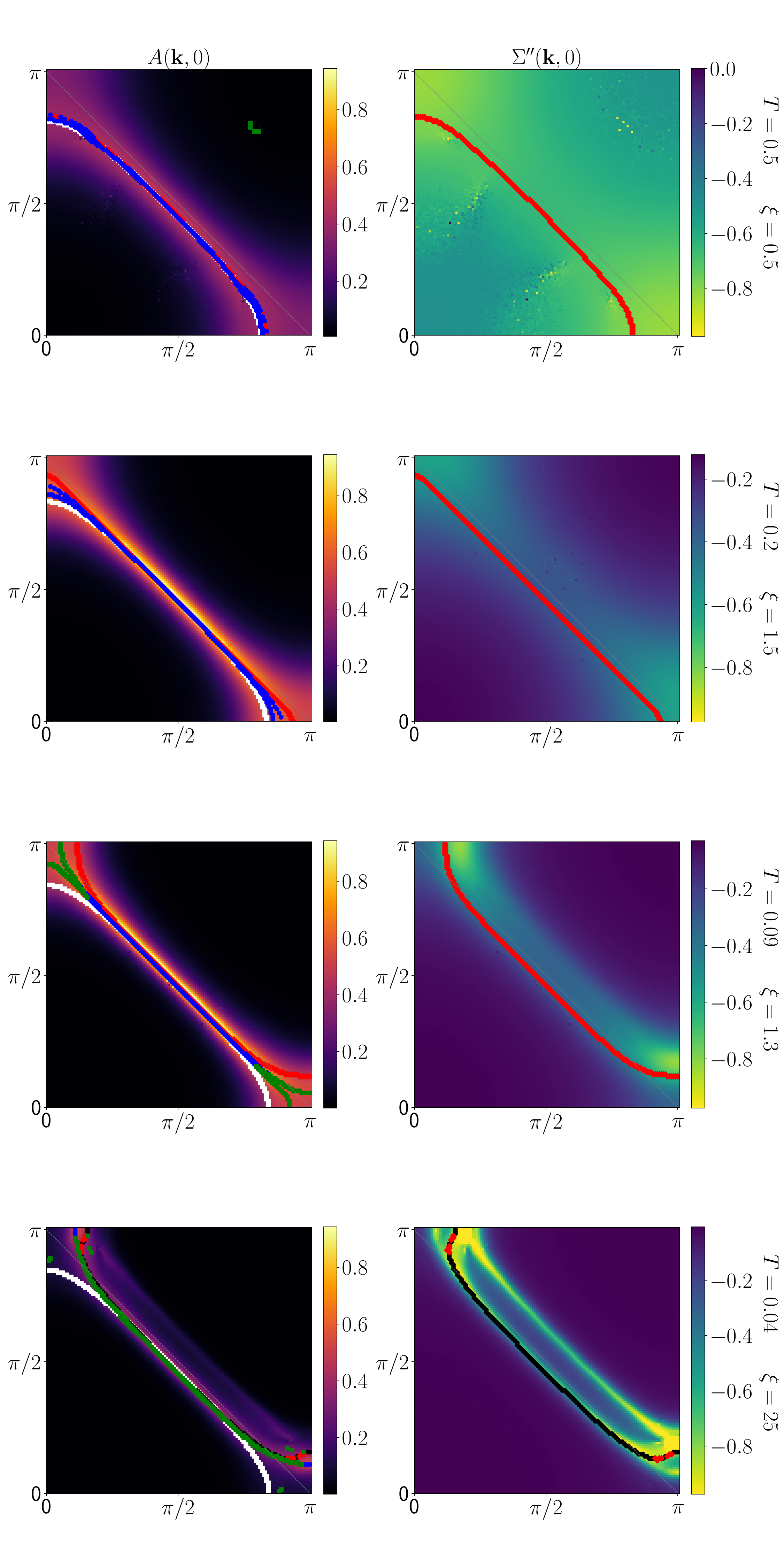}
\caption{Temperature evolution of $A(\mathbf{k},0)$ (left) and $\Sigma''(\mathbf{k},0)$ (right), with temperature decreasing from top to bottom ($U=6$, $\delta=0.12$). The notation follows \fref{fig:Akw_panel_heatmap_vers_T_n_0_913}. For $T=0.5, 0.2, 0.09$, spin fluctuations are commensurate and the behavior is similar to Case $P_1$. At $T=0.04$, $\Sigma''(\mathbf{k},0)$ develops four high-intensity regions associated with incommensurate peaks $\mathbf{Q}_i$ in $\chi_{sp}(\mathbf{q},0)$, indicating incommensurate shadow Fermi surfaces.}
\label{fig:Akw_panel_heatmap_vers_T_n_0_88}
\end{figure}

\fref{fig:Akw_panel_heatmap_vers_T_n_0_88} shows the temperature evolution of the MDC spectral function $A(\mathbf{k},0)$ (left) and the imaginary part of the self-energy $\Sigma''(\mathbf{k},0)$ (right). The notation follows \fref{fig:Akw_panel_heatmap_vers_T_n_0_913}. At higher temperatures ($T=0.5, 0.2, 0.09$), spin fluctuations remain commensurate, and the evolution of the MDCs, Fermi surfaces, and ZEQ lines is qualitatively similar to Case $P_1$.

At the lowest temperature, $T=0.04$, the behavior changes qualitatively. The map of $\Sigma''(\mathbf{k},0)$ exhibits four high-intensity regions associated with the incommensurate peaks $\mathbf{Q}_i$ in $\chi_{sp}(\mathbf{q},0)$. These define four shadow Fermi surfaces:

\begin{equation}
\epsilon(\mathbf{k}+\mathbf{Q}_i)=\mu_1,
\end{equation}
with $\mathbf{Q}_1=(\pi,(1-\alpha)\pi)$, $\mathbf{Q}_2=(\pi,(1+\alpha)\pi)$, $\mathbf{Q}_3=((1-\alpha)\pi,\pi)$, and $\mathbf{Q}_4=((1+\alpha)\pi,\pi)$. 

The Fermi surface is also qualitatively different (left panel): it is predominantly false (green), becoming true only near the BZ edge. This contrasts with the commensurate case, where true Fermi arcs appear near the diagonal and false segments emerge near the edge.

\subsubsection{EDCs at representative momentum points on the interacting Fermi surfaces and their temperature evolution}
\label{sec:Case_P_2b}

\begin{figure}[htbp] % Force it to a single float page
\centering
\includegraphics[width=\columnwidth]{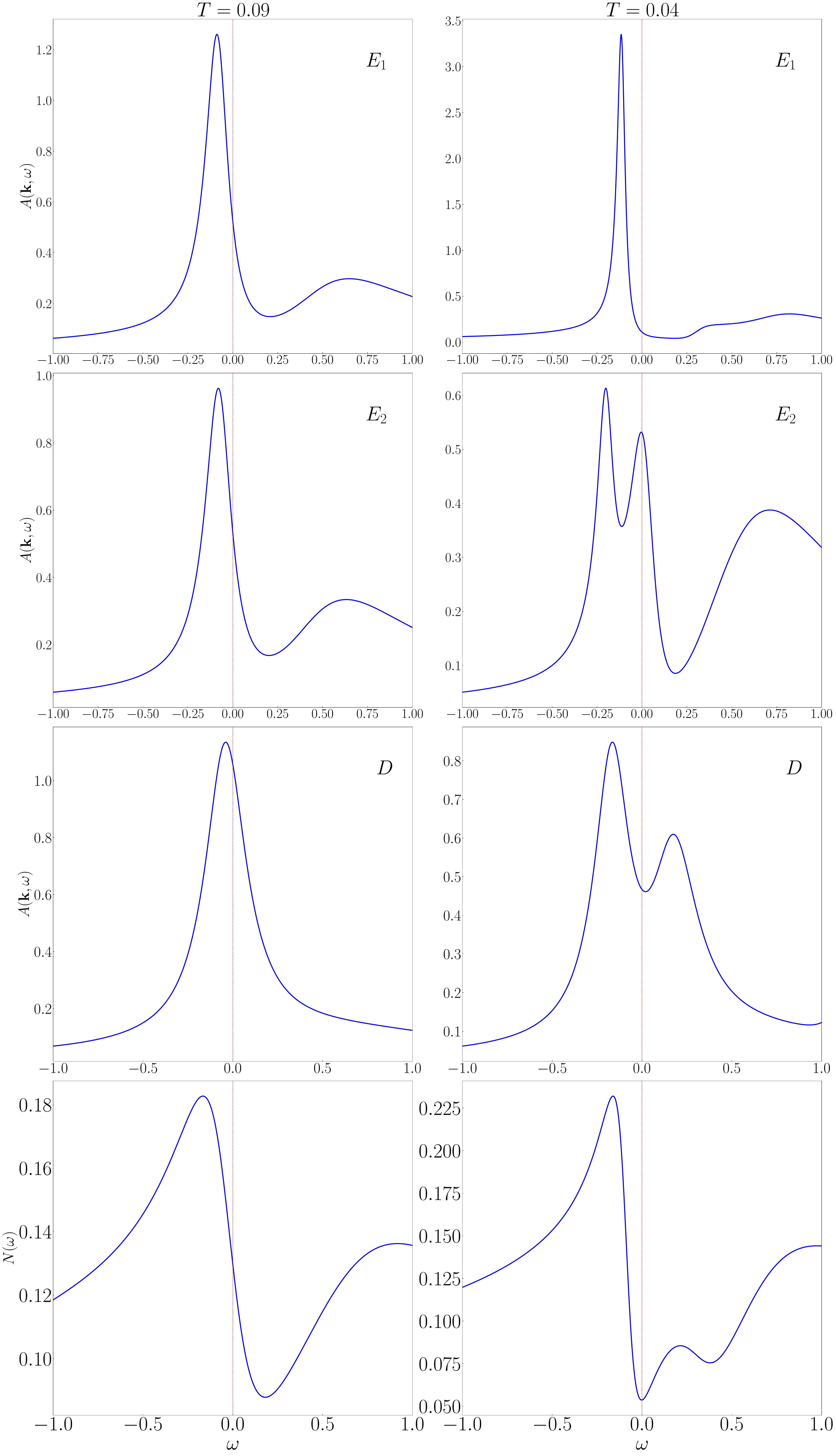}
\caption{EDCs at representative momenta ($U=6$, $\delta=0.12$). The layout and notation follow \fref{fig:EDC_panel_n_0_913_T_0_125_0_08}, with $T=0.09$ (left) and $T=0.04$ (right). At $T=0.09$, the system remains commensurate: point $D$ lies on a true Fermi arc, while $E_1$ and $E_2$ lie on false segments. At $T=0.04$, the EDC at $D$ develops a pseudogap, while at $E_2$ a peak appears at $\omega=0$ on the hole-like Fermi surface. This peak signals a precursor of the \emph{middle} SDW band, absent in the commensurate case. While commensurate order yields only lower and upper AFM bands, the incommensurate case can produce up to four spectral peaks \cite{vilk2026pseudogap}.}
\label{fig:EDC_panel_n_0_88_T_0_09_0_04}
\end{figure}

To clarify this behavior, \fref{fig:EDC_panel_n_0_88_T_0_09_0_04} shows EDCs at representative momenta. The layout follows \fref{fig:EDC_panel_n_0_913_T_0_125_0_08}, comparing $T=0.09$ (left) and $T=0.04$ (right). At $T=0.09$, the system remains commensurate: point $D$ lies on a true Fermi arc, while points $E_1$ and $E_2$ lie on false segments.

At $T=0.04$, the EDC at momentum point $D$ develops a pseudogap, while at point $E_2$ a peak appears at $\omega=0$ on the hole-like Fermi surface. This peak signals a precursor of the \emph{middle} SDW band, which is absent in the commensurate case. While commensurate order yields only lower and upper AFM bands, the incommensurate case can produce up to four spectral peaks \cite{vilk2026pseudogap}.

\subsection{Case $P_3$ ($\delta=0.18$): Strongly incommensurate regime---no antinodal pseudogap; diagonal pseudogap for $T < T_D$.}
\label{sec:Case_P3}

\subsubsection{Evolution of the zero-energy spectral function $A(\mathbf{k},0)$ and the imaginary part of the self-energy $\Sigma''(\mathbf{k},0)$ with decreasing temperature}
\label{Case_P_3a}

\begin{figure}[!t]
\centering
\includegraphics[width=\columnwidth]{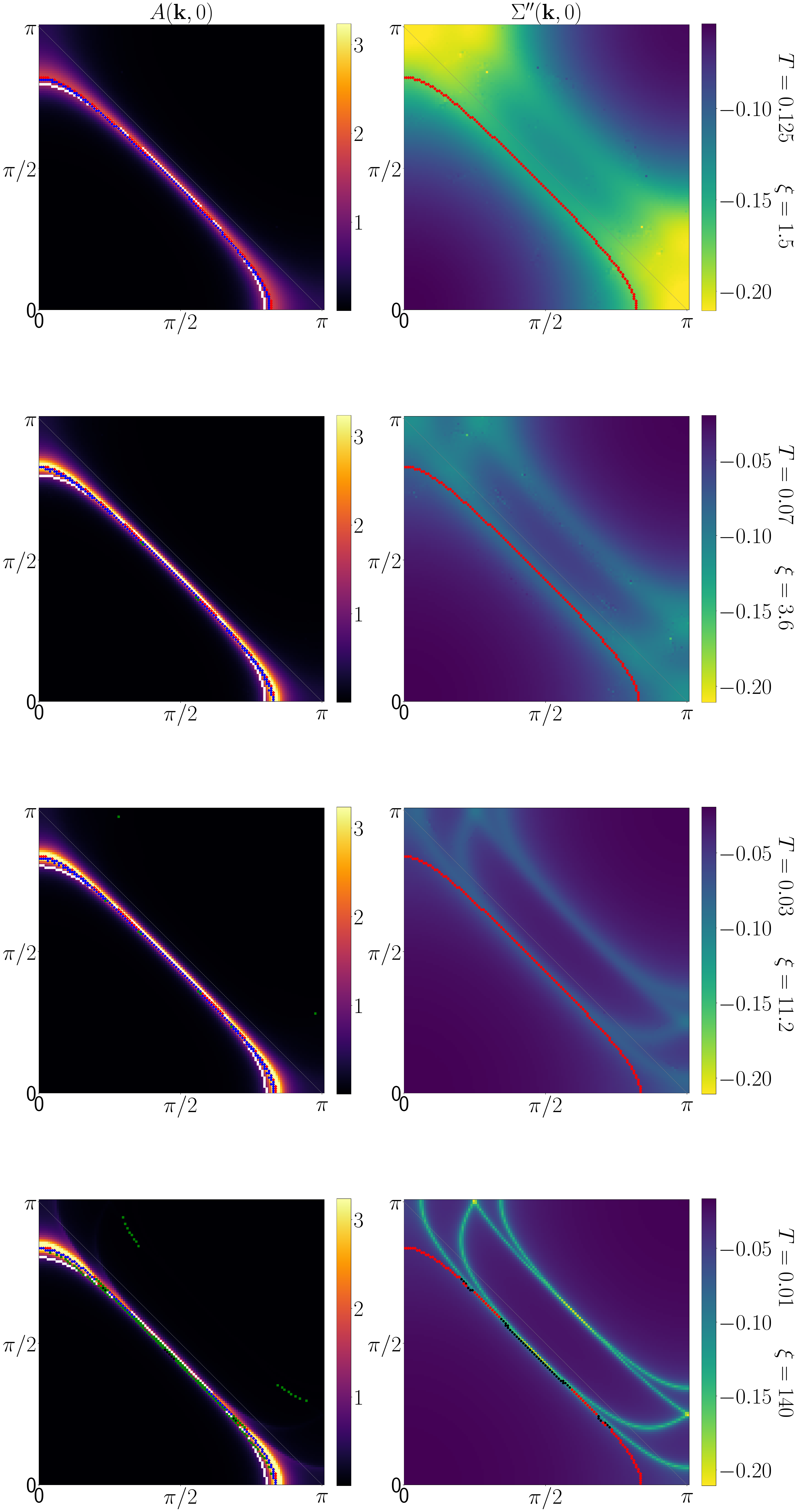}
\caption{Temperature evolution of $A(\mathbf{k},0)$ (left) and $\Sigma''(\mathbf{k},0)$ (right), with temperature decreasing from top to bottom ($U=6$, $\delta=0.18$). The notation follows \fref{fig:Akw_panel_heatmap_vers_T_n_0_913}. For all temperatures, spin fluctuations are strongly incommensurate. At low $T$, $\Sigma''(\mathbf{k},0)$ develops four high-intensity regions associated with the incommensurate peaks $\mathbf{Q}_i$ in $\chi_{sp}(\mathbf{q},0)$, indicating four shadow Fermi surfaces. Two of these nearly coincide with the interacting FS and the ZEQ line near the diagonal, where the system exhibits a pseudogap (false FS) for $T < T_D$ }
\label{fig:Akw_panel_heatmap_vers_T_n_0_82}
\end{figure}

In \fref{fig:Akw_panel_heatmap_vers_T_n_0_82}, we show the evolution of the spectral function $A(\mathbf{k},0)$ (left) and the imaginary part of the self-energy $\Sigma''(\mathbf{k},0)$ (right), with temperature decreasing from top to bottom. The notation follows \fref{fig:Akw_panel_heatmap_vers_T_n_0_913}. At this doping, spin fluctuations are strongly incommensurate across all investigated temperatures ($T=0.125, 0.07, 0.03, 0.01$).

At the relatively high temperature $T=0.125$, the interacting FS (blue line) practically coincides with the non-interacting FS (white line), consistent with Luttinger's theorem. This corresponds to a regime where $T \ll W$ but still above the crossover to the renormalized classical regime, $T > T_{RC}$. In this regime, quantum effects dominate and preserve the Fermi surface volume. As the temperature decreases below $T_{RC}$, critical thermal fluctuations become dominant, shifting the interacting FS and expanding its area, thereby violating Luttinger's theorem. This change is sometimes referred to as an interaction-driven Lifshitz transition; however, the original Lifshitz transition is purely topological and does not involve a change in Fermi volume. Moreover, this effect occurs for both weak and strong interactions, indicating that interaction strength is not the primary governing factor. Its origin lies in low dimensionality, which leads to an extended renormalized classical regime where temperature is not the smallest energy scale.

At low temperatures, $\Sigma''(\mathbf{k},0)$ develops four high-intensity regions associated with the incommensurate peaks $\mathbf{Q}_i$ in $\chi_{sp}(\mathbf{q},0)$, indicating four shadow Fermi surfaces. Two of these nearly coincide with the interacting FS and the ZEQ line over an extended region around the diagonal (see left panels). In this region, the system exhibits a diagonal pseudogap (false FS) for $T < T_D$. 

\subsubsection{EDC analysis}
\label{Case_P_3b}

\begin{figure}[!t]
\centering
\includegraphics[width=\columnwidth]{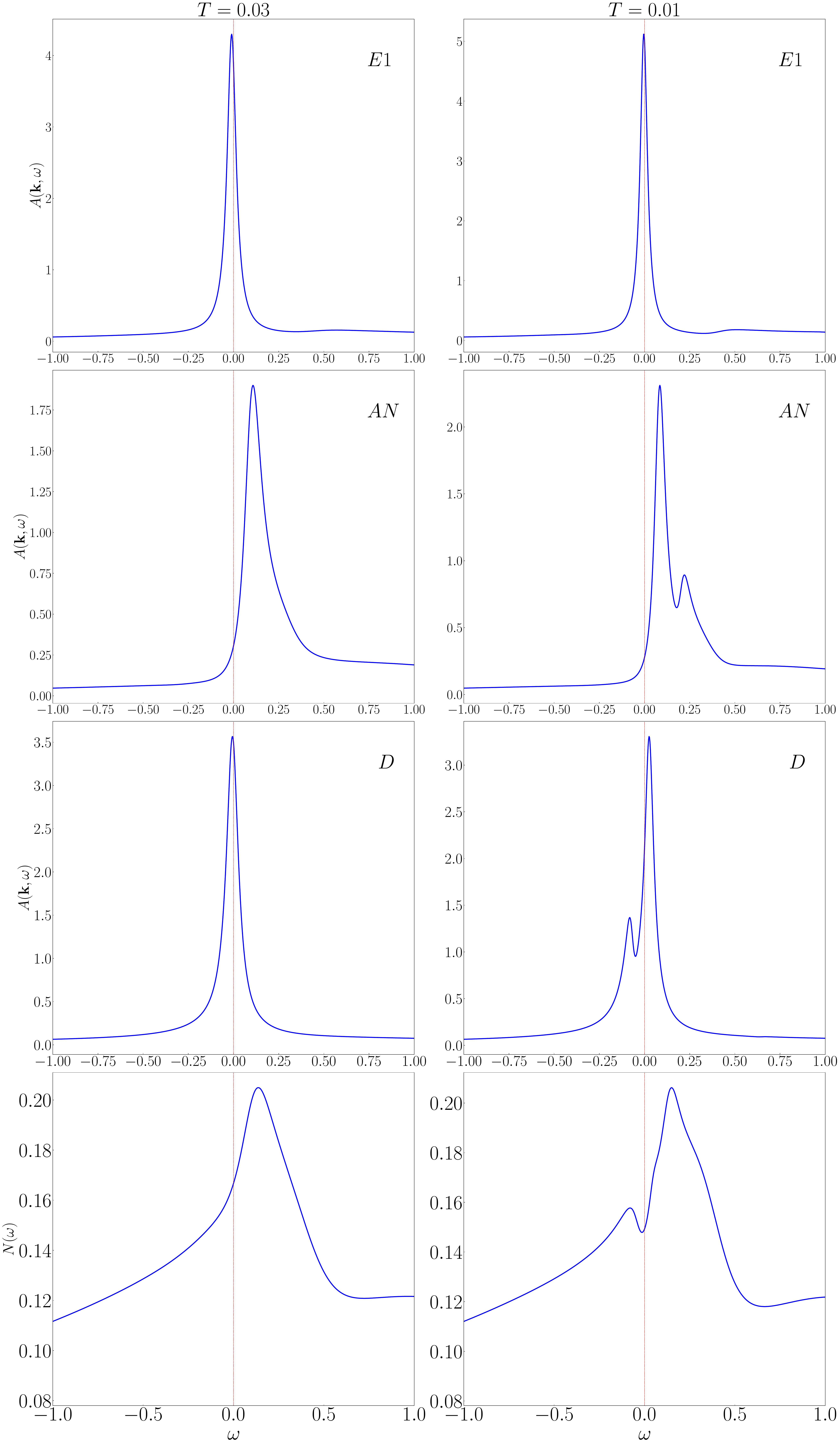}
\caption{EDCs at representative momenta ($U=6$, $\delta=0.18$). The notation follows \fref{fig:FS_ZEQ_E_D_points_notation}, comparing $T=0.03$ (left panels) and $T=0.01$ (right panels). At $T=0.03$, the EDCs for points $D$ and $E_1$ on the Fermi surface exhibit a peak at or very near $\omega=0$, indicating the absence of a pseudogap along both diagonal $(0,0)-(\pi,\pi)$ and antinodal $(0,0)-(\pi,0)$ directions. At the lower temperature $T=0.01$, the EDC at the diagonal point $D$ develops a pseudogap ($D$ lies on the false Fermi surface), while $E_1$ retains a peak at $\omega=0$ ($E_1$ lies on the true Fermi surface). For the antinodal point, no peak exists at negative frequencies, confirming the absence of an antinodal pseudogap. Nonetheless, the spectral function deviates from Fermi liquid theory, exhibiting two peaks in the unoccupied spectrum ($\omega>0$) that serve as precursors to SDW bands. The density of states $N(\omega)$ exhibits a pseudogap driven by hot spots close to the diagonal of the BZ.}
\label{fig:EDC_panel_n_0_82_T_0_03_0_01}
\end{figure}

Let us now turn to the EDC analysis for representative momentum points and the density of states. \fref{fig:EDC_panel_n_0_82_T_0_03_0_01} shows EDCs at characteristic momenta. The notation follows \fref{fig:FS_ZEQ_E_D_points_notation}, comparing $T=0.03$ (left panels) and $T=0.01$ (right panels). At $T=0.03$, the EDCs for the $D$ and $E_1$ points on the Fermi surface show a peak at or very near $\omega=0$, indicating the absence of a pseudogap in both the diagonal $(0,0)-(\pi,\pi)$ and antinodal (zone-edge) $(0,0)-(\pi,0)$ directions. The peak for the antinodal point $\mathbf{k}_{AN}=(\pi,0)$ resides at positive energy, similar to the non-interacting case. 

At the lower temperature $T=0.01$, the EDC for the diagonal point $D$ develops a pseudogap ($D$ lies on the false Fermi surface), whereas the $E_1$ point retains a peak at $\omega=0$ ($E_1$ lies on the true Fermi surface). At the antinodal point, there is no peak at negative frequencies, and thus no antinodal pseudogap opens. However, the spectral function is not consistent with a conventional Fermi or marginal Fermi liquid either: it exhibits two peaks in the unoccupied (invisible) part of the spectrum ($\omega > 0$). These are precursors of the SDW bands, which do not produce an antinodal pseudogap because the gap parameter $\Delta$ is insufficient to push the lower SDW peak below zero energy. The density of states $N(\omega)$ exhibits a pseudogap; however, contrary to the commensurate case, this pseudogap in $N(\omega)$ is driven by hot spots close to the diagonal of the BZ rather than by antinodal spectral weight suppression.

\subsubsection{Frequency dependence of the imaginary part of the self-energy}
\label{Case_P_3c}

We now analyze the imaginary part of the self-energy as a function of frequency for two key momentum points on the ZEQ line: the $D$ point on the BZ diagonal and the $E_1$ point on the $(0,0)-(\pi,0)$ edge of the BZ (see \fref{fig:FS_ZEQ_E_D_points_notation}). Note that in this case, the ZEQ line practically coincides with the interacting Fermi surface (see \fref{fig:Akw_panel_heatmap_vers_T_n_0_82}).

\begin{figure}[!t]
\centering
\includegraphics[width=\columnwidth]{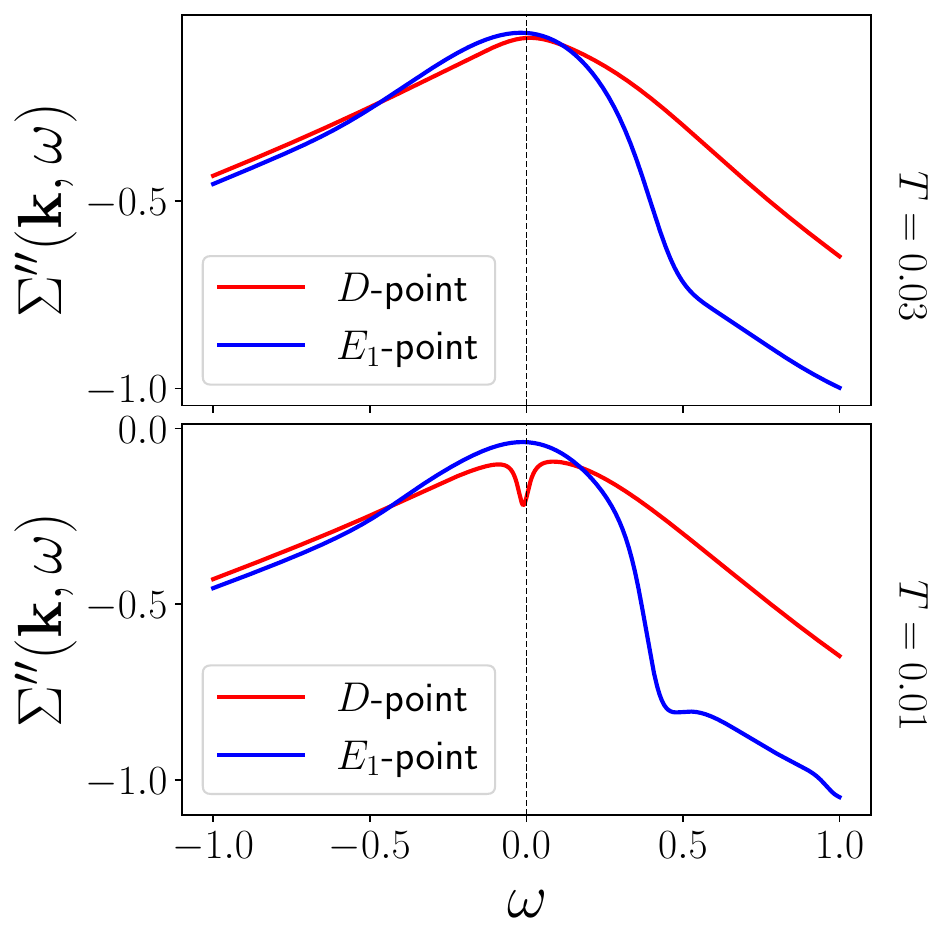}
\caption{Frequency dependence of the imaginary part of the self-energy $\Sigma''(\mathbf{k},\omega)$ for two momentum points on the ZEQ line: the $D$ point on the BZ diagonal and the $E_1$ point on the $(0,0)-(\pi,0)$ edge of the BZ (see \fref{fig:FS_ZEQ_E_D_points_notation}). Upper panel: $T=0.03$; lower panel: $T=0.01$ ($U=6$, $\delta=0.18$). For the $E_1$ point on the edge of the BZ, the low-frequency ($T < \omega \ll W$) dependence of $\Sigma''(\mathbf{k},\omega)$ exhibits a quadratic, Fermi-liquid behavior at both temperatures. For the diagonal point $D$, the low-frequency dependence is qualitatively different between the two temperatures: at $T=0.03$, $\Sigma''(\mathbf{k}_D,\omega)$ is approximately linear for $T < \omega \ll W$, indicating marginal Fermi liquid behavior, whereas at $T=0.01$, $\Sigma''(\mathbf{k}_D,\omega)$ develops a minimum at $\omega=0$, consistent with pseudogap behavior.}
\label{fig:Im_S_n_0_82_U_6_T_0_03_0_01}
\end{figure}

In \fref{fig:Im_S_n_0_82_U_6_T_0_03_0_01}, we present $\Sigma''(\mathbf{k},\omega)$ at these two momentum points for two temperatures, $T=0.03$ (upper panel) and $T=0.01$ (lower panel). For the $E_1$ point on the zone edge, the low-frequency ($T < \omega \ll W$) dependence of $\Sigma''(\mathbf{k},\omega)$ displays a quadratic, Fermi-liquid-like profile at both temperatures. In contrast, for the diagonal point $D$, the low-frequency behavior of $\Sigma''(\mathbf{k}_D,\omega)$ differs qualitatively between the two temperatures. At $T=0.03$, $\Sigma''(\mathbf{k}_D,\omega)$ is approximately linear for $T < \omega \ll W$, indicating  marginal Fermi liquid behavior. At the lower temperature $T=0.01$, $\Sigma''(\mathbf{k}_D,\omega)$ develops a minimum at $\omega=0$ rather than a maximum, consistent with pseudogap formation.

\section{Comparison of the TPSC+ with the DiagMC results}
\label{sec:Comp_with_DiagMC}

In this section, we examine the differences between the real-frequency spectral function $A(\mathbf{k},0)$ and its Matsubara proxy counterpart $A(\mathbf{k},ik_0)$. We then compare the TPSC+ results with DiagMC calculations for the Matsubara proxy $A(\mathbf{k},ik_0)$, as well as for the temperature dependence of the maxima of the spin susceptibility $\chi_{sp}(\mathbf{q},0)(T)$ and charge susceptibility $\chi_{ch}(\mathbf{q},0)(T)$. Finally, we compare our real-frequency pseudogap definition with two other criteria widely used in the literature: the maximum in the temperature dependence of the uniform spin susceptibility $\chi_{sp}(0,0)(T)$ (the Knight shift) and the criterion based on the imaginary part of the Matsubara self-energy.

\subsection{Spectral function}

As mentioned earlier, the Matsubara spectral function proxy $A(\mathbf{k},ik_0)$ is equivalent to the analytical continuation of the real zero-frequency spectral function $A(\mathbf{k},0)$ using a finite broadening $\eta=k_0=\pi T$. We thus begin by studying the effect of finite broadening on the analytical continuation.

\subsubsection{Effect of the finite broadening on the analytical continuation of the spectral function $A(\mathbf{k},0)$}
\label{sec:Effect_finite_eta}

\begin{figure}[!t]
\centering
\includegraphics[width=\columnwidth]{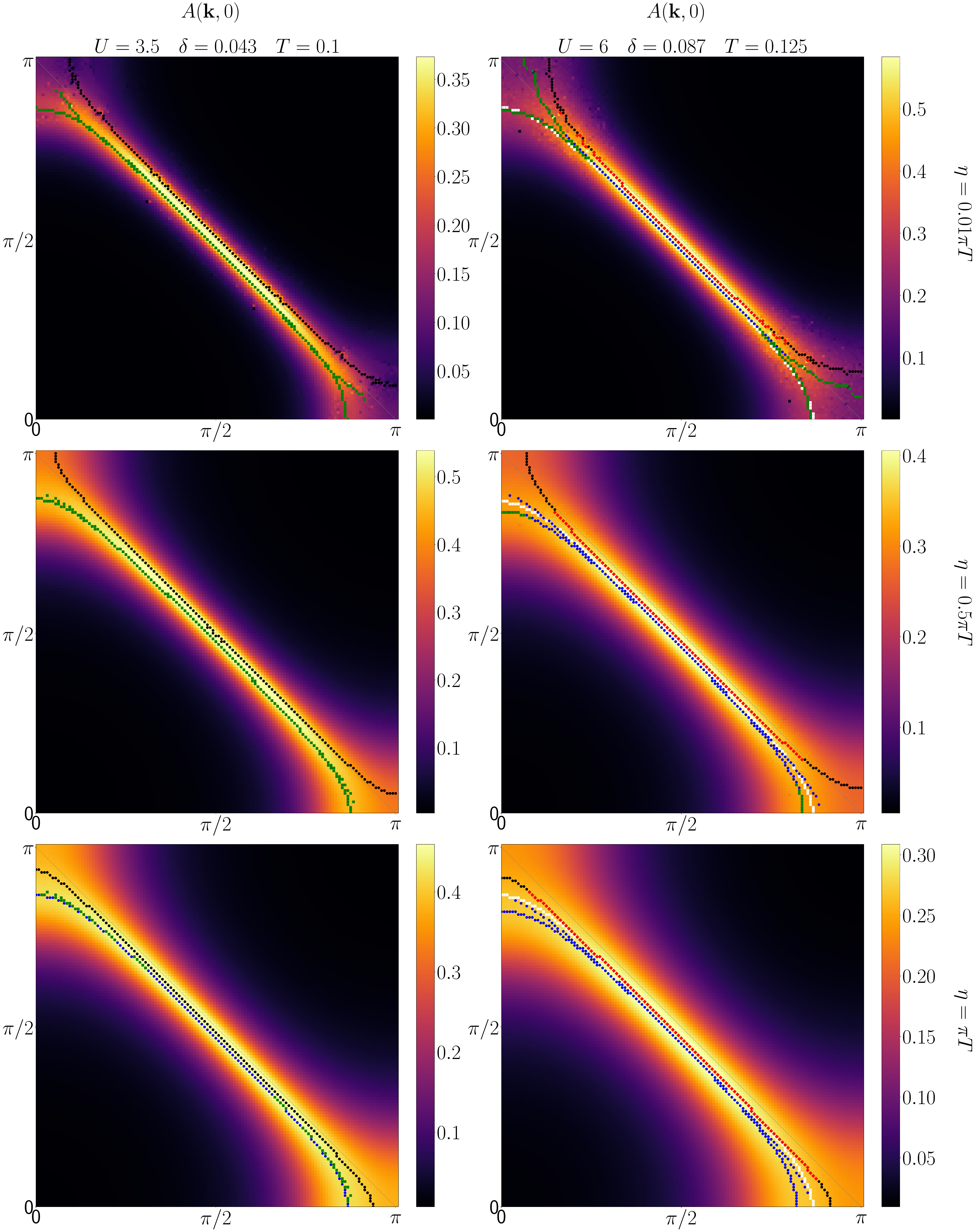}
\caption{Effect of finite broadening $\eta$ on the Padé analytical continuation of the spectral function $A(\mathbf{k},0)$. The parameters are shown in the figure. Notation follows \fref{fig:Akw_panel_heatmap_vers_T_n_0_913}. The left panels illustrate the weak-to-intermediate interaction regime, while the right panels represent the strong-to-intermediate interaction regime. In the first row, we show the result of the analytical continuation using a very small broadening $\eta=0.01 \pi T$, which is a good approximation to the theoretical limit $\eta \to 0$. In this case, $A(\mathbf{k},0)$ has a hole-like ZEQ line and two Fermi surfaces: one electron-like and one hole-like. As the broadening increases (second and third rows), the picture changes qualitatively. For $\eta=0.5 \pi T$, the ZEQ line is still hole-like, but the two Fermi surfaces practically merge and become electron-like. For an even larger $\eta=\pi T$, the ZEQ line also becomes electron-like.}
\label{fig:Akw_panel_heatmap_vers_eta_U_3_5_U_6}
\end{figure}

\fref{fig:Akw_panel_heatmap_vers_eta_U_3_5_U_6} shows the effect of finite broadening $\eta$ on the Padé analytical continuation of the spectral function $A(\mathbf{k},0)$. The parameters are shown in the figure, and the notation is the same as in \fref{fig:Akw_panel_heatmap_vers_T_n_0_913}. The left panels illustrate the weak-to-intermediate interaction regime, while the right panels represent the strong-to-intermediate interaction regime. In the first row, we show the result of the analytical continuation using a very small broadening, $\eta=0.01 \pi T$, which is a good approximation to the theoretical limit $\eta \to 0$. In this case, $A(\mathbf{k},0)$ has a hole-like ZEQ line and two Fermi surfaces: one electron-like and one hole-like. As the broadening increases (second and third rows), the picture changes qualitatively. For $\eta=0.5 \pi T$, the ZEQ line is still hole-like, but the two Fermi surfaces practically merge and become electron-like. For an even larger $\eta=\pi T$, the ZEQ line also becomes electron-like. In addition, the type of the Fermi surface changes as well. For example, for $U=3.5$ and $\eta=0.01 \pi T$, the two Fermi surfaces are both false (green lines), while for large broadening $\eta=\pi T$, which is equivalent to the Matsubara proxy $A(\mathbf{k},ik_0)$, the Fermi surface is classified as true (blue line). This occurs because, at such large broadening $\eta=0.314$, the pseudogap is filled in.

\subsubsection{Comparison of the TPSC+ results for the Matsubara proxy $A(\mathbf{k},ik_0)$ with the Monte Carlo results}
\label{sec:Comp_Ak_ik_0_with_DiagMC}

In this section, we compare the TPSC+ results for the Matsubara proxy $A(\mathbf{k},ik_0)$ with the Monte Carlo results \cite{vsimkovic2024origin} for the same quantity. Before discussing the results, we want to draw attention to the following important points:

\begin{enumerate}
\item
The data for the spectral function proxy $A(\mathbf{k},ik_0)$ are not available in the data repository for the DiagMC paper \cite{vsimkovic2024origin}. To extract the data from the heatmaps in \cite{vsimkovic2024origin}, we digitized the figures in the paper using open-source digitizing software \cite{digitize_heatmap}. The digitization works reasonably well, but there are some small artifacts, such as white pixels representing Python nan values.

\item
The results in \cite{vsimkovic2024origin} were obtained using a $64 \times 64$ momentum mesh and then extrapolated to the $256 \times 256$ lattice. The TPSC+ results, on the other hand, were obtained directly on a $256 \times 256$ lattice. As a result, they appear smoother.

\item
The Fermi surfaces and ZEQ lines were not differentiated as true or false in \cite{vsimkovic2024origin}. This differentiation requires analytical continuation to real frequencies and the corresponding EDC analysis (see \secref{sec:Characteristic_cases}), which were not performed in \cite{vsimkovic2024origin}. When comparing with the TPSC+ results, one should pay attention only to the position of the Fermi surface (green line on the DiagMC heatmap) and the ZEQ line (black line on the DiagMC heatmap), and ignore the color of the TPSC+ curves.
\end{enumerate}

\paragraph{Heatmaps comparison of $A(\mathbf{k},ik_0)$ for TPSC+ and DiagMC.}

\begin{figure}[!t]
\centering
\includegraphics[width=\columnwidth]{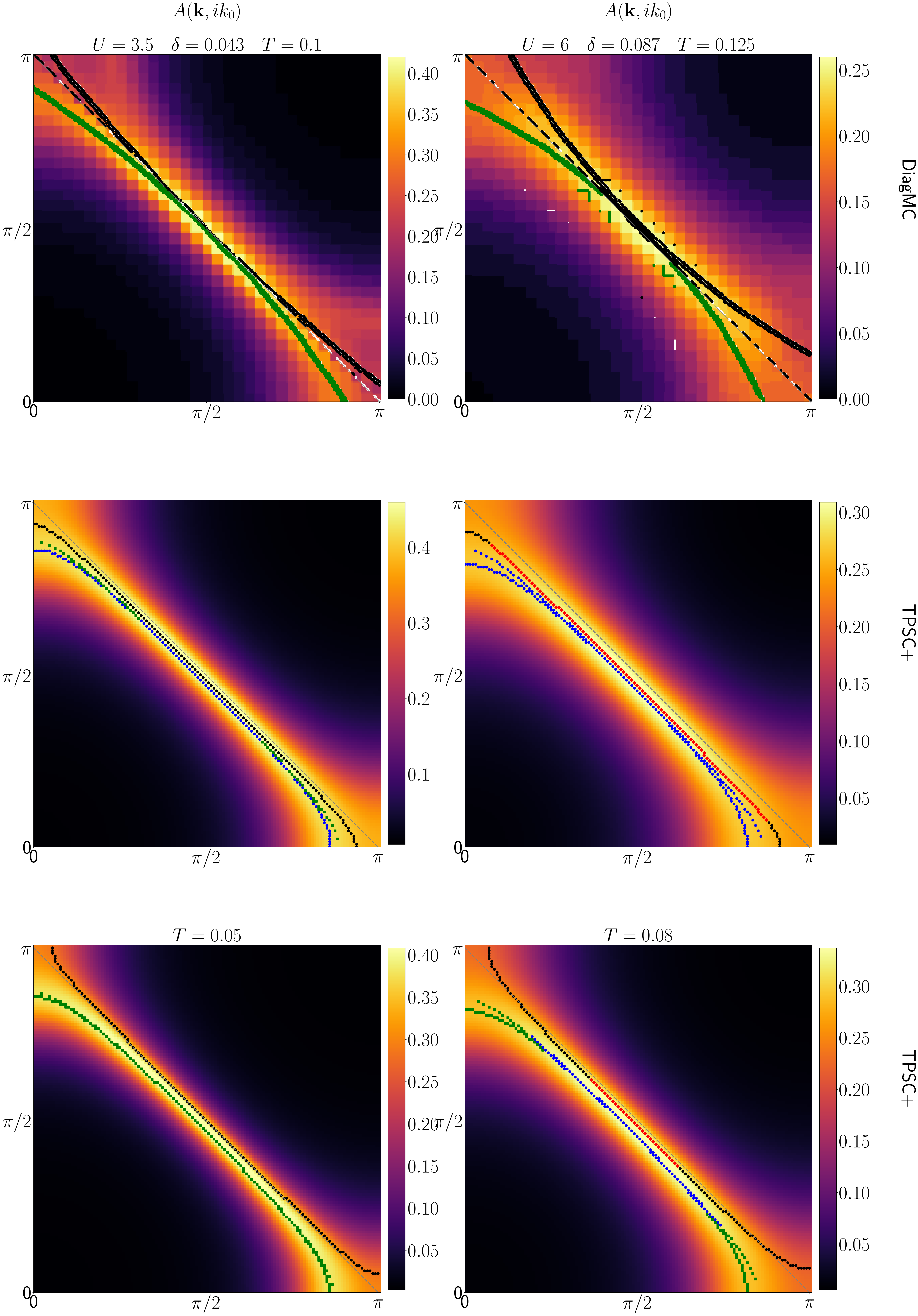}
\caption{Comparison of the TPSC+ results for the Matsubara proxy $A(\mathbf{k},ik_0)$ with the Monte Carlo results. The parameters are indicated in the figure. The DiagMC data were extracted by digitizing Fig.S2 of the Supplemental Material in Ref.~\cite{vsimkovic2024origin}. The panels on the left and right represent the weak-to-intermediate ($U=3.5$) and strong-to-intermediate ($U=6$) interaction limits, respectively. The first and second rows show the DiagMC and TPSC+ results for the same set of parameters. The third row shows TPSC+ results for the same interaction strength and doping as in DiagMC but at a lower temperature.}
\label{fig:Akw_heatmaps_comp_DiagMC_U_3_5_U_6}
\end{figure}

\fref{fig:Akw_heatmaps_comp_DiagMC_U_3_5_U_6} shows MDC heatmaps for the DiagMC and TPSC+ Matsubara spectral function proxy $A(\mathbf{k},ik_0)$. The parameters are indicated in the figure. The panels on the left and right represent the weak-to-intermediate ($U=3.5$) and strong-to-intermediate ($U=6$) interaction limits, respectively. The first and second rows show the DiagMC and TPSC+ results for the same set of parameters. The third row shows TPSC+ results for the same interaction strength and doping as in DiagMC but at a lower temperature.

Before discussing the comparison with the DiagMC results, we want to demonstrate the equivalence between the Matsubara proxy of the spectral function $A(\mathbf{k},ik_0)$ and the real zero-frequency spectral function $A(\mathbf{k},0)$ with finite broadening $\eta=k_0=\pi T$. Indeed, the MDC heatmaps in the third row of \fref{fig:Akw_panel_heatmap_vers_eta_U_3_5_U_6} are virtually identical to the MDC heatmaps in the second row of \fref{fig:Akw_heatmaps_comp_DiagMC_U_3_5_U_6}.

We now compare the DiagMC and TPSC+ results in \fref{fig:Akw_heatmaps_comp_DiagMC_U_3_5_U_6} for exactly the same parameters (the first and second rows). We can see that the DiagMC and TPSC+ heatmaps are similar in color pattern and intensity, and the Fermi surfaces are electron-like in both cases, but there is one noticeable difference. The ZEQ line has already become hole-like in DiagMC but not in TPSC+. However, as we saw in the previous subsection, the position and curvature of the ZEQ line are very sensitive to the value of the broadening $\eta$. In particular, for somewhat lower broadening $\eta=0.5 \pi T$, the ZEQ line is also hole-like in TPSC+ (see the second row of \fref{fig:Akw_panel_heatmap_vers_eta_U_3_5_U_6}).

The Matsubara proxy of the ZEQ line, $ik_0+\epsilon(\mathbf{k})+\Sigma(\mathbf{k},ik_0) - \mu=0$, eventually becomes hole-like in TPSC+ as well, but at a lower temperature. This can be seen in the third row of \fref{fig:Akw_heatmaps_comp_DiagMC_U_3_5_U_6}.

The fact that the hole-like ZEQ line appears in TPSC+ at lower broadening or lower temperature than in DiagMC suggests that TPSC+ underestimates the pseudogap effect.

\paragraph{Comparison of $A(\mathbf{k},ik_0)$ for TPSC+ and DiagMC along some high-symmetry directions.}

\begin{figure}[!t]
\centering
\includegraphics[width=\columnwidth]{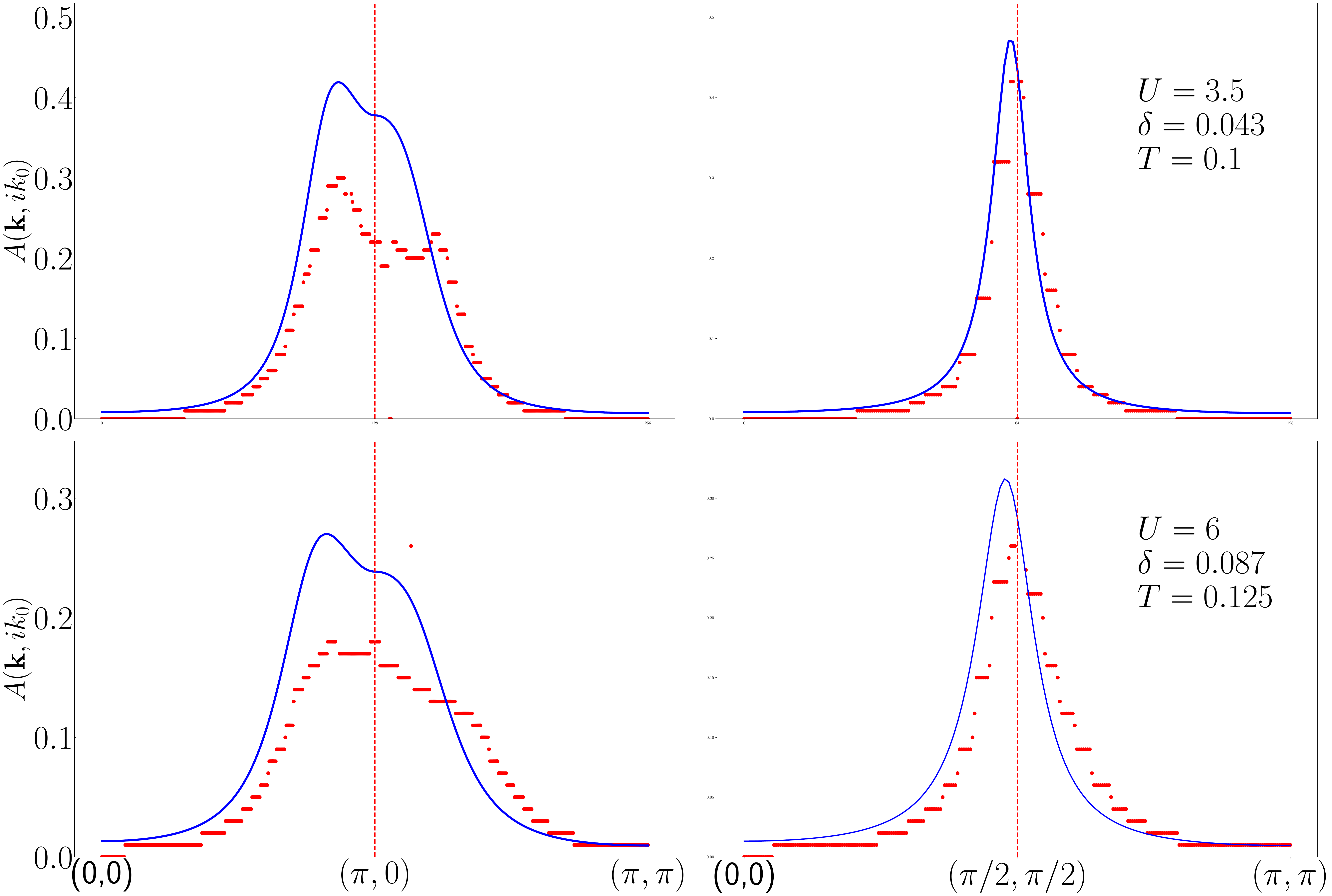}
\caption{Comparison of the TPSC+ results for the Matsubara proxy $A(\mathbf{k},ik_0)$ with the Monte Carlo results along two directions: the edge of the BZ $(0,0)-(\pi,0)-(\pi,\pi)$ and the diagonal of the BZ $(0,0)-(\pi,\pi)$. The blue lines are the TPSC+ results, and the red symbols are the DiagMC results. The DiagMC data were extracted by digitizing Fig.S2 of the Supplemental Material in Ref.~\cite{vsimkovic2024origin}. The first and second rows represent the weak-to-intermediate ($U=3.5$) and strong-to-intermediate ($U=6$) interaction limits, respectively. The TPSC+ value of $A(\mathbf{k},ik_0)$ is somewhat higher, indicating that the critical thermal spin fluctuations have a smaller pseudogap effect in TPSC+ than in DiagMC. }
\label{fig:Akw0_Edge_Diag_dir_high_comp_DiagMC}
\end{figure}

To examine this issue quantitatively, we show in \fref{fig:Akw0_Edge_Diag_dir_high_comp_DiagMC} the momentum dependence of $A(\mathbf{k},ik_0)$ along two directions: the edge of the BZ $(0,0)-(\pi,0)-(\pi,\pi)$ and the diagonal of the BZ $(0,0)-(\pi,\pi)$. The blue lines are the TPSC+ results, and the red symbols are the DiagMC results. We can see that, in both the weak ($U=3.5$) and strong ($U=6$) interaction limits, the TPSC+ value of $A(\mathbf{k},ik_0)$ is somewhat higher, which means that the critical thermal spin fluctuations have a smaller pseudogap effect in TPSC+ than in DiagMC.
           
\paragraph{Hole-like FS and ZEQ line. Difference between the spectral function $A(\mathbf{k},0)$ and its Matsubara proxy $A(\mathbf{k},ik_0)$.}

One of the noticeable artifacts of the Matsubara spectral function proxy $A(\mathbf{k},ik_0)$ is the disconnect between the temperature evolution of the Fermi surface and that of the ZEQ line. In TPSC+, for $A(\mathbf{k},ik_0)$, the former always stays electron-like, while the latter always becomes hole-like with decreasing temperature. In DiagMC, similar behavior is observed in the weak-interaction limit, and only in the strong-interaction limit can the Fermi surface become hole-like as well. By contrast, for the real zero-frequency spectral function $A(\mathbf{k},ik_0)|_{\eta \to 0}$, one of the Fermi surface lines in TPSC+ always becomes hole-like with decreasing temperature, and this occurs soon after the ZEQ line becomes hole-like.

To understand the origin of this difference in behavior, we again use the equivalence between the Matsubara spectral function proxy $A(\mathbf{k},ik_0)$ and its real-frequency counterpart with finite broadening, $A(\mathbf{k},0)|_{\eta = \pi T}$. As we saw in \secref{sec:Effect_finite_eta}, finite broadening affects both the maximal line of $A(\mathbf{k},0)$ (Fermi surface) and the ZEQ line \eref{eq:zero_energy_line}. However, its impact is stronger on the spectral function:

\begin{equation} \label{eq:Akw_eta}
A(\mathbf{k}, \omega, i \eta) =\frac{1}{\pi} \text{Im} \frac{1}{\omega +i \eta -\epsilon(\mathbf{k}) - \Sigma(\mathbf{k}, \omega, i \eta ) +\mu}.
\end{equation}

This is because $\eta$ enters both explicitly, through $\omega+i\eta$, and implicitly, through the self-energy. In contrast, the ZEQ line \eref{eq:zero_energy_line} is affected only implicitly, through $\Sigma'(\mathbf{k},0,i\eta)$. As $U$ increases, $\Sigma(\mathbf{k},0,i\eta)$ grows and eventually dominates over the explicit $i\eta$ term in \eref{eq:Akw_eta}. This likely explains why Ref.~\cite{vsimkovic2024origin} finds a hole-like maximal line of $A(\mathbf{k},ik_0)$ only at large $U$, whereas the ZEQ line is already hole-like for weaker interaction (provided $T$ is sufficiently low).

Turning back to TPSC+, the method underestimates the pseudogap effect relative to the DiagMC results in the strong-interaction regime. As a result, the explicit contribution of $i\eta=i\pi T$ in \eref{eq:Akw_eta} remains comparatively important even at large $U$. This explains why TPSC+ does not produce a hole-like maximal line for the Matsubara proxy $A(\mathbf{k},ik_0)$, although such a feature does appear in the $\eta \to 0$ limit for all $U$. Based on the behavior of the real-frequency spectral function $A(\mathbf{k},0)$, the new phase, characterized by a hole-like Fermi surface and a hole-like ZEQ line, exists for both weak and strong interactions, provided the temperature is sufficiently low.
  
\subsection{Comparison of $\chi_{sp}(\mathbf{q}_{max}, 0)(T)$ and $\chi_{ch}(0, 0)(T)$ with DiagMC results }
\label{sec:Comp_spin_charge_DiagMC}

\begin{figure}
    \centering
    \includegraphics[width=\columnwidth]{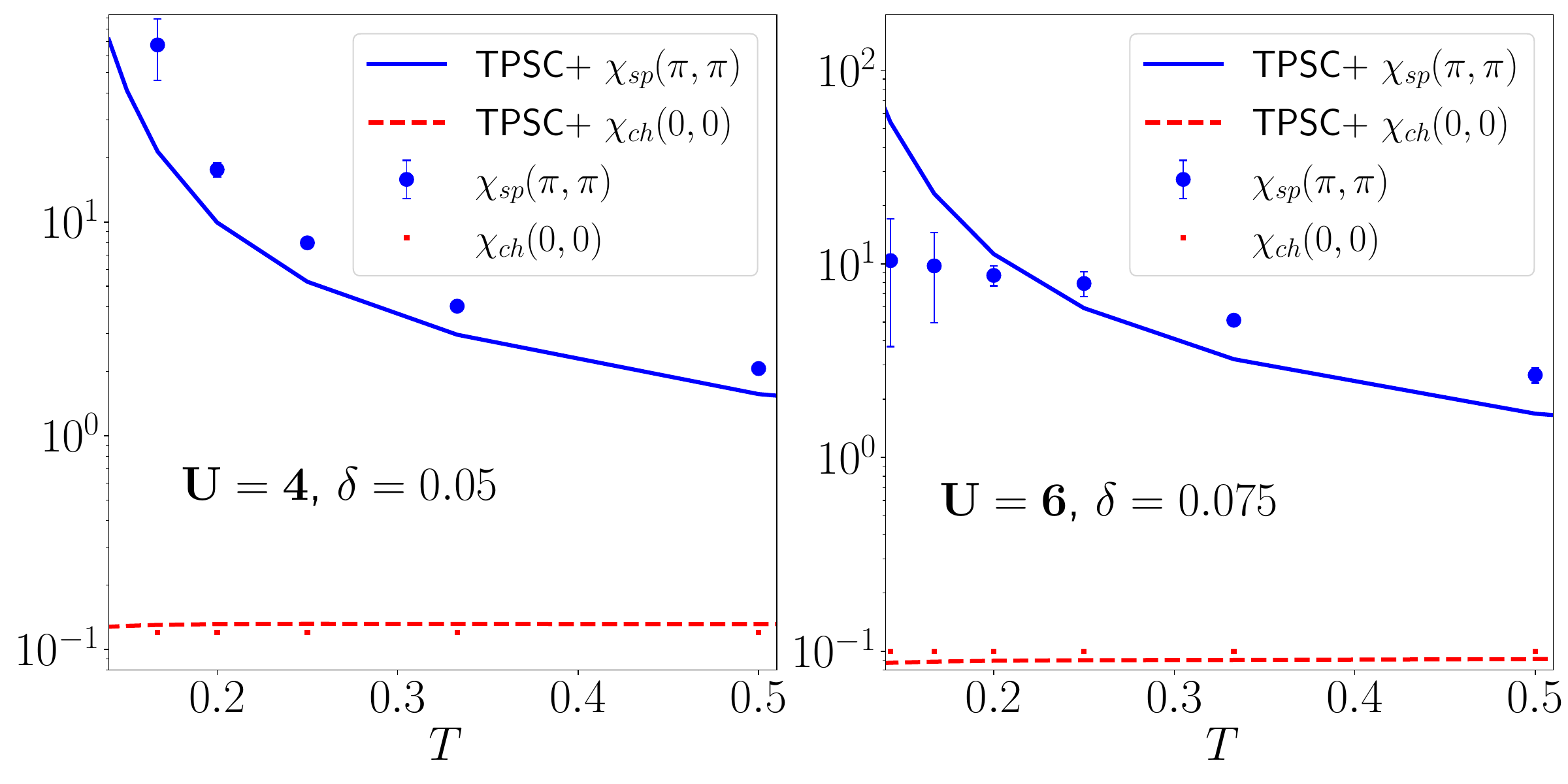} 
    \caption{The temperature dependence of the maximum values of the spin $\chi_{sp}(\mathbf{q}_{max}, 0)$ and charge $\chi_{ch}(0, 0)$ susceptibilities. The lines are TPSC+ results, and the symbols are DiagMC results from \cite{vsimkovic2024origin}. At low temperatures, the TPSC+ calculations show the onset of the renormalized classical regime with rapidly increasing $\chi_{sp}(\mathbf{q}_{max}, 0)$. For $U=4$ the DiagMC calculations also show the onset of the RC regime for $U=4$. For stronger interaction $U=6$, TPSC+ overestimates the crossover temperature to the renormalized classical $T_{RC}$. As a result, it overestimates $\chi_{sp}(\mathbf{q}_{max}, 0)$ at low temperatures}   
    \label{fig:chispmax_and_chich00_vs_T_TPSCpDiagMC}
\end{figure}

\fref{fig:chispmax_and_chich00_vs_T_TPSCpDiagMC} shows the temperature dependence of the maximum values of the spin $\chi_{sp}(\mathbf{q}_{max}, 0)$ and charge $\chi_{ch}(0, 0)$ susceptibilities. The lines are TPSC+ results, and the symbols are DiagMC results from \cite{vsimkovic2024origin}. The parameters are shown in the figures. The spin fluctuations are commensurate in this case, $\mathbf{q}_{max}=\mathbf{Q}=(\pi,\pi)$. At low temperatures, the TPSC+ calculations show the onset of the renormalized classical regime, with rapidly increasing $\chi_{sp}(\mathbf{q}_{max}, 0)$. For $U=4$, the DiagMC calculations also show the onset of the RC regime. The agreement of the TPSC+ results with the DiagMC results is fairly good in this case. For the stronger interaction $U=6$, TPSC+ overestimates the crossover temperature to the renormalized classical regime, $T_{RC}$. As a result, it overestimates $\chi_{sp}(\mathbf{q}_{max}, 0)$ at low temperatures.

\subsection{Comparison with some other pseudogap definitions used in the literature}
\label{sec:Comp_PG_def}

\subsubsection{Maximum in the temperature dependence of the uniform magnetic susceptibility (Knight shift) $\chi_{sp}(0,0)$}
\label{sec:Max_chisp00_vs_T}

One of the first indications of the pseudogap in high-$T_c$ superconductors was the observation of a maximum in the temperature dependence of the uniform magnetic susceptibility (Knight shift) $\chi_{sp}(0,0)$ \cite{johnston1989magnetic}. The temperature at which $\chi_{sp}(0,0)(T)$ reaches a maximum, $T_{max}$, was used as a proxy for the PG temperature in cold-atom and quantum Monte Carlo (QMC) simulations of the Hubbard model~\cite{chalopin2026observation}. It was also used as a secondary PG criterion in the DiagMC calculations \cite{vsimkovic2024origin}. From the ARPES perspective, the PG temperature depends on momentum, and it is therefore interesting to examine how $T_{max}$ relates to the diagonal $T_D$ and antinodal $T_{AN}$ PG temperatures.

\begin{figure}
\centering
\includegraphics[width=\columnwidth]{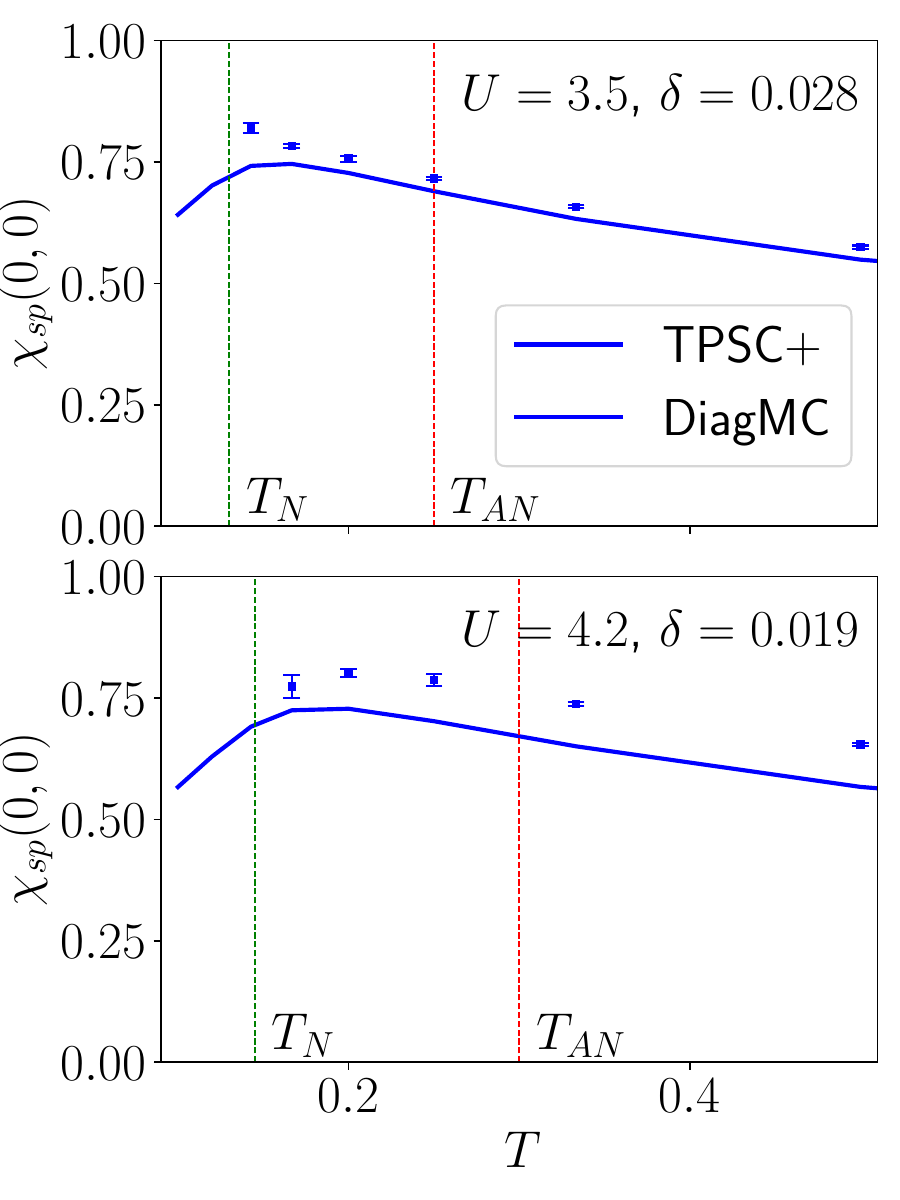}
\caption{ The temperature dependence of the uniform magnetic susceptibility (Knight shift) $\chi_{sp}(0,0)(T)$. Solid lines are TPSC+ results, the symbols are DiagMC results from \cite{vsimkovic2024origin}, and the reference lines indicate the nodal $T_D$ and antinodal $T_{AN}$ PG temperatures obtained using TPSC+ and Padé approximation for analytical continuation.}
\label{fig:chi00_vs_T_MC_and_TPSCpl_2}
\end{figure}

\fref{fig:chi00_vs_T_MC_and_TPSCpl_2} shows the temperature dependence of the uniform magnetic susceptibility (Knight shift) $\chi_{sp}(0,0)(T)$. Solid lines represent TPSC+ results, while the symbols are DiagMC results from \cite{vsimkovic2024origin}. The reference lines indicate the nodal point on the diagonal $T_N$ and antinodal $T_{AN}$ PG temperatures obtained using TPSC+ and the Padé approximation for analytical continuation. TPSC+ somewhat underestimates $\chi_{sp}(0,0)(T)$, but overall there is good agreement with DiagMC. Importantly, the maximum of $\chi_{sp}(0,0)(T)$ lies between the nodal and antinodal PG temperatures, $T_N < T_{max} < T_{AN}$. This implies that, at $T_{max}$, only part of the Fermi surface is gapped. For both sets of parameters, $T_{max}$ is closer to $T_N$ than to $T_{AN}$.

\subsubsection{Comparison with the pseudogap criterion based on the imaginary part of the Matsubara self-energy}
\label{sec:criterion_imag_Mats_Self}

A widely used momentum-resolved definition of the pseudogap is based on a criterion for the imaginary part of the Matsubara self-energy $\Sigma(\mathbf{k}, ik_0)$ \cite{Schaefer2021,vsimkovic2024origin}. According to this criterion, a pseudogap is present when
\begin{equation}
\label{eq:Mats_PG_criterion}
|\text{Im} \Sigma(\mathbf{k}, ik_0)| > |\text{Im} \Sigma(\mathbf{k}, ik_1)|
\end{equation}

For the diagonal pseudogap, Ref.~\cite{vsimkovic2024origin} selects the momentum point $\mathbf{k}_D$ where $|\text{Im} \Sigma(\mathbf{k}, ik_0)|$ is maximal along the BZ diagonal, while for the antinodal case they use the point $\mathbf{k}_E$ where it is maximal along the BZ edge.

As shown in \cite{vilk2024antiferromagnetic}, this Matsubara criterion significantly underestimates the pseudogap temperature in the weak-interaction case $U=2$ at half-filling. In both DiagMC \cite{Schaefer2021} and TPSC+, it yields antinodal $T_{AN}=0.0654$ and nodal $T_{N}=0.0625$, whereas the real-frequency spectra $A(\mathbf{k}, \omega)$ in TPSC+ indicate pseudogap formation at significantly higher temperatures, $T_{AN}=0.1$ and $T_{N}=0.08$.

\begin{figure}
\centering
\includegraphics[width=\columnwidth]{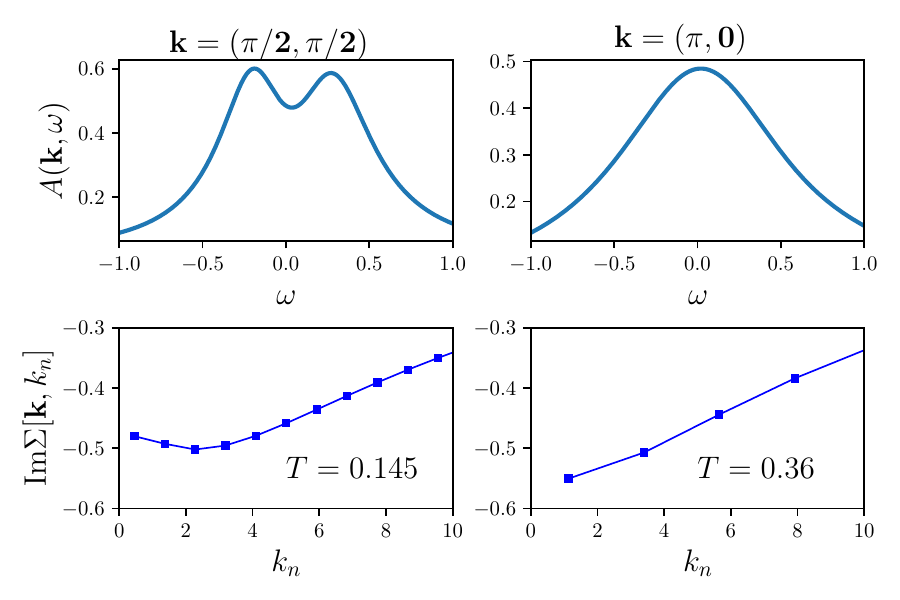}
\caption{Upper panels: spectral function $A(\mathbf{k}, \omega)$ at the nodal (left) and antinodal (right) points. Lower panels: imaginary part of the Matsubara self-energy $\Sigma(\mathbf{k}, ik_n)$ at $D$ (left) and $E$ (right) (see text). Parameters: $U=4.2$, $\delta=0.019$. In this case, the Matsubara PG criterion \eref{eq:Mats_PG_criterion} yields a false negative for the diagonal PG  and a false positive for the antinodal PG.}
\label{fig:Akw_Sigma_om_n_U_4_2_n_0_981}
\end{figure}

We now compare the Matsubara criterion with real-frequency spectra at larger interaction strength and away from half-filling within TPSC+. The upper panels of \fref{fig:Akw_Sigma_om_n_U_4_2_n_0_981} display $A(\mathbf{k}, \omega)$ at the nodal (left) and antinodal (right) points, while the lower panels show $\text{Im}\,\Sigma(\mathbf{k}, ik_n)$ at $D$ (left) and $E$ (right). The parameters match those of the lower panel in \fref{fig:chi00_vs_T_MC_and_TPSCpl_2}: $U=4.2$, $\delta=0.019$.

At the nodal point, $A(\mathbf{k}, \omega)$ exhibits a pseudogap at $T=0.145$, below the renormalized classical crossover $T_{RC} \approx 0.3$. However, the Matsubara criterion \eref{eq:Mats_PG_criterion} does not detect a pseudogap. Conversely, at the antinodal point and $T=0.36 > T_{RC}=0.3$, $A(\mathbf{k}, \omega)$ shows a single peak at $\omega=0$, while the Matsubara criterion signals a pseudogap. Thus, in this regime, the Matsubara criterion yields a false negative for the diagonal pseudogap ($T=T_D < T_{RC}$) and a false positive for the antinodal case ($T > T_{RC}$).

These results indicate that the Matsubara criterion primarily reflects the influence of thermal spin fluctuations on the Matsubara self-energy, rather than providing a direct measure of pseudogap formation.

\section{Analytical results for the self-energy and the spectral function in the renormalized classical regime}
\label{sec:Analytical_results}

In the Renormalized Classical (RC) regime, the frequency and momentum dependence of the self-energy is dominated by the thermal classical contribution \cite{Vilk1997}. In the following analysis, we use the asymptotic results obtained within this approximation.

While quantum contributions to the self-energy are not negligible, they exhibit comparatively weak dependence on momentum and frequency and can therefore be absorbed into the chemical potential. Consequently, in the asymptotic expressions below, the chemical potential in the full Green function $G$ should be understood as being shifted by the local self-energy effects due to quantum fluctuations:

\begin{equation}
\tilde{\mu}_2 =\mu - \Sigma_l(0)
\label{eq:renorm_mu}
\end{equation}
where $\Sigma'_l(0) = \sum_{\mathbf{k}} \Sigma'(\mathbf{k},0) $.  The chemical potential $\tilde{\mu}_2$ in $\mathcal{G}^{(2)}(\mathbf{k},ik_n)$ fixes the occupation $n$ so that it is identical to the occupation obtained from $\mu_1$ in $\mathcal{G}^{(1)}(\mathbf{k},ik_n)$.

Above the renormalized classical regime,

\begin{equation}
\tilde{\mu}_2 \approx \mu_1 .
\label{eq:renorm_mu_above_RC}
\end{equation}
This is because the crossover temperature to the RC regime is much smaller than the bandwidth, $T_{RC} \ll W$. As a result, Luttinger's theorem is approximately satisfied for $T_{RC} <T \ll W$, and the volume of the Fermi surface is approximately the same as that of the non-interacting system. Moreover, as we found numerically (see \secref{sec:Characteristic_cases}), the shape of the Fermi surface does not change significantly either, implying very weak momentum dependence of the self-energy above the renormalized classical regime. Only below $T <T_{RC}$ does the self-energy $\Sigma(\mathbf{k},0)$ develop quasi-singularities in momentum space, due to the thermal critical spin fluctuations. In this regime, the absolute value $|\tilde{\mu}_2|$ becomes larger than $|\mu_1|$, but the two remain comparable.

The analytical expression for a regular (non-Van Hove) momentum, $v_{\mathbf{k}+\mathbf{Q}} > \gamma \xi^{-1}$, has the form \cite{Vilk1997,vilk1997shadow}:

\begin{equation} \label{eq:sEnergy_st_Reg_R}
\begin{split}
\Sigma^R(\mathbf{k},\omega) & = \frac{\tilde{g} T U}{2 \pi \xi_0^2 \sqrt{\tilde{\omega}^2 +v_{\mathbf{k}+\mathbf{Q}}^2 \xi^{-2}}} \\
&\left[ \ln \frac{\tilde{\omega}+\sqrt{\tilde{\omega}^2 +v_{\mathbf{k}+\mathbf{Q}}^2 \xi^{-2}}}{\tilde{\omega}-\sqrt{\tilde{\omega}^2 +v_{\mathbf{k}+\mathbf{Q}}^2 \xi^{-2}} } \right]
\end{split}
\end{equation}

Here:

\begin{equation} \label{eq:tilde_omega}
\tilde{\omega} = \omega_k +i \eta
\end{equation}

and

\begin{equation} \label{eq:omega_k}
\omega_k = \omega - \epsilon(\mathbf{k}+\mathbf{Q}) +\mu_1 .
\end{equation}

At zero frequency, this expression for the self-energy $\Sigma(\mathbf{k},0)$ has a quasi-singularity in momentum space at the shadow Fermi surface $\epsilon(\mathbf{k}+\mathbf{Q}) =\mu_1$.

In the RC regime, the correlation length grows very rapidly, and the energy scale $v_{\mathbf{k}+\mathbf{Q}} \xi^{-1}$ becomes very small. In this case, expression \eref{eq:sEnergy_st_Reg_R} can be simplified further when one moves slightly away from the shadow Fermi surface:

\begin{equation} \label{eq:high_freq_cond}
|\omega - \epsilon(\mathbf{k}+\mathbf{Q}) +\mu_1|=|\omega_k| \gg v_{\mathbf{k}+\mathbf{Q}} \xi^{-1}
\end{equation}

\begin{equation} \label{eq:ReSigma_N_high}
\begin{split}
\Sigma^R(\mathbf{k},\omega) &=\frac{\Delta^2}{\omega_k } \left[1 -i \frac{\pi \operatorname{sign}(\omega_k)}{2\ln(\xi)} \right]
\end{split}
\end{equation}
where $\Delta^2$ is given by:

%%%%%%%%%%%%%%%
\begin{align}
\Delta^2=\frac{\tilde{g} U T \ln{\xi}}{\pi \xi_0^2} .
\label{eq:Delta_0_sigma_2}
\end{align}
%%%%%%%%%

For the antinodal point, the Van Hove point $v=0$, expression \eref{eq:sEnergy_st_Reg_R} contains a logarithmic divergence, rendering the expression invalid at this specific point. The analytical expressions for the Van Hove point were obtained in \cite{vilk2023criteria,vilk2025_crossovers}. In the limit $|\tilde{\omega}| \gg \gamma \xi^{-2}$, the expression is similar to \eqref{eq:ReSigma_N_high} but includes an additional factor of $1/2$ for the imaginary part \cite{vilk2023criteria}. This difference affects only points very close to the Van Hove point, $v \ll \gamma \xi^{-1}$, and is unimportant for qualitative analysis; therefore, in the analytical results below, we apply \eref{eq:ReSigma_N_high} for all momenta.

The above expressions are for the case of commensurate critical spin fluctuations. In the incommensurate case, the self-energy is the sum of four terms of the above form, with the AFM vector $Q$ replaced by the incommensurate vectors $Q_i$. Further analytical results in this section are restricted to the commensurate case and to the model with nearest-neighbor-only hopping.
The expression for the Green function is given by:

\begin{equation} \label{eq:Green_func_Cl}
G^R(\mathbf{k},\omega)
=\frac{\omega_k}
{(\omega -E^{+}) (\omega  -E^{-})+ i \pi \Delta^2 \operatorname{sign} (\omega_k)/(2 \ln \xi)}.
\end{equation}
The energies $E_{\mathbf{k}}^{\pm}$ are defined as
%%%%%%%%%%%%%%%%
\begin{equation}
E_{\mathbf{k}}^{\pm} = \frac{1}{2}\left ( \epsilon^{(2)}_{\mathbf{k}} + \epsilon^{(1)}_{\mathbf{k}+\mathbf{Q}} \pm \sqrt{\left( \epsilon^{(2)}_{\mathbf{k}}  - \epsilon ^{(1)}_{\mathbf{k}+\mathbf{Q}}\right)^2+4\Delta^2} \right ).
\label{eq:eplusmoins}
\end{equation}
%%%%%%%%%%%%%%%%
%
Here, $\epsilon_{\mathbf{k}}^{(1)}=\epsilon_{\mathbf{k}} -\mu_1$ and $\epsilon_{\mathbf{k}}^{(2)}=\epsilon_{\mathbf{k}} -\tilde{\mu}_2$.

In the model with nearest-neighbor hopping, $\epsilon(\mathbf{k}+\mathbf{Q})=-\epsilon(\mathbf{k})$ in the commensurate case, and the above expression can be simplified as:

\begin{equation}
E_{\mathbf{k}}^{\pm} = -\frac{\tilde{\mu}_2+\mu_1}{2}  \pm \sqrt{ \left(\epsilon_{\mathbf{k}} +\frac{\mu_1-\tilde{\mu}_2}{2} \right)^2+\Delta^2} .
\label{eq:eplusmoins2}
\end{equation}

The energies $E_{\mathbf{k}}^{\pm}$ determine the position of the poles in the Green function at low temperatures and thus determine the dispersion of the precursors of the lower and upper AFM bands. If neither of these bands crosses the zero-energy level, we have a pseudogap in all directions, and the system heads toward an insulating state as $T \to 0$.

Let us now consider the condition under which this occurs. The precursor of the upper AFM band $E_{\mathbf{k}}^{+}$ can never cross the zero-energy level for a hole-doped system, because $\tilde{\mu}_2, \mu_1 <0$ in this case. The lower band $E_{\mathbf{k}}^{-}$ can reach zero energy only when the gap parameter $\Delta$ decreases with doping below:

\begin{equation}
\label{eq:AFM_metal_condition}
\Delta <-\frac{\tilde{\mu}_2+\mu_1}{2} .
\end{equation}

If this occurs, the system would become an AFM metal. However, condition \eref{eq:AFM_metal_condition} was derived assuming commensurate spin fluctuations. In reality, as doping increases, the system enters the renormalized classical regime with incommensurate spin fluctuations before condition \eref{eq:AFM_metal_condition} is satisfied. The incommensurate spin fluctuations lead to the appearance of hot spots close to the BZ diagonal. At the hot spots, the pseudogap always exists in the RC regime, because in this case the singularity in the self-energy occurs directly on the Fermi surface. On the other hand, the antinodal point $(\pi, 0)$ is located away from the hot spots, and the behavior of the spectral function there is qualitatively similar to the commensurate case considered above. This explains why the pseudogap at the hot spots close to the diagonal of the BZ persists all the way to the quantum critical point, while the antinodal pseudogap disappears before that point is reached, $\delta_{PG\_AN} < \delta_{QC}$.

Let us now consider the opposite case of low doping, when the spin fluctuations in the RC regime are commensurate and $\Delta > -\frac{\tilde{\mu}_2+\mu_1}{2}$. In this case, the equation $E_{\mathbf{k}}^{\pm}=0$ has no solution. At sufficiently low temperature, the pseudogap is present everywhere, and the true Fermi surface does not exist. This, however, does not mean that the maximum line of the spectral function $A(\mathbf{k},0)$ does not exist. Consider half-filling. In this case, $\tilde{\mu}_2=\mu_1=0$, and the spectral function $A(\mathbf{k},\omega)=-(1/\pi) \operatorname{Im} G(\mathbf{k},\omega)$ has a simple form in the $\eta \to 0$ limit:

\begin{equation} \label{eq:A_k_om_delta_0}
A(\mathbf{k},0)
=\frac{\Delta^2}{2 \ln \xi}\frac{|\epsilon(\mathbf{k}))|}
{[\epsilon^2(\mathbf{k}) +\Delta^2]^2} .
\end{equation}
This expression has a maximum at:

\begin{equation} \label{eq:False_FS_eq_delta_0}
\epsilon(\mathbf{k})= \pm \frac{\Delta}{\sqrt{3}} .
\end{equation}
This results in two false Fermi surfaces that are mirror images of one another relative to the magnetic zone boundary (the $(\pi,0)-(0,\pi)$ line). One of these FS is hole-like and the other electron-like (see plots in \secref{sec:Case_P_0}). With hole doping, this symmetry is broken, but one of the false Fermi surfaces remains electron-like ($\epsilon(\mathbf{k}) < 0$) while the other becomes hole-like ($\epsilon(\mathbf{k}) > 0$) (see \secref{sec:Case_P1}). 

We can obtain an analytical result at small doping, $|\tilde{\mu}_2|,|\mu_1| \ll \Delta$. For simplicity, we also neglect $|\tilde{\mu}_2-\mu_1| \ll \Delta$ (i.e., $\tilde{\mu}_2 \approx \mu_1$). In this approximation:

\begin{equation} \label{eq:A_k_om_delta_sm}
A(\mathbf{k},0)=\frac{\Delta^2}{2 \ln \xi}\frac{|\epsilon(\mathbf{k})+\mu_1|} 
{[\epsilon^2(\mathbf{k}) +\Delta^2]^2}.
\end{equation}

The maximum lines of $A(\mathbf{k},0)$ are then given by:

\begin{equation} \label{eq:False_FS_eq_delta_sm}
\epsilon(\mathbf{k})= \pm \frac{\Delta}{\sqrt{3}}-\frac{2 \mu_1}{3}= 0 .
\end{equation}
Since $\mu_1 <0$, this results in a shift of the two false Fermi surfaces toward $(\pi,\pi)$ compared to the half-filled case. This is what we found numerically at small hole doping (see plots in \secref{sec:Case_P1}).

The above results describe the actual spectral function at zero real frequency, $A(\mathbf{k},0)|_{\eta \to 0}$. The behavior of the spectral function proxy $A(\mathbf{k},ik_0)$ studied in \cite{vsimkovic2024origin} differs in several ways, which we discuss below.

Analytical results for finite $\eta = \pi T$ are not possible in the general case; however, we can gain insight into the effect of finite $\eta$ by considering the extreme case $\eta \gg \frac{\Delta}{\ln \xi}$, i.e., when the broadening dominates the actual imaginary part of the self-energy $|\Sigma''(\mathbf{k},0)|$ due to thermal fluctuations. In this case, the self-energy away from the shadow FS, $|\omega_k| \gg v_{\mathbf{k}+\mathbf{Q}} \xi^{-1}$, has the form
\begin{equation} 
\begin{split}
\Sigma(\mathbf{k},\omega) &=\frac{\Delta^2}{\omega_k + i\eta} .
\label{eq:Sigma_large_w}
\end{split}
\end{equation}
The Matsubara spectral function proxy is particularly simple at half-filling, where it takes the following form
\begin{equation} \label{eq:A_k_0}
A(\mathbf{k},ik_0)
=\frac{1}{\pi}\frac{\eta} {[\epsilon^2(\mathbf{k}) +\tilde{\Delta}^2]} ,
\end{equation}
where $k_0=\eta=\pi T$ and $\tilde{\Delta}^2=\Delta^2+\eta^2$.
The above expression has a maximum in momentum space along a single line coinciding with the magnetic zone boundary $\epsilon(\mathbf{k})=0$. Thus, the effect of finite $\eta$ is to reduce the separation between the two false Fermi surfaces, bringing them both toward the $(\pi,0)-(0,\pi)$ line.

Now let us consider the same limit, $\eta \gg \frac{\Delta}{\ln \xi}$, but at small doping, $|\tilde{\mu}_2|,|\mu_1| \ll \Delta$, $|\tilde{\mu}_2-\mu_1| \ll \Delta$ and $\tilde{\mu}_2\sim\mu_1$.
In this case, the Matsubara spectral function proxy can be approximated as:

\begin{equation} \label{eq:A_k_0_delta_small}
A(\mathbf{k},ik_0)
=\frac{\eta}{\pi}\frac{(\epsilon(\mathbf{k})+\mu_1)^2 +\tilde{\Delta}^2} {[\epsilon^2(\mathbf{k}) +\tilde{\Delta}^2]^2+(2\mu_1\eta)^2} .
\end{equation}
In the above limit, this has a single solution, $\epsilon(\mathbf{k})=\mu_1$, for the maximum line of the spectral function proxy. Thus, the effect of finite broadening at small doping is to bring the two false Fermi surfaces together, close to the non-interacting, electron-like Fermi surface (see plots in \secref{sec:Effect_finite_eta}). 
On the other hand, when $\eta < \frac{\Delta}{\ln \xi}$, which is likely to be the case for large $U$ (large $\Delta$), the proxy $A(\mathbf{k},ik_0)$ behaves more like the true real-frequency spectral function $A(\mathbf{k},0)$, which does have a hole-like Fermi surface at low temperature.

Using \eref{eq:Sigma_large_w} and the approximations that led to \eref{eq:A_k_0_delta_small}, the ZEQ line equation is given by:

\begin{equation} \label{eq:ZEQ_line_large_eta}
\epsilon(\mathbf{k})-\mu_1 = - \Delta^2 \frac{\epsilon(\mathbf{k})+\mu_1} {[\epsilon(\mathbf{k})+\mu_1]^2 +\eta^2} .
\end{equation}
At low temperature, as $\eta=\pi T \to 0$, the right-hand side develops a discontinuity (a vertical line) at $\epsilon(\mathbf{k})=-\mu_1$, and thus the solution of this equation converges to $\epsilon(\mathbf{k})=-\mu_1$, i.e., the shadow Fermi surface (see also \fref{fig:Sigma_vs_ek_dif_T_n_0_913}). This explains the disconnect between the behavior of the Fermi surface and that of the ZEQ line for the Matsubara proxies when the broadening $\eta=\pi T$ dominates. The former is electron-like and close to the non-interacting Fermi surface, while the latter is hole-like and close to the shadow of the non-interacting Fermi surface (see plots in \secref{sec:Comp_with_DiagMC}).
 
\section{Summary and conclusions}
\label{sec:Summary}

In this paper, we used the TPSC+ approach to study the evolution of the pseudogap with temperature and doping in the single-band Hubbard model with nearest-neighbor hopping on the square lattice. Within this framework, precursors of AFM/SDW bands and the pseudogap arise from thermal critical spin fluctuations below the renormalized classical crossover temperature $T_{RC}(\delta)$, which replaces the finite-temperature phase transition in two dimensions. This is an intrinsic effect of low dimensionality, present for both weak and strong interaction, and it becomes more pronounced as $U$ increases.

To characterize this pseudogap, we simultaneously analyzed energy and momentum distribution curves directly in the real-frequency domain. We found that, in the pseudogap regime, a maximum in the momentum dependence of the spectral function $A(\mathbf{k},0)$ can coincide with a minimum or strong suppression at zero frequency in the energy distribution curves $A(\mathbf{k}_{\mathrm{max}},\omega)$. This is the opposite of the behavior expected in a Fermi liquid or marginal Fermi liquid, rendering standard Fermi-liquid terminology inadequate. Motivated by this observation, we introduced the concepts of false Fermi surfaces and false ZEQ lines, which provide a consistent framework for describing the pseudogap regime. The false Fermi surface has been observed experimentally in electron-doped high-$T_c$ materials~\cite{xu2023bogoliubov,kokkinis2025pseudogap}, where it was referred to as a “gossamer Fermi surface.”

We then analyzed the transformation of the topology, shape, and character (true vs. false) of the Fermi surface and the ZEQ line as functions of doping and temperature. Above the renormalized classical regime ($T_{RC} < T \ll W$), Luttinger's theorem holds approximately, and the real part of the self-energy $\Sigma'(\mathbf{k},0)$ depends only weakly on momentum; consequently, the interacting FS practically coincides with the non-interacting one (the quantum fluctuation regime). As the system is cooled below $T_{RC}$, classical thermal spin fluctuations dominate, leading to the development of a quasi-singularity in the self-energy along the shadow Fermi surface.

The effects of these classical thermal spin fluctuations on the single-particle spectra differ qualitatively, depending on whether the fluctuations are commensurate or incommensurate.

In the commensurate case, the antinodal pseudogap always opens first, well before the pseudogap develops along the Brillouin zone diagonal. As temperature decreases, the segment of the Fermi surface near the BZ boundary becomes false and splits into two lines: one electron-like and one hole-like. At intermediate temperatures, true Fermi arcs persist around the zone diagonals, but as the system cools further, these too eventually turn into false Fermi surfaces.

The ZEQ line undergoes a profound transformation as well. It starts as an electron-like contour that coincides with the non-interacting Fermi surface (the Luttinger-theorem regime), then flattens, and ultimately becomes hole-like. In the limit $\xi \to \infty$, it converges asymptotically to the shadow FS, along which the real part of the self-energy exhibits a discontinuity. During this transformation, the nature of the ZEQ line changes from true to false, beginning at the BZ boundary and propagating toward the zone diagonal.

In the incommensurate case, the hierarchy between the antinodal and diagonal pseudogaps reverses. The antinodal pseudogap temperature $T_{AN}(\delta)$ vanishes with decreasing doping well before the quantum critical point is reached ($\delta_{AN} < \delta_{\mathrm{QCP}}$). Below $\delta_{AN}$, the spectral function at the antinode remains non-trivial, exhibiting two precursor AFM/SDW bands; however, both peaks lie at positive energies ($\omega > 0$) and are thus invisible to ARPES. Conversely, the diagonal pseudogap persists for $\delta > \delta_{AN}$. This behavior occurs because incommensurate fluctuations produce hot spots close to the diagonals, where the pseudogap survives all the way down to the QCP.

The existence of a hole-like FS and ZEQ line in the $t'=0$ model, whose non-interacting FS is electron-like, is a rather remarkable consequence of thermal spin fluctuations in two dimensions. Depending on doping and temperature, the system can be either metallic or insulating, with a complex FS and ZEQ line structure.
\begin{enumerate}
\item
At low doping and low temperatures, both hole-like FS and ZEQ line are false (see \fref{fig:Akw_panel_heatmap_vers_T_n_0_913}, panel for $T=0.08$). The system is insulating because there is a pseudogap everywhere near $\omega=0$, implying an increasing resistivity with decreasing temperature.
\item
At somewhat higher temperature or lower doping, the system is metallic, with true Fermi arcs close to the diagonal and false hole-like Fermi surface segments close to the edge of the BZ (see \fref{fig:Akw_panel_heatmap_vers_T_n_0_913},  panel for $T=0.125$). At even higher doping, the true hole-like ZEQ line can extend all the way to the edge of the BZ (see \fref{fig:Akw_panel_heatmap_vers_T_n_0_88}, panel for $T=0.09$).
\item
Finally, in the incommensurate case at low temperature, the system has a false FS and ZEQ line close to the diagonal, but a true segment of the FS and ZEQ line at the edge of the BZ (see \fref{fig:Akw_panel_heatmap_vers_T_n_0_88}, panel for $T=0.04$). The latter segment is due to the middle SDW band in the EDC spectra, which is absent in the commensurate case, where only the lower and upper precursors of the AFM bands are present.
\end{enumerate}

In addition to studying the real-frequency spectral function $A(\mathbf{k},\omega)$ and the density of states, we compared our TPSC+ results for the Matsubara proxy $A(\mathbf{k},ik_0)$ with unbiased DiagMC calculations \cite{vsimkovic2024origin}. We found that TPSC+ tends to underestimate the strength of pseudogap signatures in $A(\mathbf{k},ik_0)$ in the strong-interaction regime compared to DiagMC, but gives a qualitatively similar description. Specifically, TPSC+ produces heatmaps similar to those of DiagMC at somewhat lower temperatures or reduced doping. TPSC+ also reproduces qualitatively similar spectral features at the same temperature and doping when evaluated with a slightly smaller artificial broadening parameter (e.g., $\eta=0.5 k_0=0.5\pi T$).

We also identified significant differences between the real-frequency spectral function $A(\mathbf{k},0)$ and its Matsubara proxy $A(\mathbf{k},ik_0)$ in the pseudogap state, arising because $k_0=\pi T$ is not the smallest energy scale in this regime. Most prominently, in the Matsubara proxy $A(\mathbf{k},ik_0)$, the hole-like FS appears only in the strong-interaction limit, whereas in the real-frequency spectral function, a hole-like FS emerges across all interaction strengths. In other words, the phenomenology identified in Ref.~\cite{vsimkovic2024origin} exists for both strong and weak interaction, provided the temperature is sufficiently low.

We conclude by discussing the relevance of our findings to high-$T_c$ cuprates. When the temperature drops below the diagonal pseudogap temperature $T_{D}$, the pseudogap suppresses the portion of the Fermi surface with the largest Fermi velocity, which should promote insulating-like behavior with increasing resistivity. While this suppression might be modified by pairing instabilities in the particle-particle channel, our results remain indirectly relevant to experiments on high-$T_c$ materials in strong magnetic fields, where superconductivity is quenched. Under such conditions, an upturn in resistivity has indeed been widely observed \cite{boebinger1996insulator,ono2000metal, Ando_2004, laliberte2016origin, Proust_Taillefer_2019}. However, a realistic description of cuprate physics requires incorporating finite next-nearest-neighbor hopping $t'$ since this can strongly modify the shape of the non-interacting Fermi surface, hence the location of hot spots ~\cite{Kyung_Hankkevych_2004, senechal2004hot} and the intersection with the AFM Brillouin zone. This will be investigated in future work.

\section*{Acknowlegments}
We thank A.-M. S. Tremblay for stimulating discussions, suggestions and comments on the manuscript. We are indebted to Chloé Gauvin-Ndiaye and Camille Lahaie, who wrote the initial version of the Python program that we used.

\bibliography{Biblio.bib}	
\end{document}